%% file: eses.tex
\documentclass[twocolumn,resetfootnote]{aastex701}
\usepackage{amsmath}
\usepackage{amssymb}
\usepackage{tabularx}
\usepackage{hyperref}

\shorttitle{VAST sample of ESEs}
\shortauthors{Anumarlapudi et al.}
\graphicspath{{./}{figures/}}

\begin{document}

\title{How common are extreme scattering events?: First large sample from the ASKAP VAST survey}

\correspondingauthor{Akash Anumarlapudi}
\email[show]{akasha@unc.edu}

\author[orcid=0000-0002-8935-9882,sname='Anumarlapudi ']{Akash Anumarlapudi}
\affiliation{Department of Physics and Astronomy, University of North Carolina at Chapel Hill, 120 E. Cameron Ave, Chapel Hill, NC, 27599, USA}
\email{akasha@unc.edu}  

\author[orcid=0000-0003-3236-8769]{Dylan~L.~Jow}
\affiliation{Kavli Institute for Particle Astrophysics \& Cosmology, P.O. Box 2450, Stanford University, Stanford, CA 94305, USA}
\email{dylanjow@stanford.edu}  

\author[]{Artem Tuntsov}
\affiliation{Manly Astrophysics, 15/41-42 East Esplanade, Manly, NSW 2095, Australia}
\email{artem.tuntsov@manlyastrophysics.org} 

\author[orcid=0000-0001-6247-3071]{Hayley Bignall}
\affiliation{Manly Astrophysics, 15/41-42 East Esplanade, Manly, NSW 2095, Australia}
\affiliation{Visiting Scientist, CSIRO, Space and Astronomy, P.O. Box 1130, Bentley, WA 6102, Australia}
\email{hayley.bignall@manlyastrophysics.org} 

\author[orcid=0000-0001-6295-2881]{David L. Kaplan}
\affiliation{Department of Physics and Astronomy, University of Wisconsin-Milwaukee, P.O. Box 413, Milwaukee, WI 53201, USA}
\email{kaplan@uwm.edu}

\author[0000-0003-0699-7019]{Dougal Dobie}
\affiliation{Sydney Institute for Astronomy, School of Physics, University of Sydney, NSW 2006, Australia}
\affiliation{ARC Centre of Excellence for Gravitational Wave Discovery (OzGrav), Australia}
\email{d.dobie@sydney.edu.au}

\author[0000-0002-3382-9558]{B. M. Gaensler}
\affiliation{Department of Astronomy and Astrophysics, University of California, Santa Cruz, 1156 High Street, Santa Cruz, CA 95064, USA}
\affiliation{Dunlap Institute for Astronomy and Astrophysics, University of Toronto, 50 St. George Street, Toronto, ON M5S 3H4, Canada}
\affiliation{David A.\ Dunlap Department of Astronomy and Astrophysics, University of Toronto, 50 St. George Street, Toronto, ON M5S 3H4, Canada}
\email{gaensler@ucsc.edu}

\author[orcid=0000-0002-8977-1498]{Igor~L.~Andreoni}
\affiliation{Department of Physics and Astronomy, University of North Carolina at Chapel Hill, 120 E. Cameron Ave, Chapel Hill, NC, 27599, USA}
\email{Igor.Andreoni@unc.edu}  

\author[0000-0002-4405-3273]{Laura Driessen}
\affiliation{Sydney Institute for Astronomy, School of Physics, University of Sydney, NSW 2006, Australia}
\email{laura.driessen@sydney.edu.au}

\author[0000-0002-2686-438X]{Tara Murphy}
\affiliation{Sydney Institute for Astronomy, School of Physics, University of Sydney, NSW 2006, Australia}
\affiliation{ARC Centre of Excellence for Gravitational Wave Discovery (OzGrav), Australia}
\email{tara.murphy@sydney.edu.au}

\author[0000-0003-1575-5249]{Joshua Pritchard}
\affiliation{Australia Telescope National Facility, CSIRO Space and Astronomy, PO Box 76, Epping, NSW 1710, Australia}
\email{joshua.pritchard@csiro.au}


\begin{abstract}
Compact radio sources are occasionally observed to undergo extreme scattering events (ESEs): large, chromatic flux-density variations, often by more than a factor of two, over a few months. These were interpreted to be a smooth plasma lensing phenomenon due to $\sim {\rm AU}$-structures in the ISM. Although ESEs were discovered decades ago, observationally, only a few tens of ESEs have been reported, often from individual sources monitored at very high cadences. As such, the nature of the plasma lenses generating such events remains elusive. In this article, we present the most comprehensive sample of ESEs to date, using the ASKAP Variables and Slow Transients (VAST) survey. We report the discovery of a three-tier sample --- 72 sources that show promising ESE events (gold), 76 sources that are likely ESE-exhibiting (silver), and an additional 88 sources where ESE events are one plausible explanation (bronze). We show that the light curves demonstrate varied morphologies, which can be interpreted in terms of the source size and the strength of the lens in the paradigm of a simple one-dimensional Gaussian lens model. The spatial distribution of ESEs shows an appreciable excess in the direction of the Galactic center and along the Galactic loops. Using this sample, we infer an overall rate of 0.012 such events per year per deg$^2$ (or 0.006 events per source per year) for our (gold+silver) sample, for sources with mean flux density above 5\,mJy at 888\,MHz (0.01\% of all isolated point sources detected in VAST).
\end{abstract}

\section{Introduction} 
Radio waves undergo a variety of propagation effects as they encounter ionized plasma in the interstellar medium (ISM), which induces both temporal and spectral variations on the incident radiation. These include \textit{dispersion} \citep{hewish_observation_1968}, \textit{interstellar scintillation} \citep{scheuer_amplitude_1968, rickett_radio_1990}, and \textit{angular/temporal broadening} \citep{angularbroadenind,bhat_multifrequency_2004}. Dispersion, commonly observed in pulsars, is predominantly seen as a characteristic spectral sweep due to a frequency-dependent refractive index. Scintillation can and is seen to cause moderate flux variations (often around 10\%--100\%) in the observed light curve. Angular broadening refers to a change in the angular size of the background source, while temporal broadening refers to an increase in the temporal scale of the emission\footnote{This is noticeable easily in pulsars and fast radio bursts.}. Both scintillation and angular/temporal broadening are caused by scattering of the radio waves in the ionized ISM. Another particularly striking and extreme manifestation of radio scattering comes in the form of \textit{extreme scattering events} (ESEs), where extreme changes in flux density are seen over a few days to months. Despite being relatively understudied, ESEs are broadly thought to be caused by smooth plasma lenses at $\sim$AU scales in the ISM. To date, only a few tens ($\approx$20) of candidate ESE events, confirmed and speculative, have been reported \citep{fiedler_extreme_1987,cognard_extreme_1993,fiedler_summary_1994,fey_vlbi_1996,lazio_extreme_2000,Senkbeil2008,pushkarev_vlba_2013,coles_pulsar_2015,bannister_real-time_2016,Kerr_2018}. 

ISM propagation effects at radio wavelengths arise due to the frequency-dependent refractive index of the ionized medium: $n (\omega) = \sqrt{1 - \frac{\omega_p^2}{\omega^2}}$, where $\omega_p^2=n_e e^2/m_e\epsilon_0$ is the plasma frequency given the local free-electron density $n_e$, and $\omega$ is the observing frequency. Dispersion arises as the lower frequencies propagate more slowly through an ionized medium, leading to the characteristic frequency sweep. Inhomogeneities in the local electron density (and hence the refractive index) result in rays arriving at the observer via different paths, leading to a chromatic interference pattern at the observer. This chromatic interference pattern is referred to as scintillation. The term \textit{diffractive interstellar scintillation} (DISS) is often used to refer to the fast ($\approx$min--hours), large-amplitude (mean of 100\%), narrow-band variations in the observed flux. The same scattering material responsible for scintillation can also lead to angular and temporal broadening of the source. If the turbulent ISM is isotropic and volume-filling over large length scales, observations of pulsars suggest that the spatial scales responsible for DISS may extend down to $\mathcal{O}{(10^{-5})}\,{\rm AU}$  \citep{narayan_physics_1992,bhat_multifrequency_2004}. \textit{Refractive interstellar scintillation} (RISS), on the other hand, refers to long ($\sim$days to months), small-amplitude (few tens of percent), wide-band variations in the observed flux density. RISS depends on the distribution of similar density inhomogeneities that deflect the rays into/out of the line of sight, and likely probe these on larger spatial scales ($\sim$AU scale). However, it has also been argued that at least some pulsar scintillation observations which might be categorized as DISS are the result of smooth, intermittent density inhomogeneities on $\sim$AU scales more commonly associated with RISS \citep{Brisken2010,pen_pulsar_2014,Jow2023regimes}.

 In the case of ESEs, the nature of the underlying density perturbations and their characteristic sizes is even less well established. Simple causal arguments from the transitional timescales\,---\,i.e the timescales on which extreme flux density variations are observed\,---\,imply a perturbation scale of $\gtrsim$1\,AU, similar to the case of refractive scintillation. However, regular, large-amplitude ($>$ a factor of 2) variations in the ESE light curves strongly suggest the presence of coherent focusing and de-focusing regions in the intervening medium. Hence, while the length scales are similar, these structured variations make the underlying density perturbations act as a single smooth lens, as opposed to the stochastic variations seen in refractive scintillation.

Despite having been discovered almost four decades ago, the physical origin of ESEs remains mysterious. Simple spherical lens models of the plasma lenses require that the electron density of the structures be many orders of magnitude larger than the ambient warm ionized medium (WIM) \citep{fiedler_extreme_1987,romani_radio_1987,Clegg1998}. However, it is argued that such over-pressured structures are not common enough and can not survive long enough compared to the observed ESE timescale to explain the ESE phenomenon. Various proposals have been put forward to solve this over-pressure problem, including gravitationally-stable dense clouds \citep{walker_extreme_1998,walker_interpretation_2001}, elongated clouds/structures \citep{romani_radio_1987,dong_extreme_2018}, two-dimensional current sheets viewed edge on \citep{goldreich_folded_2006,pen_refractive_2012,pen_pulsar_2014}, and more complex surface geometries \citep{jow_cusp_2024}. However, given the limited number of discoveries, the origin of ESEs still remains poorly understood. Addressing this could have important implications for our understanding of the ISM, as ESEs can uniquely probe the turbulent medium on these tiny AU scales, which are difficult to probe otherwise. 

One of the reasons for the very limited sample size of ESEs is the lack of repeated scans of large parts of the extragalactic sky with a high cadence at radio wavelengths. Currently, almost the entire sample of known ESEs was found in the direction of select extragalactic compact sources or individual pulsars that were observed on a daily/weekly cadence. While this has produced extremely well-sampled light curves, this limitation has left questions about the nature and prevalence of such events unanswered. Current wide-field, all-sky radio surveys like the Variables and Slow Transients survey \citep[VAST;][]{murphy_vast_2013,murphy_askap_2021} at the Australian SKA Pathfinder  \citep[ASKAP;][]{askap} are in a prime position to discover a large sample of ESEs \citep{murphy_vast_2013}. VAST is a survey science project that scans the Galactic sky every 2\,weeks and the extragalactic sky once every 2\,months down to a sensitivity of 250\,$\mu$Jy at 888\,MHz. While most of the targeted surveys for ESEs and the discoveries themselves are performed at GHz frequencies (2.7\,GHz or higher, \citealp{fiedler_summary_1994}), VAST opens up the sub-GHz frequency range, where ESEs are expected to be more pronounced and may be long-lived compared to few-GHz frequencies.

In this article, we conduct a search for new ESE candidates using the extragalactic component of the ASKAP VAST data. This article is organized as follows: \S\ref{sec:what_are_eses} motivates the defining characteristic of ESEs from an observational point of view and outlines our selection criteria. A brief summary of our data set, search techniques, and methods is given in \S\ref{subsec:sample}--\S\ref{subsec:spectra}. We provide the results in \S\ref{sec:results} and discuss various implications of these results in \S\ref{sec:discussion}, before concluding in \S\ref{sec:conclusions}.

\section{Defining ESE\MakeLowercase{s}} \label{sec:what_are_eses}

From a theoretical point of view, ESEs can be cleanly described as plasma lensing events. Observationally, discrete sampling at a single frequency over a limited bandwidth means the temporal signature of ESEs can easily be contaminated by other phenomena, such as refractive scintillation and/or intrinsic variability from the background source. As a result, most ESE candidates so far have been visually selected by looking at long-term (multi-year) light curves \citep{fiedler_extreme_1987,fiedler_summary_1994,lazio_extreme_2000}. These were identified as flux depressions in the light curve that are \textit{unique} and \textit{atypical} compared to long-term variations. The presence of such features at multiple frequencies and their relative behavior was used to infer the chromatic nature of these events and has been understood as different from regular scintillation effects/AGN variability. However, a precise quantitative definition (e.g, the amplitude of the variations, their timescale, etc) is lacking, partially due to the incomplete understanding of such events and the plasma lenses that cause them. Current and upcoming all-sky surveys, which have the potential to discover a large sample of such events, are/will be performed at a single observing frequency over limited bandwidths and hence will not have such wide (1--10\,GHz) frequency coverage in most cases. In addition, the cadence of these surveys will be sparser than the historical observations (daily/weekly) which will make it difficult to isolate extreme scattering from other contaminants. The lack of multi-year light curve history for these sources will also mean that the \textit{uniqueness} of the identified features can't be readily established. This will have heavy implications for any ``unbiased'' sample selection. Hence, we will first discuss common sources of confusion and motivate a set of criteria that we used to filter ESE candidates.

\subsection{Sources of confusion}\label{subsec:confusion}
The primary sources of confusion for ESEs are interstellar scintillation and intrinsic AGN variability. Over the timescales corresponding to the VAST cadence, RISS is the propagation effect that offers a source of confusion. Flux density variations from RISS are usually modest (i.e. a few tens of percent; \citealp{stinebring_five_2000,lovell_first_2003,rickett_interstellar_2006}).  Extreme variability, observed over a few months, can point to ESE candidates. Hence, a basic selection criterion should include one of large flux density variation. Here we require the ratio of maximum to minimum flux density being $>$2, in order to distinguish ESEs from RISS. It is worth noting that ESEs and RISS may represent a continuous spectrum of plasma-lensing-induced variability, rather than two distinct phenomena. Very long baseline interferometric observations of pulsar scintillation have challenged the simple picture that diffractive interstellar scintillation (DISS) is sourced by fluctuations on distinct scales from RISS \citep{Brisken2010}. It is possible that both scintillation and ESEs represent a spectrum of lensing signatures from the same underlying physical structures, which would make a theoretical description complete. But this poses challenges observationally, problematizing simple observational distinctions between the phenomena.

In the case of intrinsic variability, variable radio emission in AGNs usually arises from the jet core \citep{marscher_models_1985}, the inner jet (usually $<$\,pc, \citealp{hughes_polarized_1985,zensus_parsec-scale_1997}), and the jet knots \citep{marscher_inner_2008}. Emission from more extended kpc jets and lobes \citep{celotti_kinetic_1993,blandford_relativistic_2019} is not expected to show significant variability \citep{ulrich_variability_1997}, i.e $>$100\%, particularly on sub-year timescales. A limit on the emitting region can also be inferred from causal arguments\,---\,2$\Gamma^2ct$, where $\Gamma$ is the bulk Lorentz factor (typically $\approx$10; \citealp{lister_mojave_2019}), $c$ is the speed of light, and $t$ is the time difference between observations. This yields a limit of $\approx$10\,pc that can produce temporal variations on the timescale of VAST cadence. 

Within the inner pc scales, emission from lower frequencies is proposed to come from larger radii \citep{marscher_models_1985}, since it is self-absorbed close to the jet core. As a result, light curve variability is expected, but found to be less pronounced (or at similar levels) at lower frequencies, $\sim$\,GHz, compared to higher frequencies \citep{aller_pearson-readhead_1992}. Hence, a pronounced variability at lower frequencies ($\approx$\,GHz) compared to higher frequencies (a few GHz) more likely points towards a propagation effect rather than an intrinsic origin, and we include this as a selection criterion.

\subsection{Selection criteria}\label{subsec:fps}

While the two criteria mentioned above --- large flux density variations and pronounced low-frequency variations --- can help in narrowing down plausible ESE candidates, we also require that the temporal and spectral evolution follow the current understanding. In the current ESE sample, the light curves of the ESE candidates are reasonably described by a simple 1-D Gaussian lens model, particularly in highlighting the key features in the light curves (a detailed description of the lens model is provided in \S\ref{subsec:lc_model}). Hence, we require that the light curve is well described by a 1-D Gaussian lens. While a Gaussian lens qualitatively captures the essential features, we note that there is no robust evidence or strong expectation for all the plasma lenses to be 1-D Gaussians. Hence, strict quantitative thresholds on goodness of fit can often reject promising ESE candidates. Given the limitations of the theoretical models, we use a combination of quantitative metrics and qualitative inspection to select ESE candidates.

In addition, spectral variations are expected alongside temporal variations as the background source undergoes the scattering event. While the bandwidth of VAST is too narrow (288\,MHz) to conclusively establishing underlying spectral variations, the presence of simultaneous spectral variations is used as a source of additional confidence for the ESE candidates (detailed description is provided in \S\ref{subsec:spectra}). With this, we use the following four criteria to select promising ESE candidates:

\begin{enumerate}
    \item \textbf{Large flux density variations.} The ratio of maximum to minimum flux density should be $\geq$2. This is to preferentially select against refractive scintillation.
    \item \textbf{More pronounced behavior at 888\,MHz compared to GHz frequencies.} Given the lack of multi-epoch data from ASKAP at higher frequencies, we compare the 888\,MHz VAST data with 3\,GHz high-frequency VLASS data to select ESE candidates\footnote{We also note the limitation that the VLASS data is sparser (3\,yr cadence) than VAST data, but this is unavoidable given the lack of other data sets.}. While the exact ratio of the fractional variability (VAST to VLASS) depends on complex factors, including intrinsic variability, the source's spectral behavior, and the behavior of plasma lenses at different frequencies, we simply require that this is $>1$.
    \item \textbf{Agreement with a simple Gaussian lens model.} We select candidates in which the light curve shows key features that are captured by this model. As described in \S\ref{sec:results}, we use a combination of quantitative cuts and qualitative assessment to filter sources.
    \item \textbf{Simultaneous spectral variations}. We preferentially select candidates that show spectral variations as the candidate undergoes the scattering event. As described above, the precise spectral evolution depends on the nature of the background source and the lens model. Hence, we proceed model-agnostically by comparing the characteristic scale of variation in the spectral index time series. Structured variations will have an intermediate length scale compared to stochastic (rapidly varying) and absence of variations (very long-term varying). 
\end{enumerate}

While this will not result in a truly unbiased sample selection, our goal here is to select all the candidates for which a plasma lensing event can be reasonably established. This will prioritize reliability over completeness. More advanced model-dependent filtering to uncover a complete sample of ESEs is left to future work.

\section{Methods}\label{sec:methods}
\subsection{Initial sample selection}\label{subsec:sample}
The primary data set used in this article stems from the extragalactic data taken as part of the ASKAP VAST survey \citep{murphy_vast_2013,murphy_askap_2021}. All VAST data were processed using \textsc{ASKAPsoft} \citep{guzman_askapsoft_2019} to produce continuum images, retaining full polarization information. \textsc{selavy} \citep{selavy}, a dedicated source finder for ASKAP data, was used to extract sources detected in the images with $>5\sigma$ significance. Sources detected in individual epochs and images were collated by \textsc{vasttools} \citep{vasttools_2024} to perform source association and produce light curves. A detailed description of these methods is given by \cite{de_ruiter_askap_2026}.

The total number of un-vetted sources in the first data release is 2,279,209. We use a variety of quality cuts\,---\,i) minimum signal-to-noise ratio (SNR $>$8), ii) ratio of integrated to peak flux density (\textit{compactness}$<$1.5), and iii) the proximity to other sources ($>$30\arcsec)\,---\,to select bright, isolated point sources, resulting in a sample of 24,590 sources. The next step is to look for variability in the light curve. Traditional methods often employ point statistics such as $\eta$ (the reduced $\chi^2$) and $V$ (fractional modulation) statistics \citep{swinbank_lofar_2015}, which quantify the deviation of a light curve from the steady source model. However, these often fail to distinguish between stochastic variations and structured variations. Hence, we employed a machine learning (ML) classifier to preferentially filter out stochastic variations and yield a sample of systematically varying sources. A detailed description of our ML pipeline is provided in Appendix~\ref{sec:app_ml}. This yielded an initial sample of 4,446 sources. As a first pass, we visually selected highly variable light curves that \textit{qualitatively} resemble ESEs (based on the expectations from a Gaussian lens). This yielded an initial sample of 304 candidates. Further classification and filtering involve multi-frequency light curves as well as light curve and spectral modeling, which we detail in the following sub-sections \ref{subsec:lcs}-\ref{subsec:spectra}. A summary of this is provided in Table~\ref{tab:filt}.

\begin{deluxetable}{lr}
    \centering
    \caption{Summary of the filtering steps leading to the initial sample of ESE candidates.\label{tab:filt}}
    \tablehead{\colhead{Filtering step} & \colhead{Sample remaining}}
    \startdata
    Total number of sources & 2,279,209 (100\phantom{.0}\%)\\
    Isolated, point sources & 1,480,705 (\phantom{0}65\phantom{.0}\%)\\
    Bright, isolated sources & 24,590 (\phantom{00}1.1\%)\\
    ML filter & 4,446 (\phantom{00}0.2\%)\\
    Initial sample & 304 (132 ppm) \\
    \enddata
\end{deluxetable}

\subsection{Multi-frequency data} \label{subsec:lcs}
While the initial sample selection was done using the extragalactic VAST data, we also use all publicly available ASKAP data, taken under other survey science projects, to augment our data set. These include data taken at frequencies spanning 855--1656\,MHz. In addition, we also use data from the Very Large Array Sky Survey \citep[VLASS; ][]{vlass} to examine the light curve behavior at 3\,GHz. For all these data, we acquire the individual images and estimate the flux density at the source's position. We fit a 2D Gaussian, using the synthesized beam information, to these point sources to extract the flux density measurements. We select sources in which the variation is more pronounced in the VAST data compared to VLASS data, thereby preferentially selecting events that are likely due to propagation effects. 

\subsection{Light curve modeling (and limitations)} \label{subsec:lc_model}

Based on the prototypical ESE \citep{fiedler_extreme_1987} and subsequent reported events \citep{fiedler_summary_1994,lazio_extreme_2001}, a few characteristic features in the light curve have been identified. This includes a pronounced flux depression flanked by enhancement on either side of it. While the exact nature of the lenses causing ESEs remains poorly understood, a simple 1D model of an over-dense plasma lens, described by \cite{Clegg1998} successfully replicates these features, identifying the depression with its ``de-focusing'' center and enhancement with ``focusing'' (and, potentially, formation of ``caustic spikes'') at its sides \footnote{While in this article, we reserve ourselves to over-dense Gaussian lens, we also note that alternative lens models (e.g., under-dense lenses) and their interpretations (e.g., over-focusing at the center) are not ruled out currently and are possible.}. Hence, we follow \cite{Clegg1998} to model the observed 888\,MHz light curves using a Gaussian lens profile. Complete details of the model can be found in the original article; a brief description and our adaptation of the model are provided in Appendix~\ref{app:lenses} for interested readers. The model consists of six parameters that can fully describe the light curve --- i) strength of the lens $\alpha$, ii) angular extent of the source $\beta_{\rm src}$, iii) relative velocity between the lens and the observer $v$, iv) fraction of the source's flux density that is unaffected by the lens $f_{\rm unlens}$, v) total flux density prior to the lensing event $F_{\nu; \rm tot}$, and vi) a reference time $T_0$. 

While we model the light curves with a 1D Gaussian lens model, there are major caveats that are worth highlighting. There is no robust observational evidence for the plasma density variations to follow a Gaussian distribution. Indeed, due to the lack of a large sample of such events, these density variations are poorly constrained currently. Precise modeling requires the knowledge of true plasma density variations in the ISM. Appendix~\ref{app:lenses} also provides alternate lens profiles that can result in different light-curve shapes and morphologies. Additionally, it is not necessary that all the plasma lenses in the ISM follow a single description, and hence not all the ESE light curves have to be self-similar. Compounding this, the Gaussian lens model described is valid only in the geometric optics regime, and hence any diffractive effects, particularly in the strong focusing regions, i.e caustics, will be completely unmodeled. 

These limitations in theoretical models and the presence of additional variability mean that simple statistical tests based on the goodness of the fit of these models will be inadequate in many cases and can result in rejecting sources that are promising candidates. To address this, we assume that the data are described by the combination of a lens model and correlated noise and fit both of them simultaneously (more details about the fit are provided in Appendix~\ref{app:lc_fit_noise}). This correlated noise can represent deviation from a perfect Gaussian lens, scintillation, intrinsic variability, or any combination of these. The amplitude (1-$\sigma$) of this correlated noise is assumed to be less than 2\% of the maximum flux density, and the length scale is assumed to be less than 120\,days (span of 3 VAST observations). These assumptions were made since such a red-noise component is complementary to extreme longer-term ESE variations, and allowing for large amplitude and/or longer-term red noise will absorb the power from the ESE (see Figure~\ref{fig:vastj2011-3358} for example). We then quantitatively describe the model accuracy by using the likelihood in the presence of this correlated component. 

While we quantitatively select candidates using the above-mentioned criterion, we also select candidates where the light curve shows the expected features but has a lower likelihood value. From our inspection, this comes primarily due to unmodeled caustics or outliers prior to the ESE event that are larger than the assumed noise model. 

\subsection{Spectral variations} \label{subsec:spectra}

To explore the presence of spectral variations, we generated time- and frequency-resolved spectra\,---\,\textit{dynamic spectra} (DS)
---\,for all observations. Starting from calibrated visibilities, we first generated the background sky model. We used \textsc{Wsclean} \citep{offringa-wsclean-2014,offringa-wsclean-2017} to model the field around the source of interest and generate clean components. In select cases, we cleaned a small region around very bright ($>$ a few Jy) and distant (a few degrees) sources, which can still produce significant contamination at the target's location. After the initial round of cleaning, we carried out one cycle of self-calibration to correct for small errors in the phases and then produced a deep-cleaned model of the field. We then subtracted these model visibilities from the observed visibilities to produce background-subtracted visibilities. Finally, we shifted the phase center to the target's location and averaged the emission over all the baselines to produce the DS. We then averaged the data over the temporal axis to generate 1D frequency spectra for all the individual observations.

In general, the bandwidth is sufficiently small (288\,MHz) that most of the spectra from the VAST data can be modeled as power-law functions. Hence, we fit a power law, $S_{\nu} \propto \nu^{\alpha}$, to extract the power-law index, $\alpha$, as a proxy to monitor changes to the frequency spectra as the source undergoes an ESE\footnote{Such spectral index time series measurements are not available by default and the generating these for every source is computationally very expensive. Hence, these were generated only for the times when the source was identified to be going through a potential scattering event.}. As mentioned in \S\ref{sec:what_are_eses}, variations in spectral time series are heavily dependent on the lens model and other competing effects; hence, we do not perform a model-based fit for spectral variations. Rather, we simply look for the presence of structured variations in the potential ESE candidates. Using the extracted spectral index time series, we try to estimate the characteristic length scale over which these variations occur using a Gaussian process (more details on the fit are provided in Appendix~\ref{app:spec_fit_gp}). This will help us distinguish whether the spectral index varies stochastically or in a structured manner. We normalize the inferred time scales to the time-span of the actual observation, so that a time scale of $>1$ refers to variability on a time scale greater than the duration of the observation itself, which we consider to be an absence of variation. We also filter out time scales of  $<0.05$ (rapid variations). The lower limit of 0.05 is chosen given the cadence (60 days) and the data span ($\approx$ 1200 days) as they represent sub-epoch-to-epoch variations.

\section{Results} \label{sec:results}

Given the initial sample of 304 sources, we apply all four criteria described in \S\ref{sec:what_are_eses} to select promising ESE candidates and divide the sample into three different sub-samples, which we call the \textit{gold}, \textit{silver}, and \textit{bronze} samples. The gold sample consists of sources in which all four criteria are satisfied. The silver sample has sources in which the light curve has all the necessary features expected from a plasma lens, but does not show strong evidence for spectral variations in the VAST data --- satisfying only three of the four criteria. Sources in which the light curve has additional variability that cannot be simply explained by a single lens are classified under the bronze sample. Below, we expand on this classification.

\subsection{Gold sample}
The gold sample represents sources in which an ESE interpretation can be best established, as both temporal and spectral variations are seen. However, as described in \S\ref{subsec:lc_model}, a fully quantitative approach can result in biases given the limitations in lens models. Hence, we adopt a combination of a quantitative approach and a qualitative approach to select sources in the gold sample.

\subsubsection{Quantitative selection}\label{subsec:quant}

\begin{figure}
    \centering
    \plotone{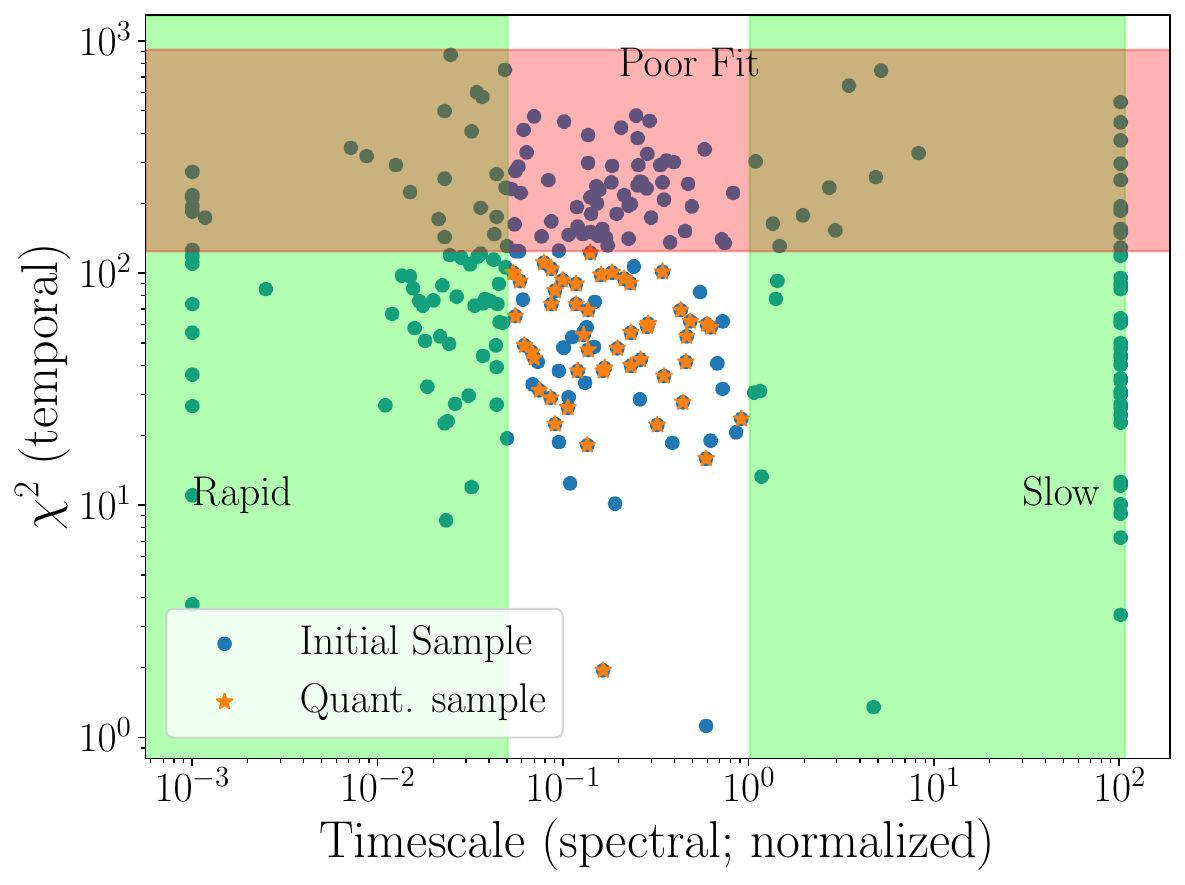}
    \caption{Temporal and spectral properties of the initial ESE sample: The abscissa shows the characteristic timescale in the spectral index time series (see \S\ref{subsec:spectra}), and the ordinate shows the $\chi^2$ (correlated) of the model fit to the light curve. The timescale of the spectral data is normalized between 0 and 1. The timescale of the kernel is capped at 100. The orange points show the quantitative sample described in \S\ref{subsec:quant}.}
    \label{fig:ese_quant_sep}
\end{figure}

\begin{figure}
    \centering
    \includegraphics[width=1\linewidth]{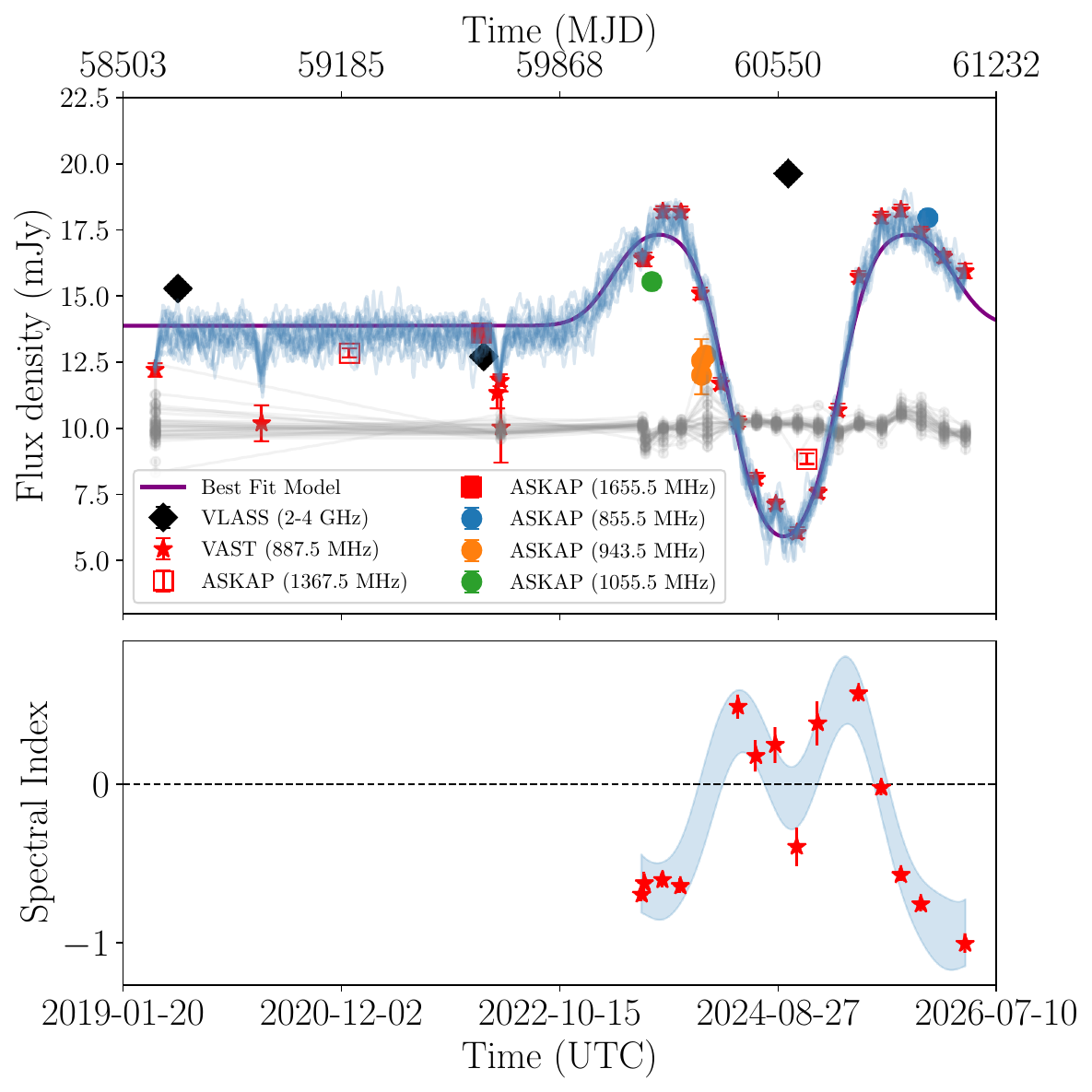}
    \caption{ESE candidate ASKAP J201134$-$335857: \textit{Top:} Shown are the multi-frequency light curves using the observations from VAST, VLASS, and other contemporaneous ASKAP surveys. The solid purple line shows the best-fit 1D-lens model fit to the data. The blue curves show 20 different random samples from a joint (lens+noise) model fit (described in \S\ref{subsec:lc_model}). Also shown as the gray scatter in the background are the light curves for various bright ($>$ 15\,mJy) sources that are within a degree of the target, to rule out instrumental variations. \textit{Bottom:} The evolution of the spectral index as the source goes through the scattering event. The light blue region (1-$\sigma$ error region) represents a non-parametric fit to the variation (described in \S\ref{subsec:spectra}).}
    \label{fig:vastj2011-3358}
\end{figure}

Figure~\ref{fig:ese_quant_sep} shows the temporal and spectral properties of the candidates in the initial sample, and everything that is in the regions shaded in lime and/or red is quantitatively filtered out. Sources with characteristic timescales between [0.05, 1], and with $\chi^2<125$ (translating to a reduced $\chi^2$ of $\lessapprox$3) relative to the best-fit lens model are quantitatively chosen. An example of this, ASKAP J201133.7$-$335857, is shown in Figure~\ref{fig:vastj2011-3358}, where we show all publicly available VAST data and the VLASS data. The characteristic ``U'' shaped light curve is visible and is indicative of a scattering event. Comparing the light curve with those of bright ($>$15\,mJy) nearby ($<$1\,deg) sources rules out an instrumental origin and implies that the variation in the light curve is astrophysical and specific to the source. 

The solid purple curve in Figure~\ref{fig:vastj2011-3358} shows the best fit model with the parameters: $\alpha=8.21(1)$, $\beta_{\rm src}=2.2(1)$, $f_{\rm unlens}=0.38(1)$, $F_{\nu; \rm tot}=13.33(3)$\,mJy, $T_0=60562(1)$\,d, and $v=167(5)$\,km/s $\times \left(\frac{a}{10\, \rm AU}\right)$. If the data are assumed to be completely independent, this results in a poor goodness-of-fit estimate. However, under the assumption of the presence of correlated noise, it can be seen that this can explain the variations in the light curve.

We extracted the dynamic spectra and the time-averaged frequency spectrum for observations in the light curve during which the source undergoes the potential scattering event. Figure~\ref{fig:vastj2011-3358} (bottom panel) shows the spectral index for these observations from a power-law fit to the individual spectra. A clear evolution can be seen, with the spectral index switching sign and becoming positive, and gradually relaxing as the source undergoes a potential scattering event, as expected in the paradigm of a 1D Gaussian lens. This gives confidence that the changes in the spectral shape of ASKAP J201134$-$335857 are indeed due to plasma lensing. 

This selection has resulted in 25 sources, which we call the quantitative sample. Table~\ref{tab:ese_prop} lists all the candidates in this subsample alongside their best-fit lens properties. Light curves of the remaining sources are presented online and can be accessed \href{https://vast-survey.org/eses/}{here}.

\subsubsection{Qualitative selection}\label{subsec:qual}
As described in \S\ref{subsec:lc_model}, unmodeled caustics and the presence of very few outliers (outside the ESE event) in bright sources can bias the likelihood estimate since they are neither accounted for in the lens model nor the noise model. In such sources, exclusion of these few data points makes them quantitatively aligned with the previous sample both in terms of the goodness of fit and the presence of spectral variations. Hence, in addition to the quantitative sample, we also include this \textit{qualitative sample}. This resulted in the selection of 47 candidates. Jointly, we refer to these 72 (25+47) sources as the gold sample. Table~\ref{tab:ese_prop} lists the best-fit model parameters for all the sources in the gold sample, and the light curves are available \href{https://vast-survey.org/eses/}{here}.

\subsection{Silver sample}\label{subsec:ext}

In addition to this gold sample, we also highlight 76 additional candidates in which the light curves resemble lensing events, but no simultaneous spectral variations are seen. However, given the limitations in the bandwidth of VAST data, a scattering-event interpretation remains plausible in these sources. However, establishing this requires additional lines of evidence. Hence, for readers interested in advanced lens modeling, the silver sample is equally valuable as the gold sample. Table~\ref{tab:ese_prop_ext} lists the sources that fall into the silver sample alongside their best-fit model parameters. The light curves are shown \href{https://vast-survey.org/eses/}{here}.

\subsection{Bronze sample}\label{subsec:amb}
Finally, we highlight 88 other sources in which the light curve shows variability and expected features, but also has additional variability that cannot be simply explained by an isolated scattering event. Establishing an ESE interpretation in such sources requires long-term historical knowledge to see how unique the observed dip is or more sophisticated lens models which can capture such additional variability. 
Lacking both, we simply label them ambiguous candidates and place them in the bronze sample, as ESEs are one of the possible explanations in these sources. A more detailed modeling of all these events is beyond the scope of the current paper and will be presented in a follow-up work (Jow et al 2026, in prep). For readers interested in advanced lens modeling, the sample should be considered with caution given the additional variability.  The light curves, alongside the model fits for this sample, are shown \href{https://vast-survey.org/eses/}{here}.

\section{Discussion}\label{sec:discussion}

\subsection{Spatial distribution}\label{sec:space_dist}

\begin{figure*}[!htb]
    \centering
    \includegraphics[width=0.8\linewidth]{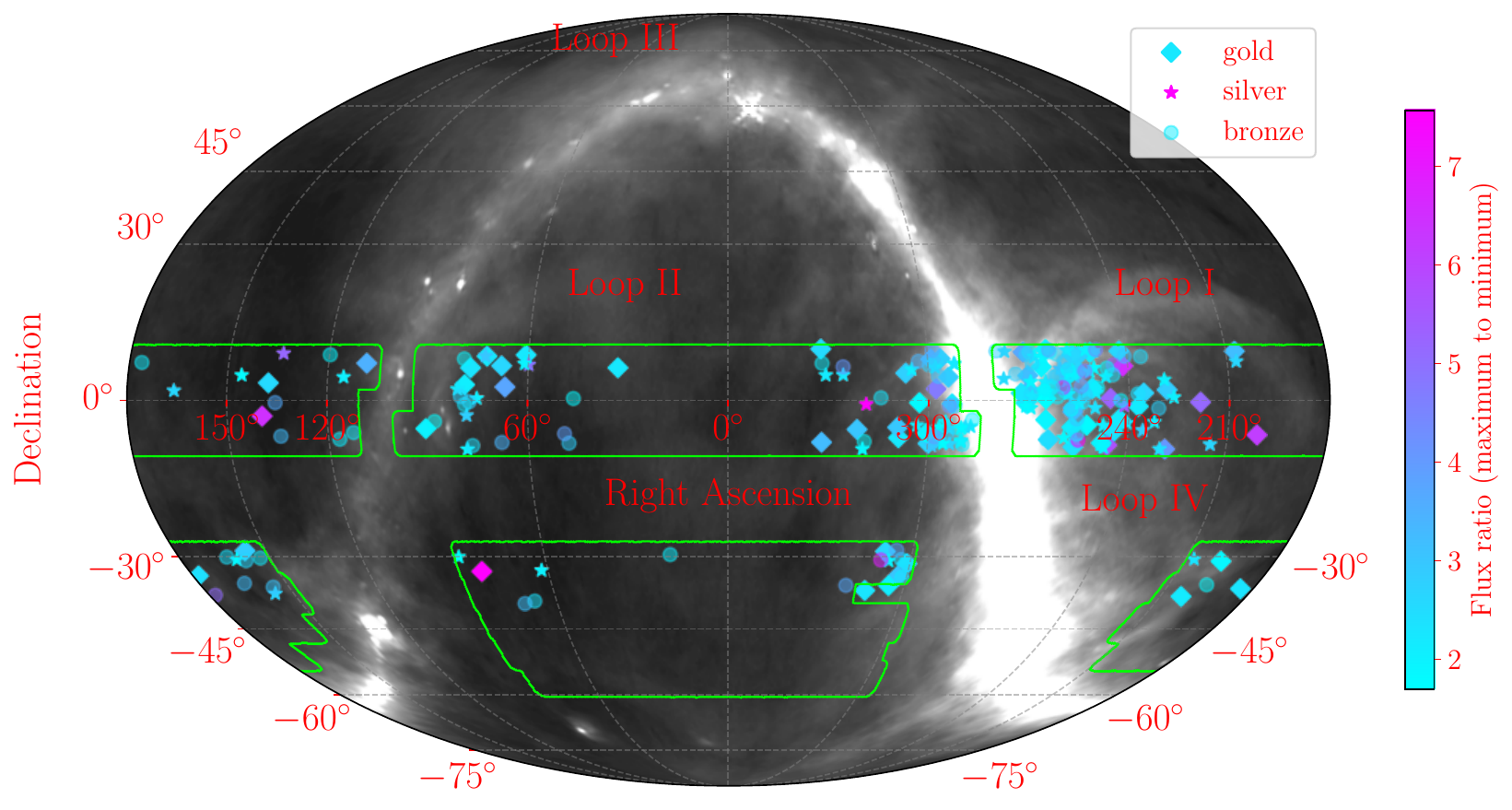}
    \caption{Spatial distribution of ESE candidates in the VAST data set. The lime border and the region inside show the footprint of extragalactic VAST data, and the scatter points (in the heatmap) show the spatial locations of the individual sources. Circles show our gold sample of sources, while stars show the silver sample. The color scale represents the ratio of the maximum to the minimum flux density. Shown in the background is a combination of the Galactic neutral hydrogen density map from the HI4PI survey \citep{H14PI} and the 408 MHz radio continuum data from \cite{haslam_408_1981}. Smoothed versions of the sky maps, for all three sub-samples (gold/silver/bronze), are available online.}
    \label{fig:ese_cands_sky_dist}
\end{figure*}

We show the spatial distribution of our samples of sources (gold, silver, and bronze samples) in Figure~\ref{fig:ese_cands_sky_dist}. The background map shows the 408\,MHz radio continuum, highlighting the Galactic plane and the four relic Galactic loops. We observe a significant increase in the number of sources undergoing potential ESEs close to the Galactic plane. A simple Anderson-Darling (AD) test \citep{Anderson-Darling} gives a test statistic of 15.5\footnote{AD test statistic close to 1 implies that the two distributions are similar.} (at a p-value of 0.01\%) for the hypothesis that our population (gold+silver+bronze) is similar to a uniform distribution. When the whole sample of 24,590 variable sources is considered, the AD test gives a statistic of 1, clearly implying that the ESE sample is not drawn uniformly from the VAST data. We see that the sources close to the Galactic center possibly trace the Galactic loops. This was tentatively proposed by \cite{romani_radio_1987} and \cite{fiedler_extreme_1987}, but with a very limited sample. While this excess is visible, we also observe a similar excess in the Southern portion of the Galactic center. Hence, the relative importance of Galactic loops vs the proximity to the Galactic center is less clear. The AD-test yields a statistic of 8.3, so we cannot conclusively establish the fact that our source population tracks the neutral hydrogen density.

Although it remains unclear, it may be that the same structures responsible for RISS also produce ESEs. \cite{lovell_micro-arcsecond_2008} analyzed variability\,---\,albeit at a much higher frequency of 5\,GHz\,---\,in extragalactic compact sources, likely due to RISS, and proposed a correlation with $\rm H_{\alpha}$ intensity \citep{haffner_wisconsin_2003}. Hence, we also test whether our source population tracks the $\rm H_{\alpha}$ intensity instead of neutral Hydrogen \citep{haffner_wisconsin_2003}. However, the AD-test yields a statistic of 11.5 for our source population to track the $\rm H_{\alpha}$ intensity, and hence we cannot conclusively establish this as well. In addition, we do not see any significant differences between the gold, silver, and bronze samples, suggesting that we are not sampling random variable candidates from the extragalactic radio population, and this is likely due to a scattering phenomenon.

\subsection{Individual light curve morphologies}\label{sec:lc_morp}
We examine the diversity of the light curves of the VAST ESE candidates in the gold and silver samples. As a general comment, we note that the scenario in which 100\% of the background source's flux undergoes scattering is quite rare, as our model fits demand the presence of an unlensed component contributing to the overall flux in a large number of sources. Under the scenario of a 1D Gaussian lens, this implies the presence of two emission components: emission from a core, which predominantly undergoes the extreme scattering, and emission from a wide region that is relatively unaffected. Such multi-component emission has been observed in many AGNs/Quasars \citep{Kellermann1998}. We note that the description used here is to match the language of \cite{Clegg1998} -- emission from the lensed component can be due to compact and extended (compared to the lens) components, both affected by the lens, and different from the unlensed component of the flux.

There are phases in the light curves that can be used to infer the scattering geometry in a relatively simple manner. This depends sensitively on the size of the background source and the strength of the lens. Below, we classify our VAST sample based on their morphology, which provides direct insight into these in the case of a 1D Gaussian lens. 

\subsubsection{Strong lens\,---\,Lensing of a compact component}
\begin{figure*}[!htb]
    \centering
    \includegraphics[width=0.325\linewidth]{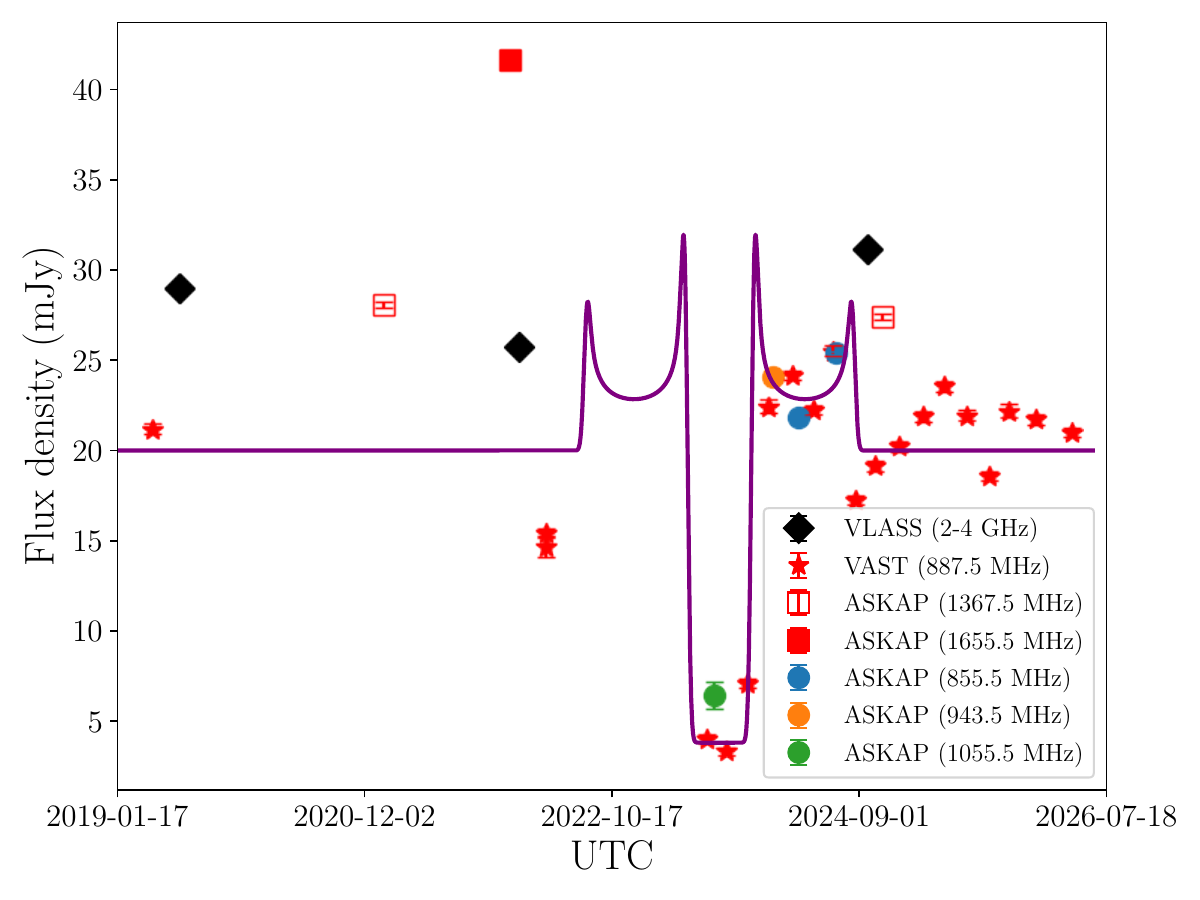}
    \includegraphics[width=0.325\linewidth]{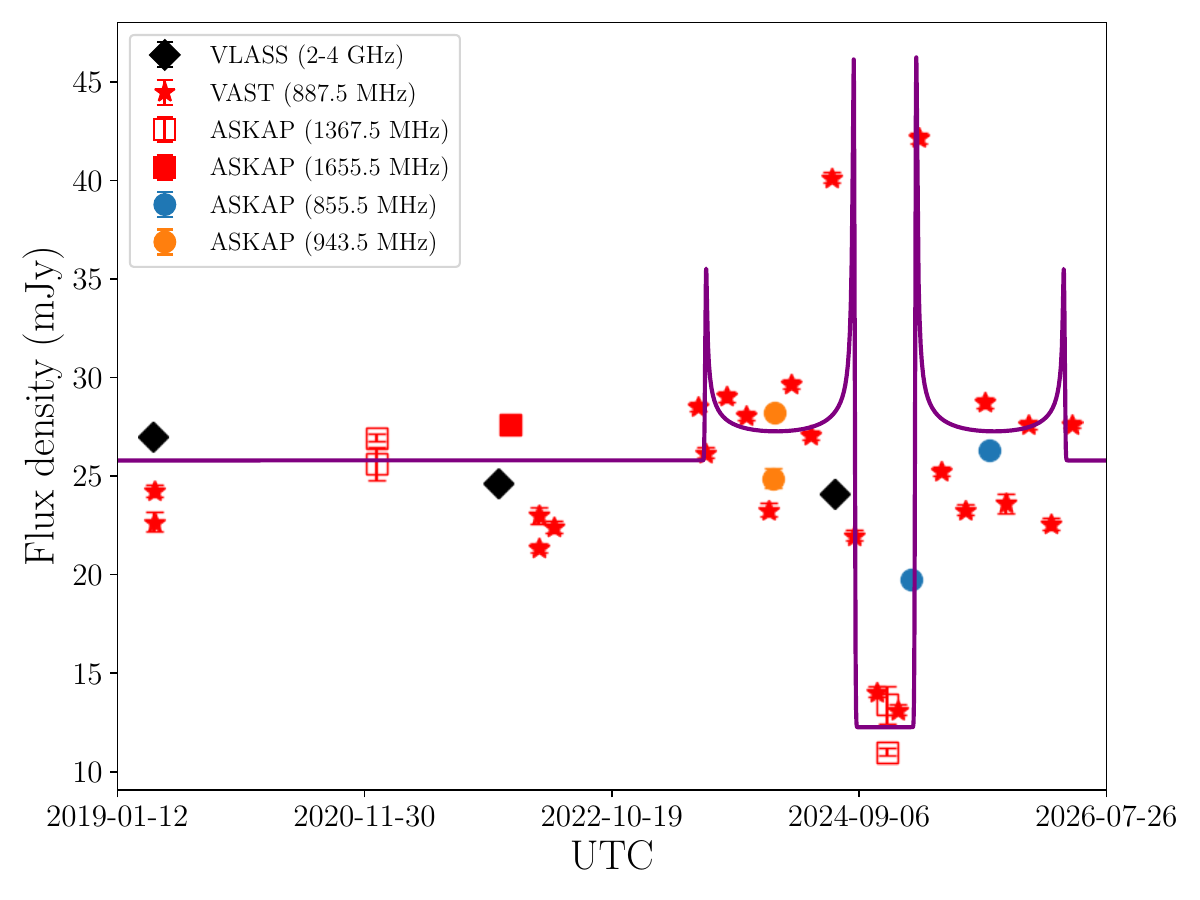}
    \includegraphics[width=0.325\linewidth]{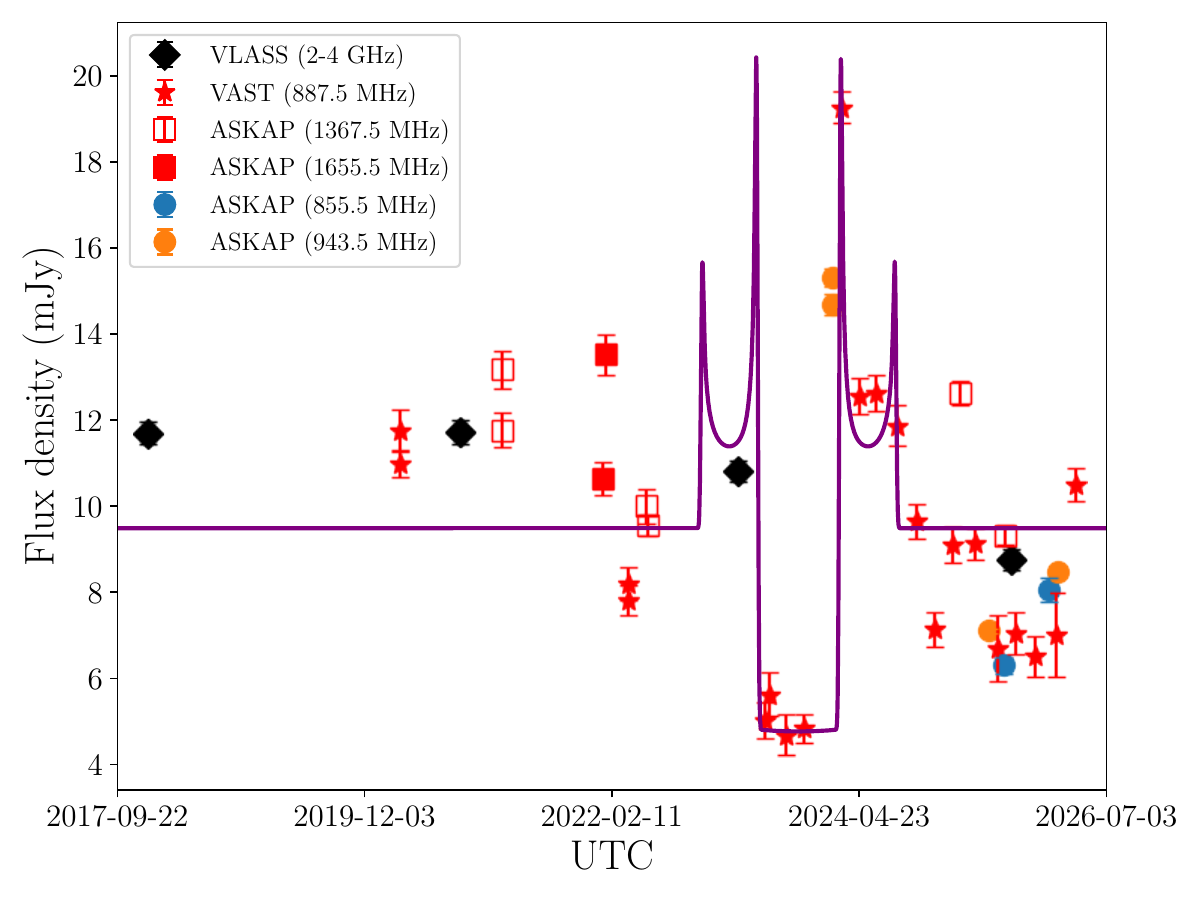}
    \caption{VAST ESE sample: Examples of compact components that are scattered by a strong lens --- ASKAP J052803.1$-$330348 (left), ASKAP J134946.1+090816 (middle), and ASKAP J15h1331.7$-$091920. The flat bottom and the surrounding caustic spikes can be easily seen.}
    \label{fig:strong_lens_point}
\end{figure*}

For a background source with a compact emitting region, a strong lens produces a light curve that exhibits a flat bottom, which represents the incident rays getting strongly de-focused. This is usually sandwiched between the two zones of ``caustic'' spikes, which are the zones of strong focusing. Figure~\ref{fig:strong_lens_point} shows a select example of three such sources (among many) in our sample. However, as can be seen in Figure~\ref{fig:strong_lens_point}, the bottom of the light curve does not reach zero, as expected for a point source. For these three sources, we find that $\sim$50\% of the total flux density is affected by the extreme scattering due to the lens, with the rest relatively unaffected (see Table~\ref{tab:ese_prop} for estimates). Under-dense lenses may be able to accommodate the more modest flux decrement in the de-focusing zone without needing to appeal to multiple emission components \citep{pen_refractive_2012}; however, restricting ourselves to a 1D Gaussian model to fit the light curves, the data is insufficient to distinguish these two scenarios (see Section~\ref{sec:underover} for a further discussion of under- vs over-dense lenses).

\subsubsection{Strong lens\,---\,Lensing of an extended component}

\begin{figure*}[!htb]
    \centering
    \includegraphics[width=0.325\linewidth]{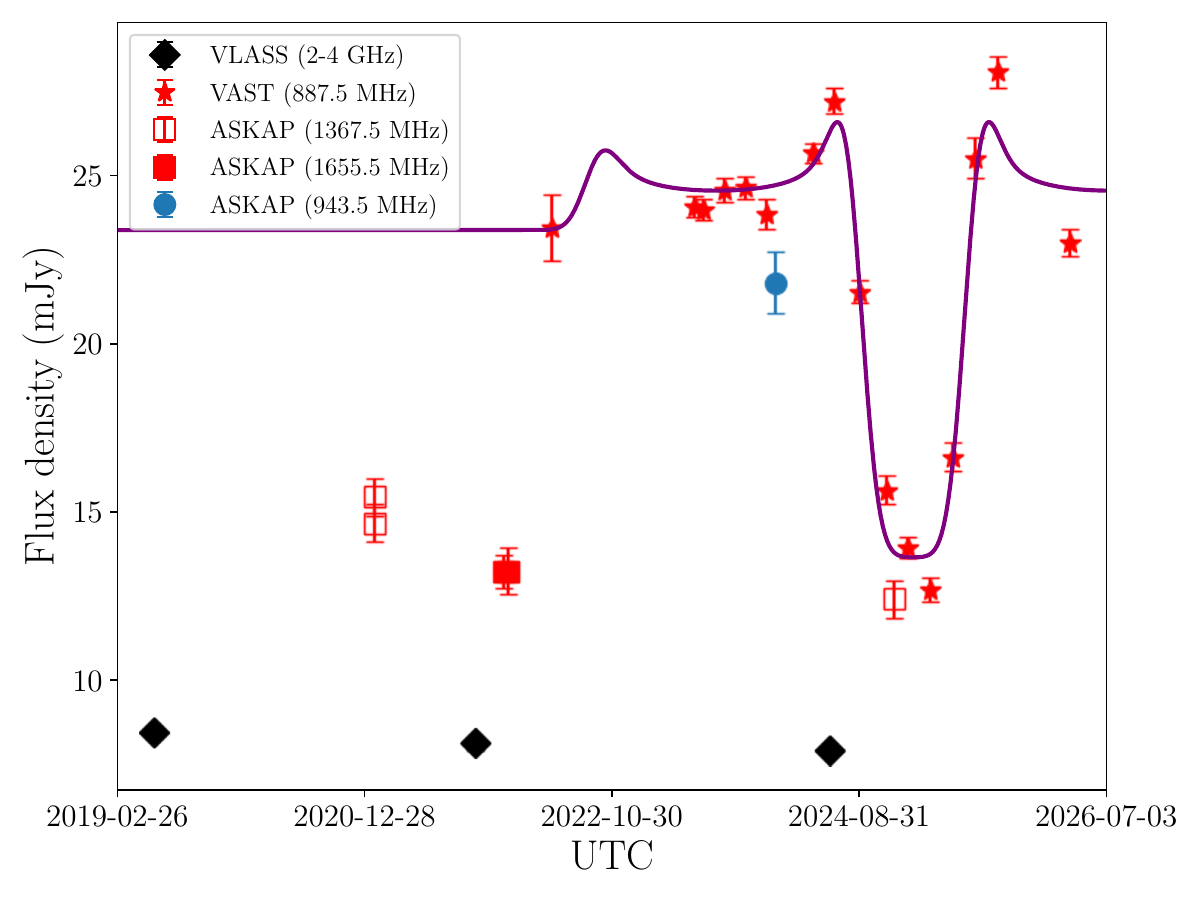}
    \includegraphics[width=0.325\linewidth]{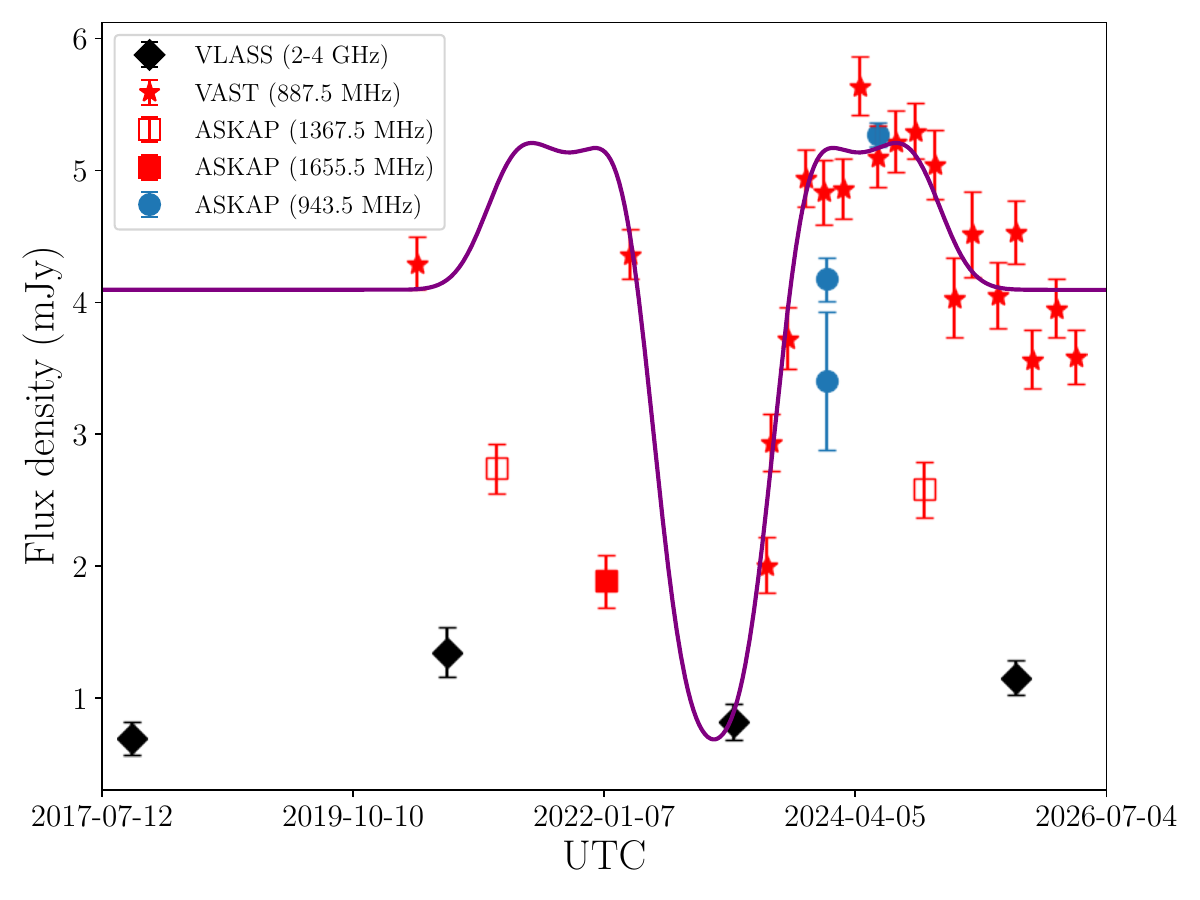}
    \includegraphics[width=0.325\linewidth]{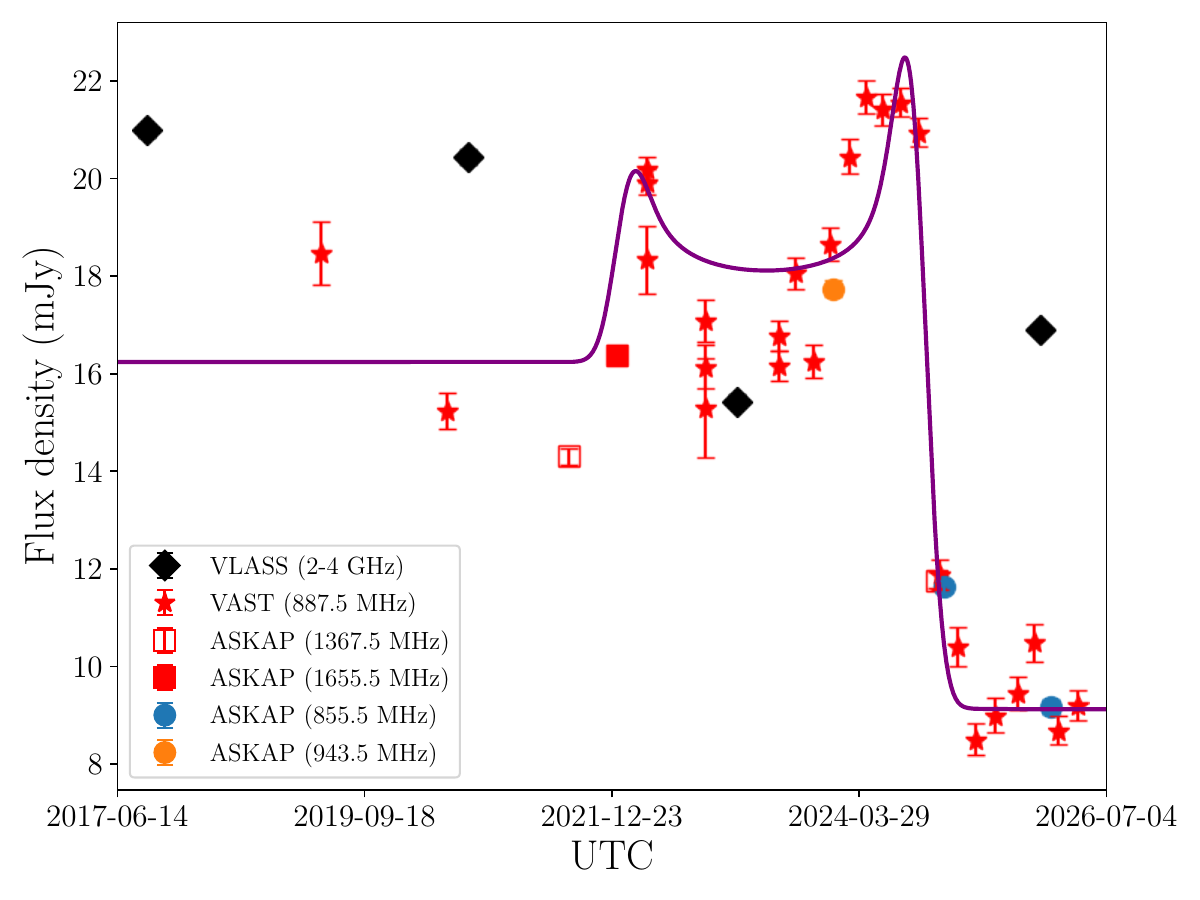}
    \caption{VAST ESE sample: Examples of extended components that are scattered by a strong lens ---  ASKAP J164405.3+021409 (left), ASKAP J194444.6+074302 (middle), and ASKAP J220755.3+093305 (right).}
    \label{fig:strong_lens_ext}
\end{figure*}

In the case where the component undergoing lensing is an extended source compared to the lens, the resultant effect is equivalent to the sum of all the point source profiles. This will result in i) smoothing out of caustic spikes so they will appear less pronounced, and ii) emission from the off-axis and extended region is refracted into the de-focusing zone, so the flat bottom of the light curves morphs into a smooth decline and rise. Figure~\ref{fig:strong_lens_ext} shows examples for this case, where both of these effects can be seen.  

\subsubsection{Weak lens\,---\,Lensing of a compact component}
\begin{figure*}[!htb]
    \centering
    \includegraphics[width=0.325\linewidth]{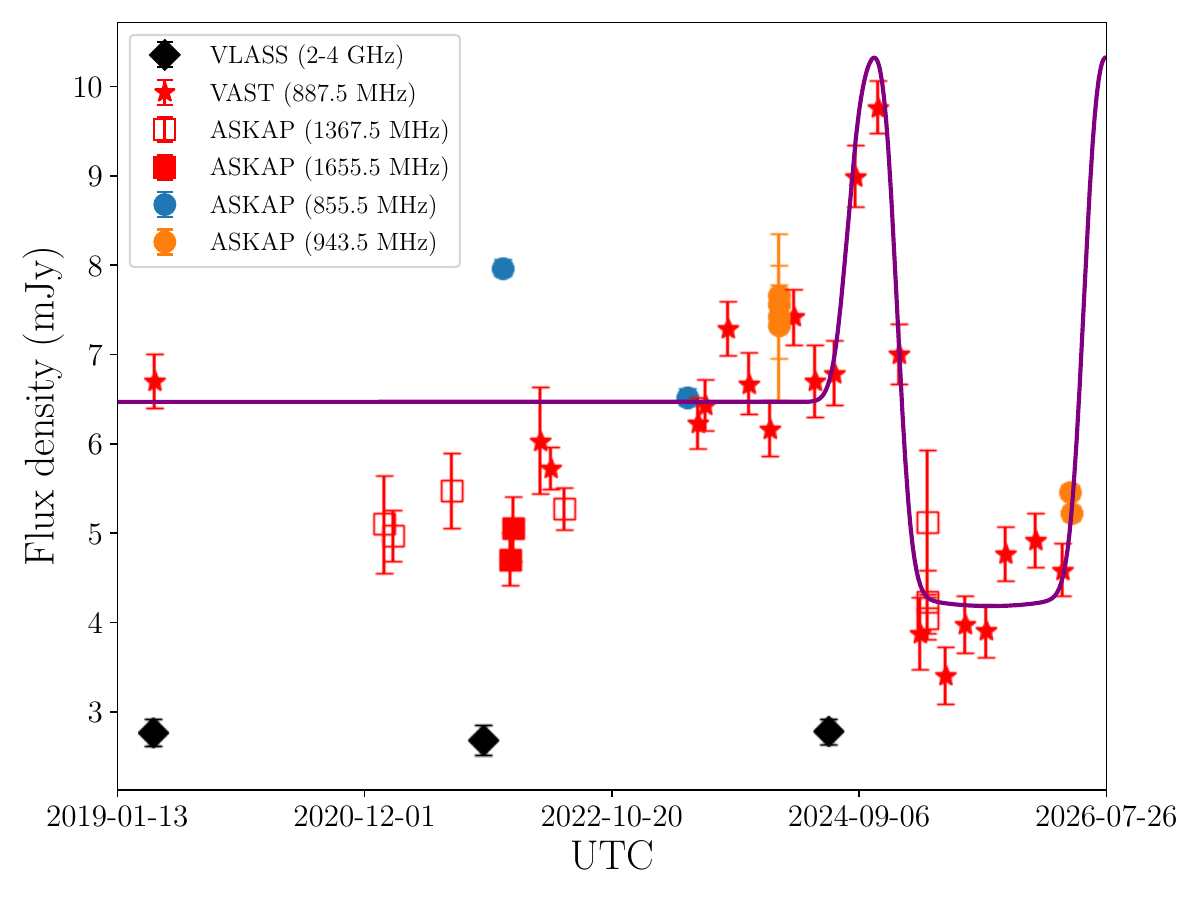}
    \includegraphics[width=0.325\linewidth]{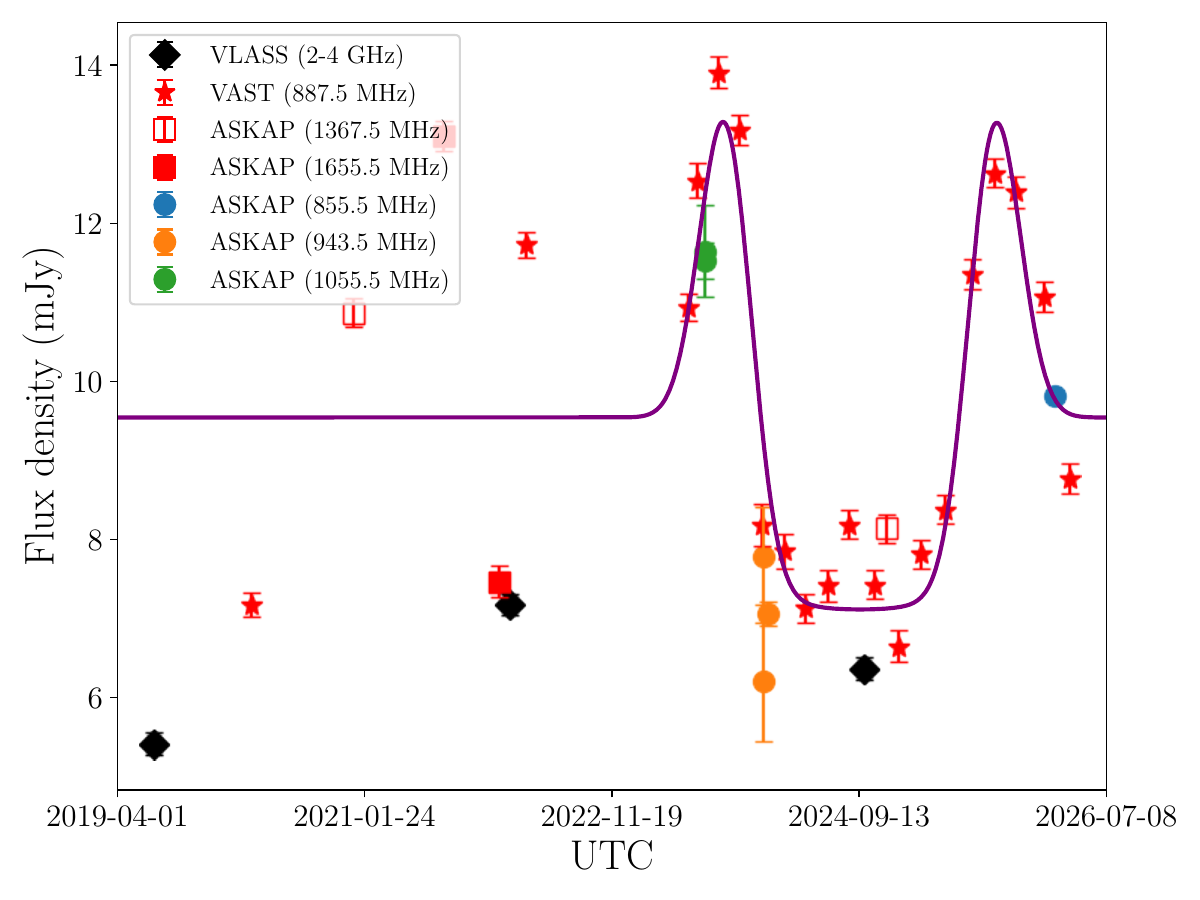}
    \includegraphics[width=0.325\linewidth]{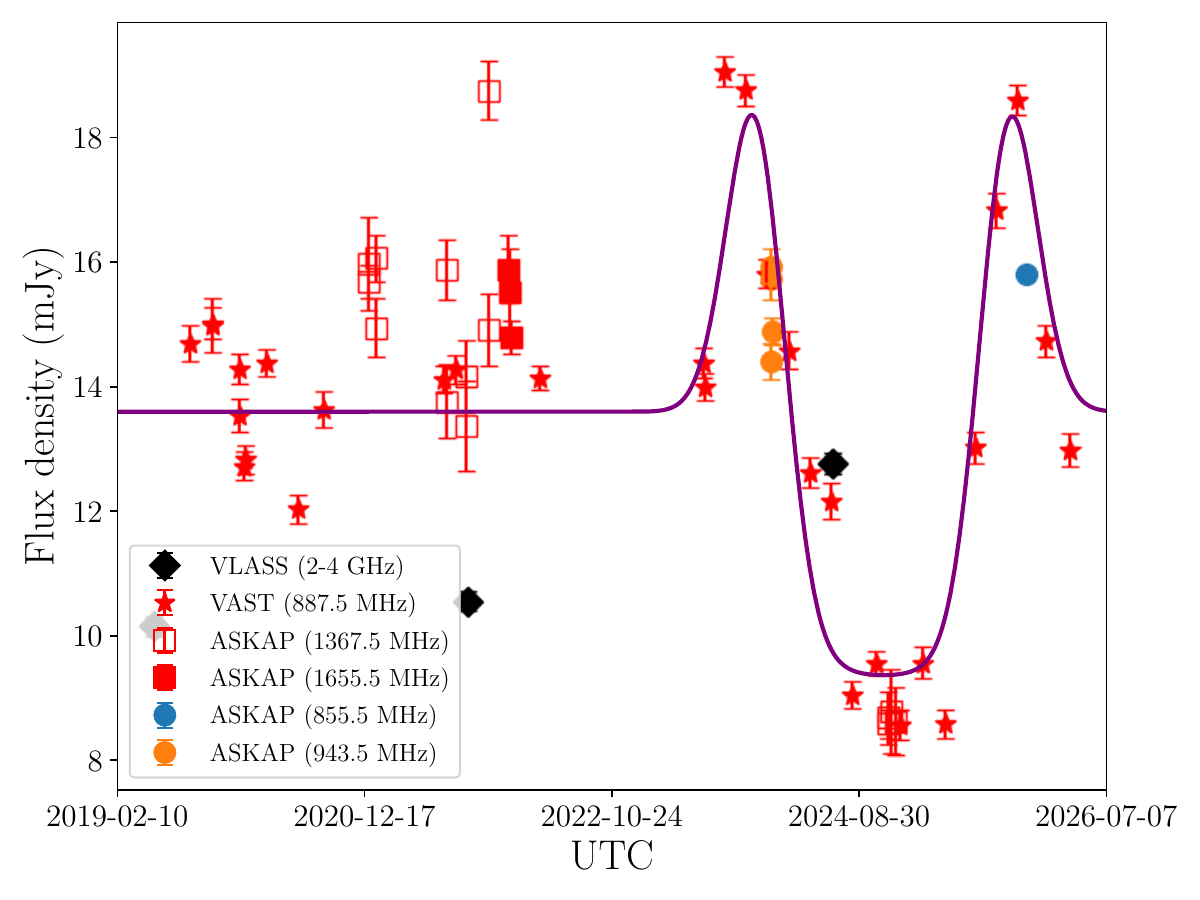}
    \caption{VAST ESE sample: Examples of compact components that are scattered by a weak lens --- ASKAP J151446.8+015128 (left), ASKAP J202108.8$-$355527 (middle), and ASKAP J203505.4$-$071131 (right).}
    \label{fig:weak_lens_point}
\end{figure*}

As the strength of the lens gets weaker, both the focusing zones (caustic spikes) and the de-focusing zones (the dip in the light curves) start to get much less pronounced for a compact source. This results in the inner and outer caustics merging with each other, and the resultant light curves exhibit a dip sandwiched between single humps on either side. Figure~\ref{fig:weak_lens_point} shows the VAST sources that exhibit this behavior.

\subsubsection{Weak lens\,---\,Lensing of a extended component}
\begin{figure*}[!htb]
    \centering
    \includegraphics[width=0.325\linewidth]{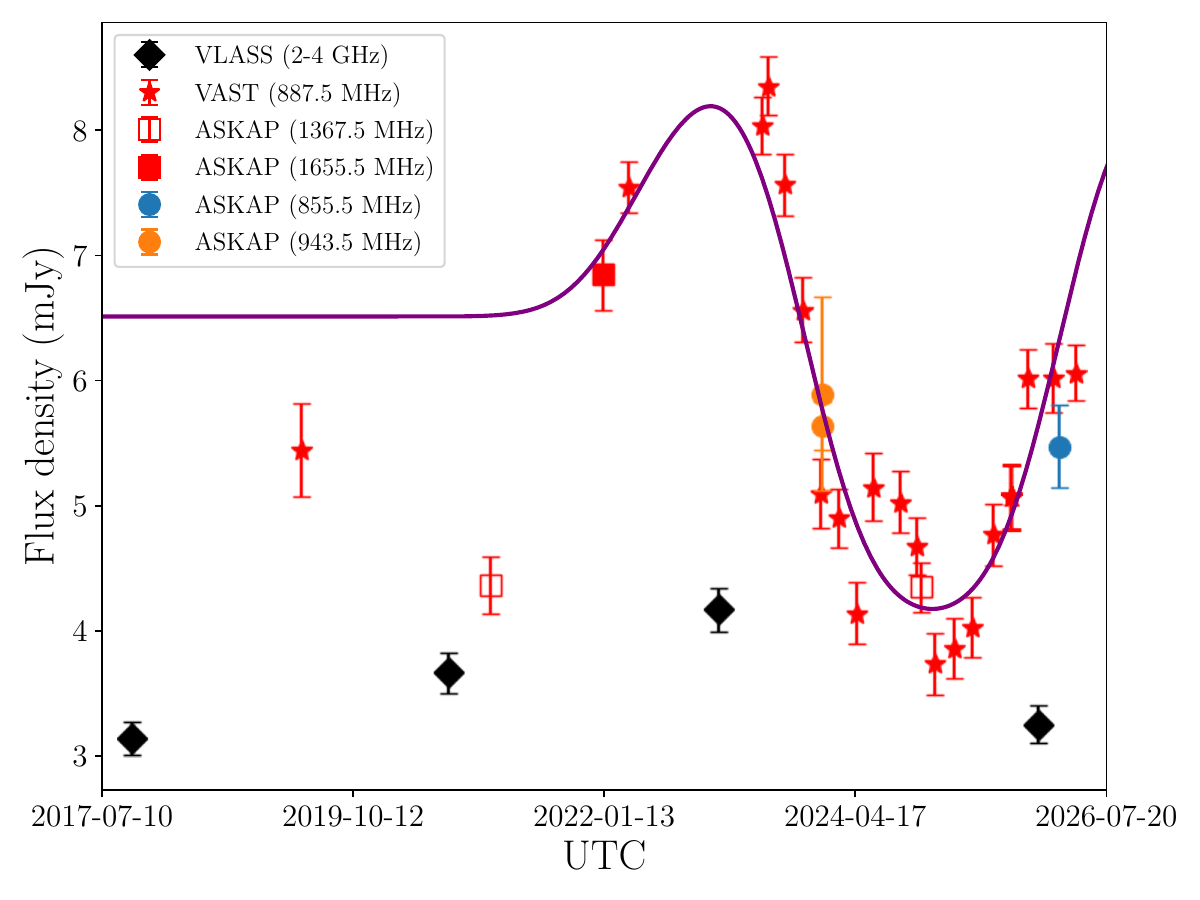}
    \includegraphics[width=0.325\linewidth]{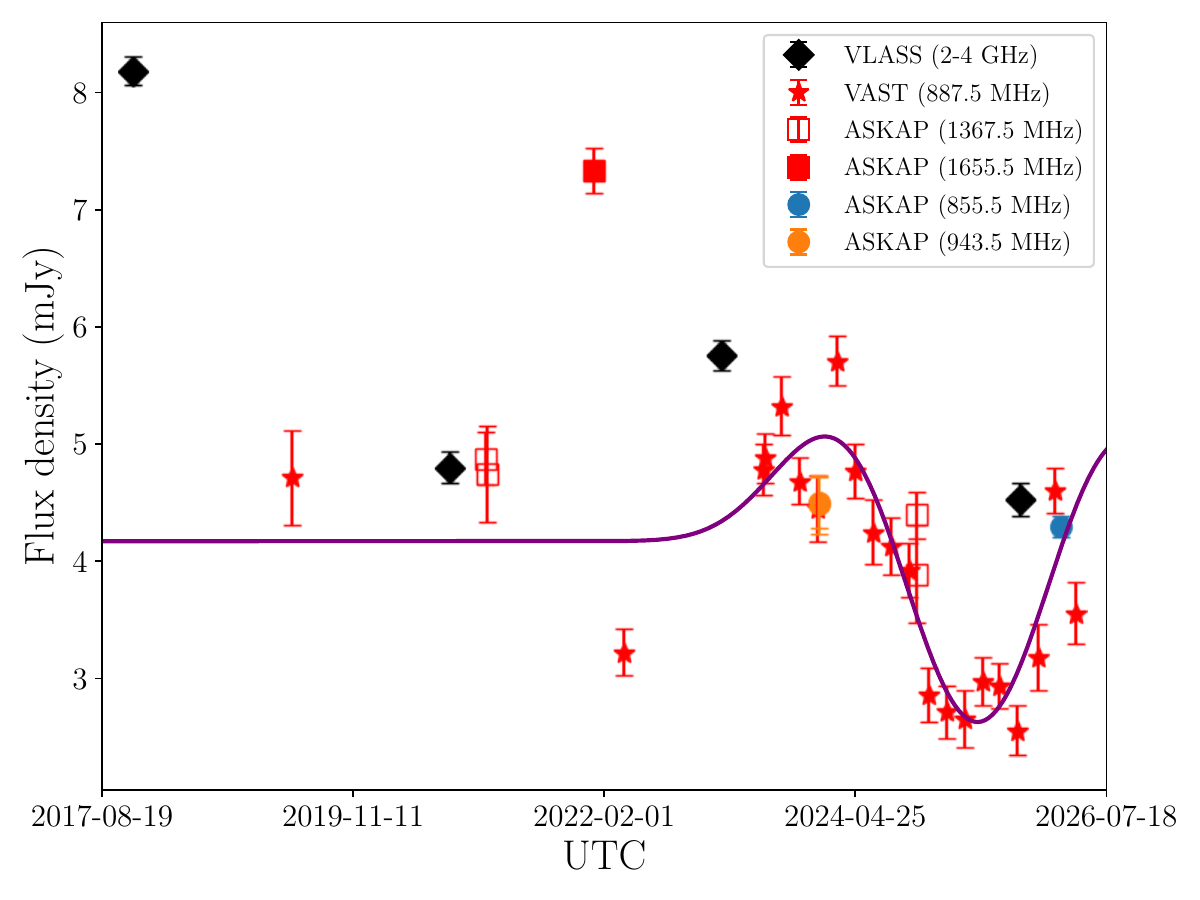}
    \includegraphics[width=0.325\linewidth]{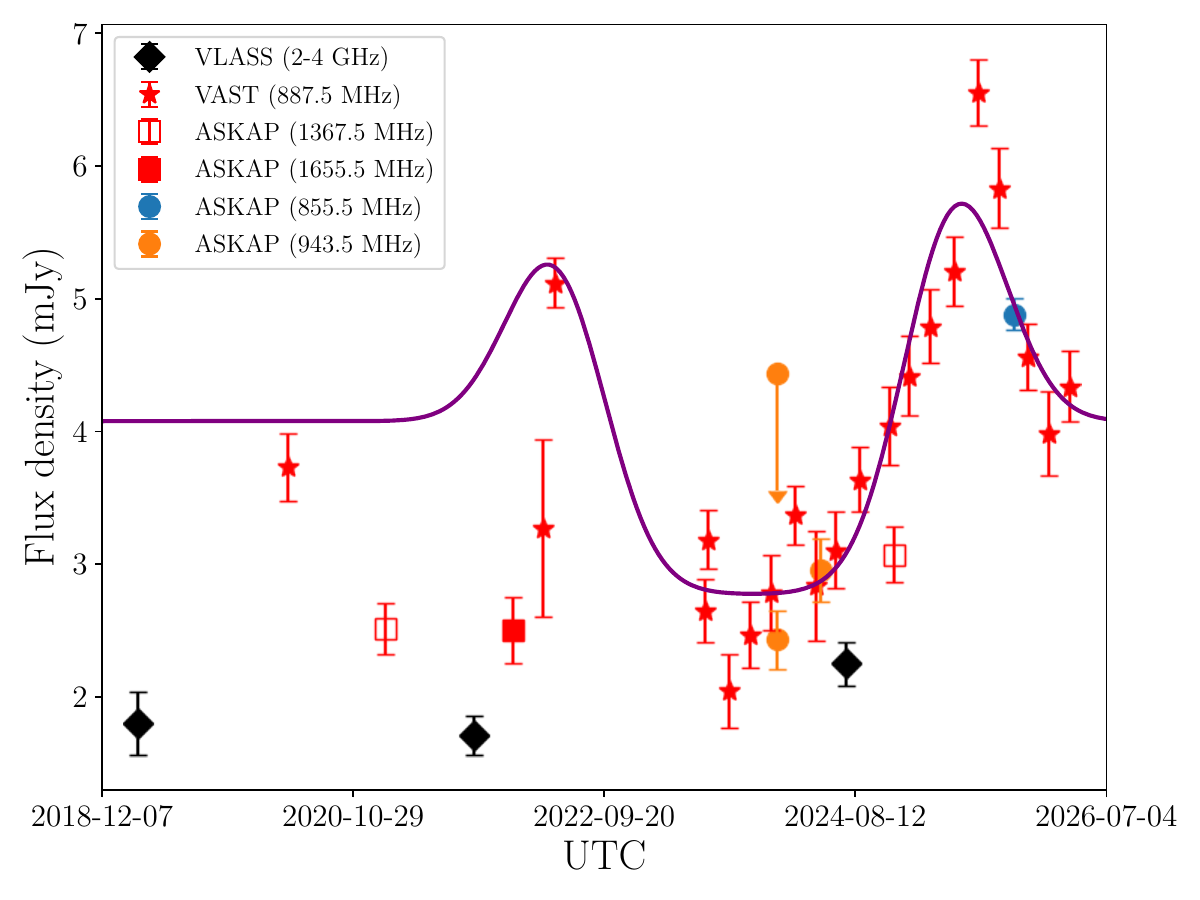}
    \caption{VAST ESE sample: Examples of extended components that are scattered by a weak lens --- ASKAP J040242.0+082524 (left), ASKAP J044937.4+081134 (middle), and ASKAP J180044.3+043159 (right).}
    \label{fig:weak_lens_ext}
\end{figure*}

As the angular extent of the component that is scattered increases compared to the lens, the light curve smooths out further, which results in a suppressed peak-to-dip ratio in addition to a lack of pronounced caustics. In the extreme case of an extended source, multiple components tend to smooth out the light curve to a point where the effects of scattering cannot be easily made out. Examples of such sources undergoing an ESE event are shown in Figure~\ref{fig:weak_lens_ext}.

\subsection{Properties of the lenses}
Even with the relatively narrow bandwidth of the VAST data, spectral changes (similar to Figure~\ref{fig:vastj2011-3358}) are seen in the gold sample. This implies that the underlying lensing mechanism is a chromatic process, rather than an achromatic one. Flux density variations at similar levels to ESEs are also observed in another class of objects, called Symmetric Achromatic Variables \citep[SAVs;][]{vedantham_symmetric_2017}. These are distinguished as sources exhibiting achromatic flux density variations over a broad wavelength range, possibly due to gravitational lensing. In contrast, the chromatic variations observed for sources in our sample favor plasma lensing over gravitational lensing. 

\begin{figure*}[!ht]
    \centering
    \includegraphics[width=\linewidth]{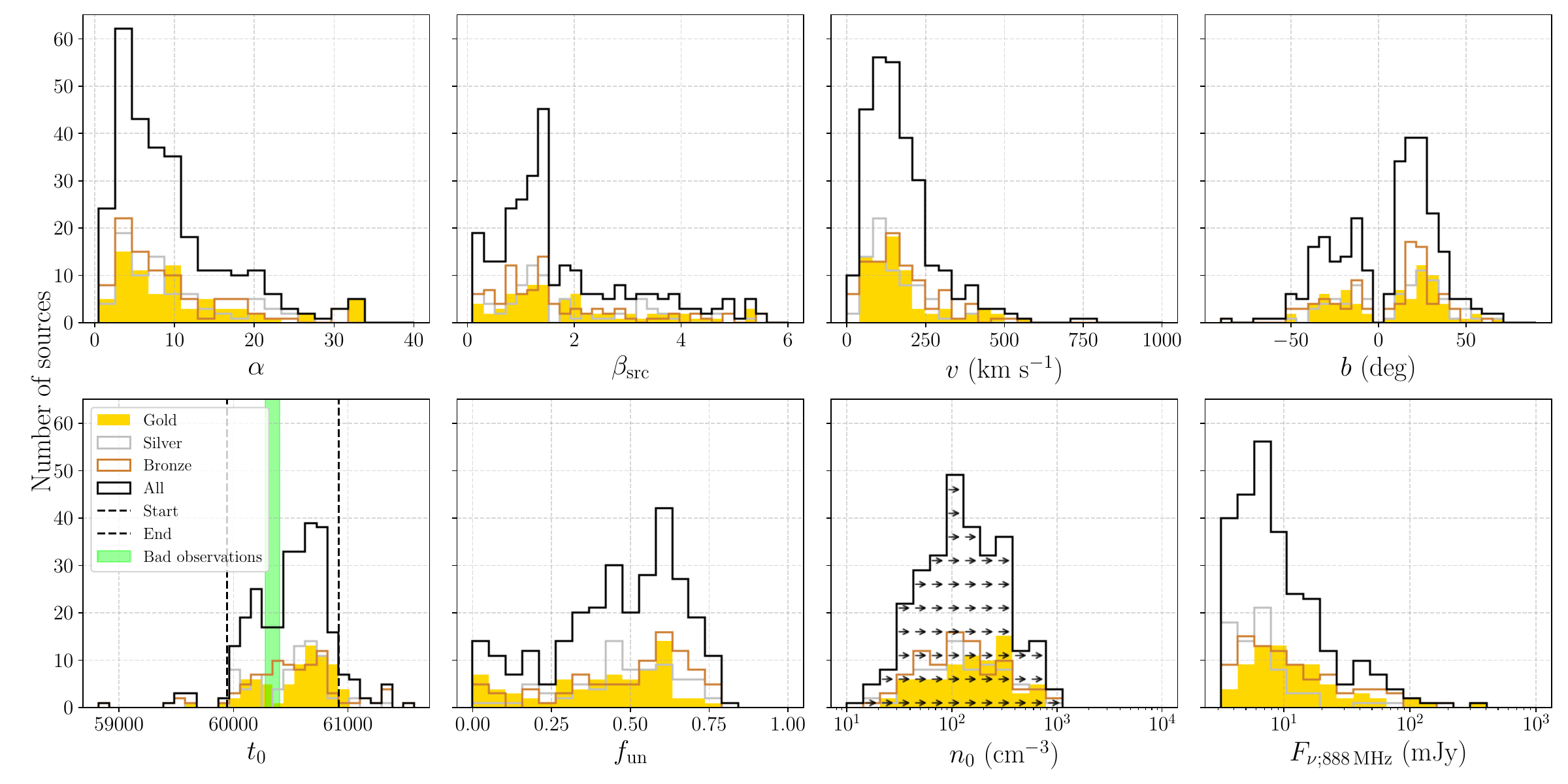}
    \caption{Gaussian lens properties: this figure shows the distribution of lens strength ($\alpha$), source size ($\beta_{\rm src}$), inferred relative velocity ($v_{\rm obs}$), galactic latitude ($b$), fraction of unlensed flux ($f_{un}$), reference time ($T_0$; MJD), inferred plasma density ($n_0$), and the peak flux density $F_{\nu; \rm 888\,MHz}$ for all the sources in gold, silver, and bronze samples. The vertical dashed lines show the start and end of the VAST data set. The lime-shaded region corresponds to the period where VAST observations were deemed bad. The arrows for the density represent lower limits. Color scheme: gold -- gold sample, silver -- silver sample, bronze -- bronze sample, black -- all samples.}
    \label{fig:ese_params}
\end{figure*}

Figure~\ref{fig:ese_params} shows the distribution of the lens and source parameters from the 1D Gaussian lens model fits. We see that many of the ESE candidates are caused by sources with angular sizes comparable to the lens and either by moderately strong or weak lenses, as evident from the light curve morphologies (see \S\ref{sec:lc_morp}). For a Gaussian lens, the strength of the lens, $\alpha \propto \lambda^2$, and the refraction angle,  $\theta_r \propto \lambda^2$, more strongly affect emission at low frequencies. Given the range of $\alpha$ at 1\,GHz, it can be seen that at 3\,GHz, the strength of the lens will be quite weak ($\alpha <1$), which supports the lack of pronounced variations in the VLASS data. For an unbiased sample, the reference time should be uniform over the time span of VAST data. However, due to data quality issues, the data between November 2024 and April 2025 are rejected, and hence any ESEs with maximum flux depression during this interval will be missed. This is also reflected in the distribution (the lime-shaded region in Figure~\ref{fig:ese_params}), which otherwise is broadly uniform.

While the 1D lens model restricts us to scenarios where the impact parameter vanishes, i.e., where the line of sight to the background source passes through the center of the lens, in realistic cases where these density fluctuations are 2D sheets or 3D structures, a zero impact parameter configuration can be relatively rare compared to occurrences with a non-zero impact parameter. Even in such cases, given the dependence of the refraction angle and the impact parameter \citep[e.g., ][]{dong_extreme_2018}, low-frequency emission will be more strongly affected. Given the lack of variation in VLASS data at a level comparable to VAST data, this is still consistent with a plasma lensing event.

As mentioned in Section~\ref{subsec:lc_model}, the degenerate nature of $N_0$, $a$, and $D$ (combined into $\alpha$) makes it difficult to disentangle the physical parameters of the lenses. However, if one assumes values for $a$, and $D$, one can estimate the electron densities that are responsible for these events. If we consider the typical scale (vertical) height of the Galaxy to be $\approx$1\,kpc, given the galactic latitudes of background sources, we estimate the upper limits on the typical distances of these lenses. Assuming the length scale of the lens to be $a\approx$1\,AU, we estimate a lower limit on the characteristic density of the lens. Figure~\ref{fig:ese_params} (bottom right) shows the lower limits on the central density of the lenses. We emphasize that these are lower limits, based on the maximum distance estimates, and hence the true density can be much higher if the plasma structures are much closer. Even with this, the lower limits on the density ($\gtrsim$100 cm$^{-3}$) imply a pressure $P/k > 10^2 T_e$\, K cm$^{-3}$ (where $T_e$ is the ionized electron gas temperature), a few orders of magnitude in excess of the expected ambient pressure in the warm (or hot) ionized medium \citep{ferriere_interstellar_2001}.

\subsection{Underdense vs Overdense lenses}\label{sec:underover}
\begin{figure*}[!htb]
    \includegraphics[width=\linewidth]{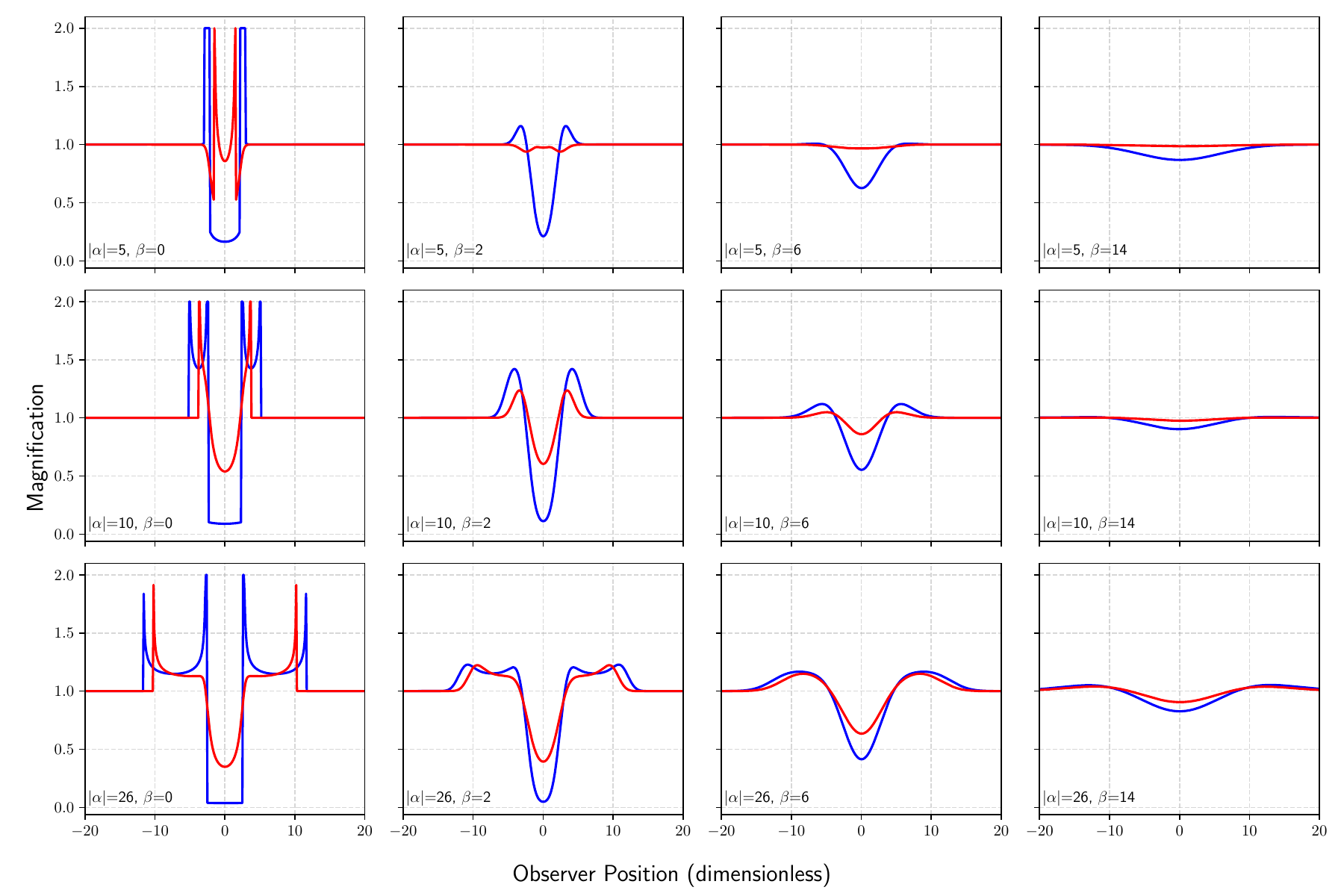}
    \caption{Figure shows the expected magnification, and hence the light curve, pattern in the case of over-dense lenses (blue curves) and convergent under-dense lenses (red curves). Different lens strengths ($\alpha$) and source sizes ($\beta$) are shown in different panels.}
    \label{fig:conv_vs_div}
\end{figure*}

While ESEs are well described by underlying plasma lenses, both over-densities and under-densities in the intervening medium can act as lenses. Evidence for over-dense lenses exists in a few individual cases \citep{bannister_real-time_2016}; but is far from established. While we can test the nature of density variations, given the cadence and the narrow bandwidth, robustly distinguishing between the two for the entire sample can be difficult. Figure~\ref{fig:conv_vs_div} shows the single-frequency light curve profiles as expected in the case of over-dense and under-dense lenses. It can be seen that a clear distinction can be made only in the case of unresolved point sources and strong lenses with higher cadence (first column in Figure~\ref{fig:conv_vs_div}). As the source becomes extended and the lens becomes weaker, the light curves start to look alike, and this can cause covariance between different configurations. 

This is also in line with what the data suggests. For our model fits, we also included models where the strength of the Gaussian lens is $<$0, reflecting under-densities. In the case of strong lenses, where both the caustics could be made, out, the posterior distributions show a clear preference for $\alpha>0$ (over-densities) compared to $\alpha<0$. However, in the cases of moderate/weak lenses and/or extended sources, we see that the posteriors for $\alpha$, and $\beta$ are almost bimodal, and temporal information alone cannot cleanly distinguish them. While spectral information is required to do so, our data are limited to narrow-band, which cannot break the degeneracies, since the spectral evolution over this small range is consistent with both scenarios. Simultaneous wide-band spectral coverage, similar to \cite{bannister_real-time_2016}, will be required to distinguish between ISM over- and under-densities. 
Hence, when reporting the best-fit values in Table~\ref{tab:ese_prop}, we limit ourselves to over-dense lenses.

\subsection{The nature of the background sources}

\begin{figure}
    \centering
    \plotone{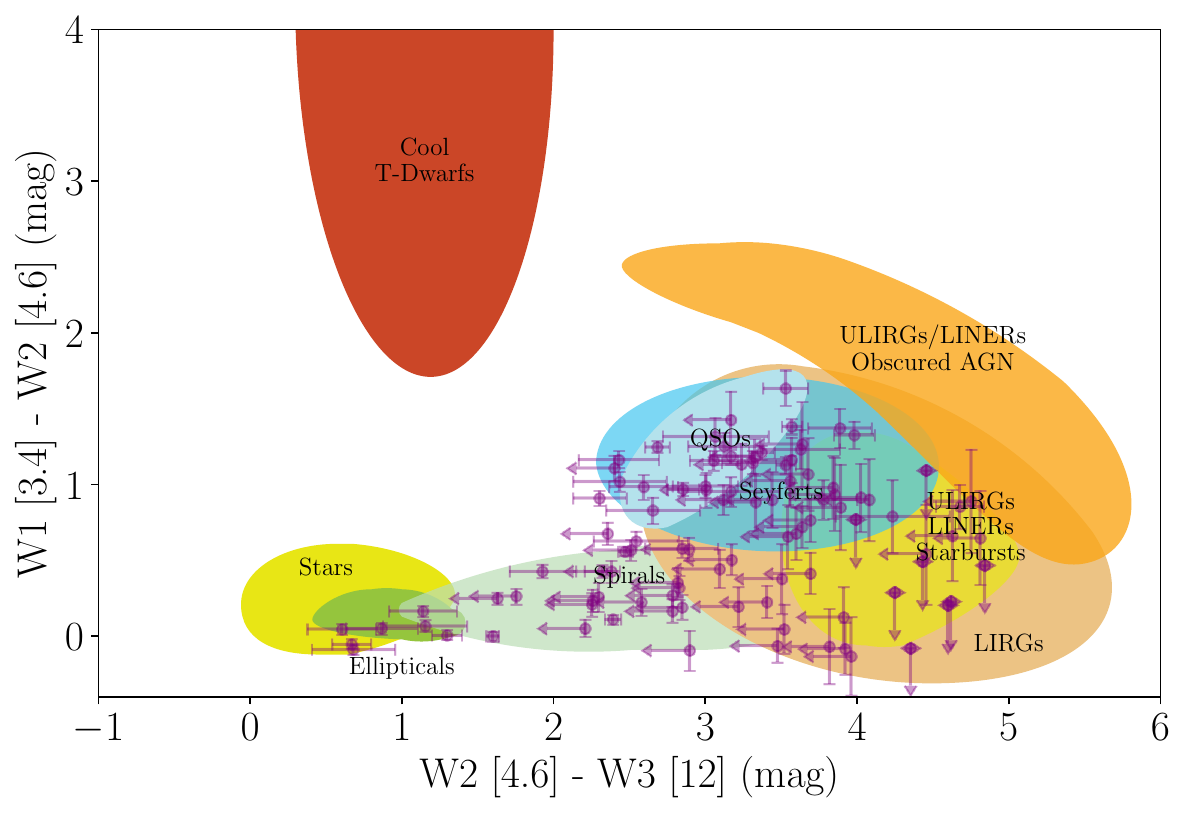}
    \caption{WISE counterparts to the VAST ESE sample: Figure shows the WISE color (W1-W2)-color (W2-W3) plot demarcating various source classes. Overplotted in purple are the counterparts to the VAST ESE sample.}
    \label{fig:wise_cc_plot}
\end{figure}

While the presence of an unresolved point source at radio wavelengths can be inferred and is required for a plasma lensing event, it is also useful to check how many of the sources are consistent with this. To do so, we cross-matched our sample of VAST sources against archival surveys to look for multi-wavelength counterparts. Using a 2.5\arcsec\, error radius, a cross-match with the known AGN catalogs \citep{allwise_agn,lqac-5,milliquas,sdss-agn} resulted in the identification of 41 AGNs. At least 15 of them are spectroscopically classified as quasars, with others photometrically classified in the Wide-field Infrared Survey Explorer (WISE)-AGN and/or the Million Quasar catalogs \citep{assef_wise_2018,flesch_million_2023}. We then looked at the infrared counterparts, primarily in the \textit{AllWISE} data \citep{cutri_vizier_2021}, to infer their nature from the infrared colors. This has resulted in the identification of 101 WISE counterparts. Figure~\ref{fig:wise_cc_plot} shows these IR counterparts on the WISE color-color diagram \citep{wright_wide-field_2010}. We infer that while a large number of these background radio sources undergoing ESE are quasars/Seyferts/AGNs, there are also sources that are localized to the centers of spiral and elliptical galaxies which do not show obvious AGN colors. Unfortunately, only a handful of these have archival optical spectra, and hence, the nature of the AGNs at the center of spiral/elliptical galaxies could not be confirmed. In total, only 109 of the 236 sources have counterparts at other wavelengths. While an unidentified/high redshift AGN provides an obvious explanation, it will be more interesting to see if there are sources in our sample whose spectra do not reveal AGN features, in which case an alternate compact radio-emitting region would be required to explain the observed ESE event.

\subsection{All-sky rates}
One of the limiting factors to derive an all-sky rate of ESE events previously was the lack of high-cadence surveys over large parts of the extragalactic sky. Hence, historically, most of the ESEs were found in very bright sources that were very frequently monitored. But, with the availability of a good cadenced survey like VAST, we can infer all-sky rates. Monitoring a sample of 300 bright sources over a decade, \cite{fiedler_summary_1994} derived a rate of 0.017 events/source/year. \cite{lazio_extreme_2001} observed a sample of 150 sources over 17 years and derived a rate of 0.007 events/source/year. It is interesting to note that a lot of the 150 sources in the \cite{lazio_extreme_2001} sample are derived from the \cite{fiedler_summary_1994} sample; hence, these are not unique sources/ESE detections. Our selected sample of ESE candidates was chosen to be $>$5\,mJy. The number of sources in the VAST data set with flux density $>$5\,mJy is 263,038, and hence a naive estimation yields an overall rate of 0.0003(0.0006) events/source/year, considering the gold and (gold, silver) samples.  However, since our search is untargeted, contrary to the previous studies, the overall derived rate is expected to be lower, given the lack of source curation. A more reasonable estimate can be derived by using our sample after selecting for variable sources (since most of the historical samples were variable AGN). Our variable sample consisted of 24,590 sources, and hence this yields a rate of 0.003 (0.006) events/source/year, comparable to the previous estimates. However, given the lack of exact criteria that were used to select historical samples, a direct comparison is difficult, but one can notice that with strict source curation, similar to previous studies, our derived rate can only be higher. With more all-sky surveys like the Deep Synoptic Array (DSA) and the Square Kilometer Array (SKA), the rate derived here will be more meaningful, since future ESE samples will be untargeted.

Converting it to the survey area, the VAST survey covers a total sky area of 12,500 deg$^2$. This yields an overall rate of 0.006 (0.012) events per year per deg$^2$ at 888\,MHz, considering the gold (gold+silver) sample. However, as can be seen in Figure~\ref{fig:ese_cands_sky_dist}, this rate is higher towards the Galactic center.

\subsection{Expectations for the upcoming surveys}
\begin{figure}[!t]
    \centering
    \includegraphics[width=1\linewidth]{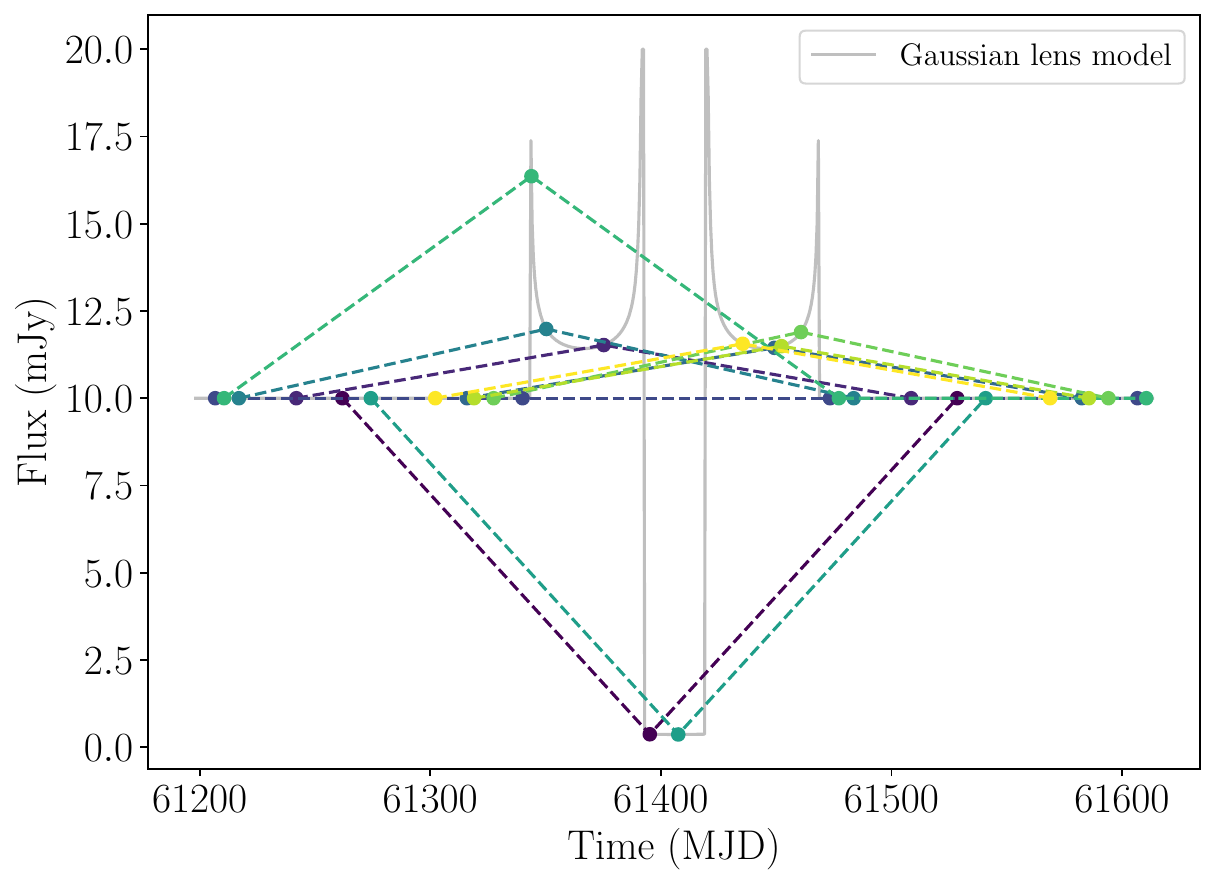}
    \caption{Figure shows an illustration of the resultant light curves in the era of DSA. The gray curve shows the expected light curve from a 1D Gaussian lens, and the colored dashed curves show different realizations of DSA light curves.}
    \label{fig:dsa_lc}
\end{figure}

With more sensitive radio arrays like the Deep Synoptic Array \citep[DSA;][]{DSA}, and the Square Kilometer Array \citep[SKA; ][]{SKA} coming online in the next few years, the number of ESE events is expected to increase enormously with the gain in sensitivity. However, the primary challenge in this case becomes the identification of ESE candidates in these surveys. We present a detection scenario for DSA here, since the all-sky survey strategy for DSA is more mature than SKA. While the DSA increases the sensitivity by at least an order of magnitude from ASKAP VAST, the proposed cadence of $\approx$4\,months will undersample the light curves to confidently establish an ESE interpretation in many sources. Figure~\ref{fig:dsa_lc} shows an example of an ESE event ($\alpha$=25, $\beta$=0, $v$=100\,km/s), and different realizations of the light curves at the DSA cadence. It can be seen that the light curve matches the theoretical expectations of an ESE, characterized by the huge decrement in flux, only if the ESE event is caught during the period when the scattering is most pronounced. In many cases, the light curves only show modest variations, which makes them extremely difficult to distinguish from refractive scintillation and/or intrinsic AGN variability. In other cases, the light curves mimic ``flares'', when the caustics are sampled, which makes ESEs a source of contamination for true astrophysical flaring events. This problem becomes worse in the case of weak lenses and extended sources. This poses a challenge, as many true ESE events may be missed if only temporal behavior is considered, as is done in current and previous studies. While a typical ISM velocity is considered here, for lower relative velocities $\lesssim10$\,km/s, the light curve sampling improves, and, hence, DSA may preferentially detect slowly evolving ESEs.

The addition of spectral information can help in alleviating the problem in some of these cases. Contrary to ASKAP, DSA has a wider frequency coverage (0.7--2\,GHz) and hence a requirement of simultaneous and systematic temporal and spectral variations can help differentiate ESE from other variable sources at radio wavelengths. However, such an approach partially depends on the theoretical models for the lenses, which themselves are currently uncertain. While a tentative set of criteria are proposed in this work, with the lack of very high ($\approx$days to weeks) cadence observations, more robust and observationally well-defined selection criteria are needed to filter ESE events in future surveys.

\section{Conclusions}\label{sec:conclusions}
Using data from the ASKAP VAST survey, we have identified a sample of 236 sources in a search for ESEs. All of these sources show a flux density variation of $>$2, which we take to distinguish the variation from `normal' refractive scintillation. All of these sources have pronounced low-frequency variations (comparing VAST to VLASS), suggesting an ISM origin. Using a simple 1D Gaussian lens model, we model the observed light curves to derive the properties of the underlying lenses. Given the current limitations of the lens models, we discuss the pitfalls of adopting a purely quantitative approach, in terms of sample selection. Using these, we divide this sample into three subsamples --- gold, silver, and bronze based on the significance of the temporal and spectral variations. We show that the VAST light curves can stem from diverse scenarios\,---\,different combinations of strong and weak lenses, and compact and extended emission regions. Overall, we derive an all-sky rate of 0.012 events per year per deg$^2$ for our (gold+silver) sample, but we note that this is heavily direction-dependent. In terms of optical depth, this reflects a rate of 0.006 events/source/year. Given the rate, we establish that these events may be more common than historically detected, increasing the historical sample by at least five-fold. This opens up a new and unique avenue to study the ISM structure and inhomogeneities at $\approx$AU scales. We discussed the potential and limitations of current and upcoming sensitive large-scale surveys in uncovering larger populations of such events.

\begin{acknowledgments}
AA and DLK were supported by NSF grant AST-1816492. DLK is further supported by NSF grant AST-2511757.  Parts of this research were conducted by the Australian Research Council Centre of Excellence for Gravitational Wave Discovery (OzGrav), project number CE230100016. 
This scientific work uses data obtained from Inyarrimanha Ilgari Bundara / the Murchison Radio-astronomy Observatory. The Australian SKA Pathfinder is part of the Australia Telescope National Facility (\url{https://ror.org/05qajvd42}) which is managed by CSIRO. Operation of ASKAP is funded by the Australian Government with support from the National Collaborative Research Infrastructure Strategy. ASKAP uses the resources of the Pawsey Supercomputing Centre. The establishment of ASKAP, the Murchison Radio-astronomy Observatory, and the Pawsey Supercomputing Centre are initiatives of the Australian Government, with support from the Government of Western Australia and the Science and Industry Endowment Fund. AA acknowledges the use of Claude Code (Sonnet 5) for making the  \href{https://vast-survey.org/eses/}{ESE webpage}.
\end{acknowledgments}

\input{ese_prop}
\input{ese_prop_extended}

\begin{contribution}


\end{contribution}

%
\facilities{ASKAP, VLA}



\appendix
\twocolumngrid
\restartappendixnumbering

\section{Machine learning filter for VAST data}\label{sec:app_ml}
Traditional scalar statistics like $\eta$ and $V$ \citep{swinbank_lofar_2015} often fail to distinguish between stochastic variability and systematic variability since they primarily rely on deviation from a constant flux density model. Hence, filtering on $\eta$ and $V$ to look for astrophysical transients can often result in significant contamination from variable (intrinsic/extrinsic) active galactic nuclei (AGN), the most common sources in the radio sky at 1\,GHz above $\approx$1\,mJy. To overcome this, we developed a Gaussian process (GP) based machine learning (ML) classifier for VAST data.

To train the classifier, we simulated a data set with a similar sensitivity to VAST data, but which can be divided into four different categories. In each category, we simulate 10,0000 sources with flux densities sampled from the observed sample of VAST sources. To reproduce stochastic variations, we simulated light curves with median flux densities following the observed VAST distribution and 1-$\sigma$ standard deviation varying from $10^{-1}$ to 0.2 times the mean flux. The 1-$\sigma$ error bar on each of these is sampled randomly between 0.25\,mJy (typical sensitivity of VAST in the extragalactic region) and 1\,mJy. Sources in the other three categories are chosen to represent interesting phenomena: i) linear increase/decrease in flux density in the light curve, with the ratio of latest to earliest flux densities varying between $[0.2, 0.5] \cup [2, 7]$, ii) periodic variation in flux densities with periods varying from [0.5, 2] $\times$ the data span, and iii) a Gaussian like rise and decay (or inverse for ESE-like events). 

We modeled the individual light curves using Gaussian process (GP) regression and fitted for the hyperparameters describing the GP. The data covariance is modeled as a sum of two kernels: a radial basis kernel (RBF; squared exponential) to capture long-term variations and a white noise kernel to absorb epoch-to-epoch low-level variations. The width of the RBF is designed to have a lower bound of 0.2 $\times$ the data span, to avoid overfitting epoch-wise low-level variation, and the power of the noise kernel is designed to have an upper bound of median absolute deviation to avoid overfitting for long-term variations. After fitting each light curve with the model described above, we extract a set of features that quantify the fit and its deviation from a steady source model. These features are given in Table~\ref{tab:gp_fit}. 

\begin{deluxetable*}{ll}[!htb]
    \caption{Feature set from the GP fit used to train the model.}
    \label{tab:gp_fit}
    \tablehead{\colhead{Feature} & \colhead{Description}}
    \startdata
    $l_{\rm RBF}$ & Length scale of the RBF kernel \\
    $A_{\rm RBF}$ & Amplitude of the RBF kernel \\
    $\sigma_{\rm WN}$ & Noise level of the White noise kernel \\
    Min-Max & Difference between the maximum and minimum of the predicted light curve \\
    Frac-diff & Min-Max relative to the median of the predicted light curve \\
    Peak-SNR & Maximum signal-to-noise of the predicted light curve \\
    $N_{\rm peaks}$ & Number of peaks in the predicted light curve (20\% above local median) \\
    slope & Slope of the linear fit to the predicted light curve \\
    $\chi^2$ & The goodness of the linear fit \\
    jitter & Difference in the fraction of the light curve points above and below the median\\
    2-point slope & The average 2-point slope between consecutive points on the light curve \\
    {skew, kurtosis} & Higher-order moments of the predicted light curve \\
    loglike & The log-likelihood of the GP fit \\
    \enddata
\end{deluxetable*}
    
We then used \texttt{xgboost} \citep{xgboost}, a tree classifier, to classify the light curves using these features and train the model. Simulating a similar data set to test the model, we achieved $>$99\% accuracy when predicting the test data set using the trained model.

\begin{figure*}
    \centering
    \plottwo{{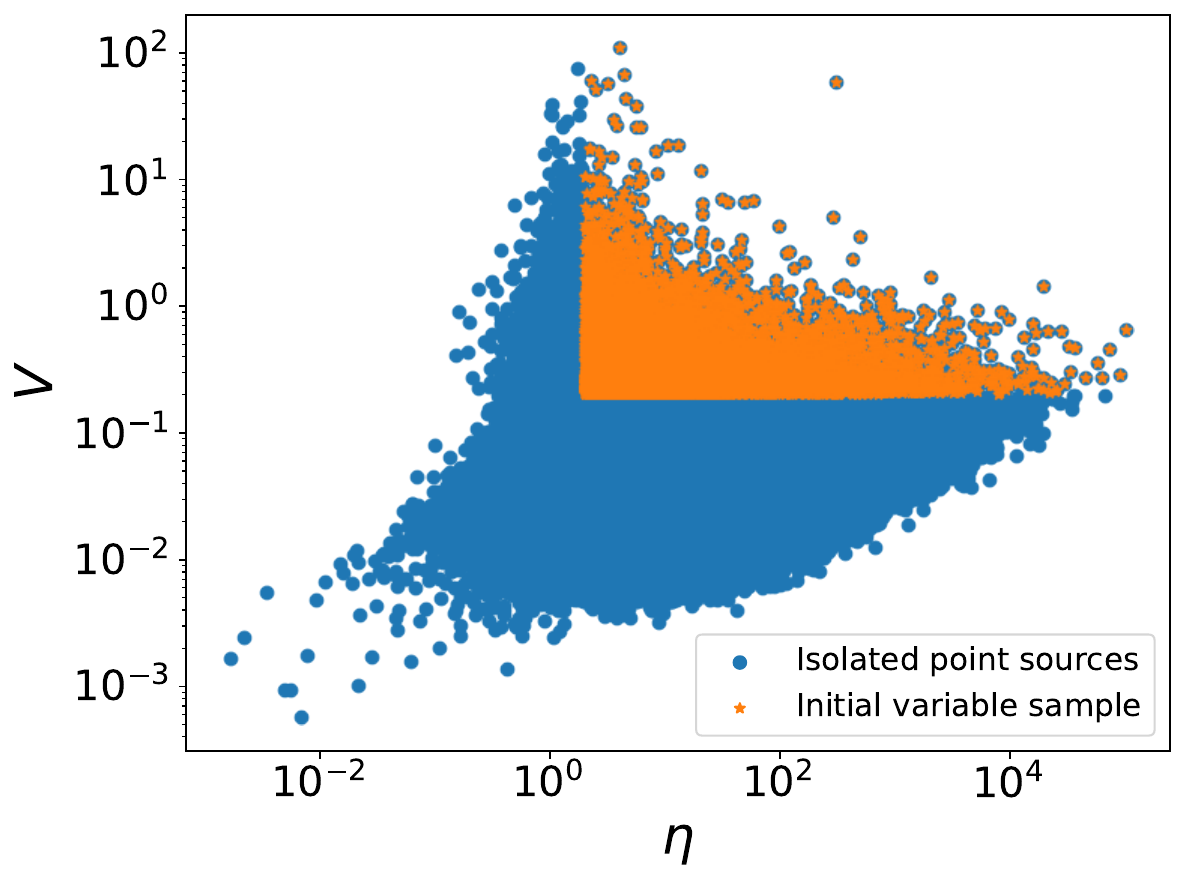}}{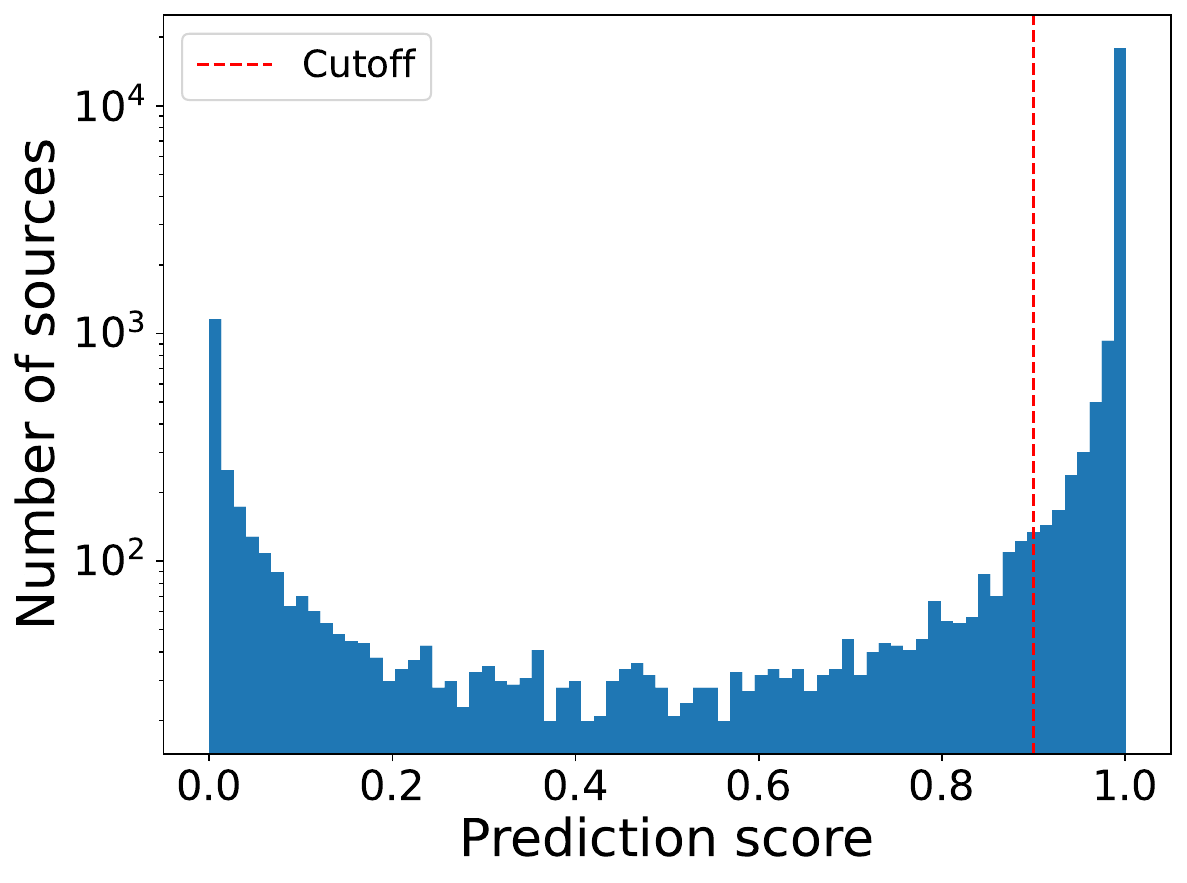}
    \caption{\textbf{\textit{Left:}} Our initial sample of variable sources selected using basic cuts of $\eta>2$ and $V>0.2$. The sample highlighted in orange is used as an input for the ML filter. \textbf{\textit{Right:}} Histogram of the probabilities for a source to be stochastically varying as opposed to any other systematic variation. The dashed red line shows our cutoff of 0.9 above which source are deemed to be stochastically varying.}
    \label{fig:gp_predict}
\end{figure*}

We modeled the observed light curves from the VAST data to extract the same set of features and used this trained model to predict and filter out light curves showing stochastic variability. The input sample is shown in the left panel of Figure~\ref{fig:gp_predict}. Since ESE events couldn't be well represented by a specific kernel, our goal was to reject stochastic variability from this initial sample rather than identify ESE behavior. Hence, we reject sources where the probability of variations being stochastic is $>90$\%. Figure~\ref{fig:gp_predict} (right panel) shows the distributions of probabilities for sources to be stochastically varying, and our cut of 90\% reliability removed 82\% of the parent sample.

\section{Lens models} \label{app:lenses}

The temporal variations in the flux density of a background source induced by an intervening lens at different frequencies can be fully described, given the electron density distribution of the intervening medium. One such prescription is given by \cite{Clegg1998}, who used a 1D Gaussian lens model to describe the observed light curves. 

The model starts with the assumption that the radiation from the background source undergoing refraction is located far from both the observer and the lens and can be well described by plane waves in the geometrical optics regime. In general, the parameters influencing this refraction at any single instance include: i) the peak integrated column density, $N_0$, ii) the transverse length scale of the lens, $a$,, iii) the distance between the lens and the observer, $D_L$, and iv) the angle of incidence with respect to the line connecting the lens and the observer, $\theta_{i}$. However, many of these are degenerate and can be compressed to two independent parameters. The first one is the strength of the lens, $\alpha= (\lambda^2 r_e N_0 D_L)/(\pi a^2)$, where $\lambda$ is the observing wavelength and $r_e$ is the classical electron radius. The second is $\beta_i= \theta_i D_L/a$\,---\,the angle of incidence in terms of the angular size of the lens, as seen by the observer. For extended sources, $\beta_i$ is integrated over the brightness profile of the background source, usually assumed to be a Gaussian profile with a characteristic scale $\theta_{\rm src}$. 

The time evolution of this refraction phenomenon, which results in the observed light curve, needs two additional parameters: $v$, the transverse relative velocity between the observer and the lens along the lens, and $T_0$, a reference time, usually describing the minimum of the light curve (when the refraction effect is most pronounced). In general, it is also possible that the emission at a particular wavelength can arise from multiple regions\,---\, a central region that undergoes lensing and a broader region that is unaffected by the lens. Here, the extent of the background source is relevant to the lens and not to the instrumental resolution. This means that a source can be ``broad'' but unresolved. To avoid confusion, in this article, we refer to these components as \textit{lensed} and \textit{unlensed} components. In this case, the emission from the lensed region will predominantly show the scattering effect, with the unlensed region largely unaffected. Hence, we use two additional parameters to account for this: $f_{\rm unlens}$, the fraction of the source's flux that is unaffected by the lens, and $F_{\nu; \rm tot}$, the total flux density of the source prior to the ESE. To summarize, the resulting model can be described by six independent parameters, \{$\alpha$, $\beta_{\rm src}$, $v$, $f_{\rm unlens}$, $F_{\nu; \rm tot}$, $T_0$\}. We note that the angular extent of the background source is wavelength-dependent for both lensed and unlensed components, with the emitting region at lower frequencies being larger \citep{Blandford1979,Lobanov1998,marscher_models_1985}. However, this will not have a meaningful impact for our data set, given the small VAST bandwidth. 

Additionally, if the transverse velocity of the lens is comparable to the velocity of the Earth around the Sun, then the motion of the Earth causes undulations in the observed light curve. However, in the case of a 1D lens, only motion along the lens profile can be constrained. Since the general path of the intervening lens, with respect to the Earth's velocity vector, is usually unknown, we cannot constrain the absolute relative velocity. Given the VAST cadence, this effect in most cases will not be noticeable; hence we mention it here for completeness, but ignore it in the model fitting.

We used \texttt{emcee} \citep{emcee}, an ensemble Monte Carlo sampler, to sample the posterior distributions of the model parameters. We initiated 32 walkers over a wide range of model parameters, and let the walkers ``burn-in'' for 500 iterations. We then let them sample for 2000 iterations and estimated the sample median and standard deviations to describe the best-fit values of the model parameters. Evaluating the light curves for individual lens and source configurations can be computationally expensive, and hence, we pre-compute these light curves on a fixed grid of $N_0$ (hence $\alpha$), and $\beta_{\rm src}$ (hence $\theta_{\rm src}$). This means that the sampler switches randomly between states discretely (on this computed grid) for $\alpha$, and $\beta$, and continuously for other variables. 

\subsection{Generic lens profiles}\label{app:gen_lenses}
This section describes the diversity of the predicted ESE light curves depending on the profile of the plasma lens. In general, the underlying lens can either be a diverging lens, a converging lens, or a combination of both. A Gaussian lens, for example, has both diverging and converging regions. In either case, the shape of the observed light curve depends on the gradient of this lens profile. While the overall characteristics, like the central flux suppression and the caustics, remain valid, the shape of the light curve varies for different profiles.

\begin{figure*}
    \includegraphics[width=\linewidth]{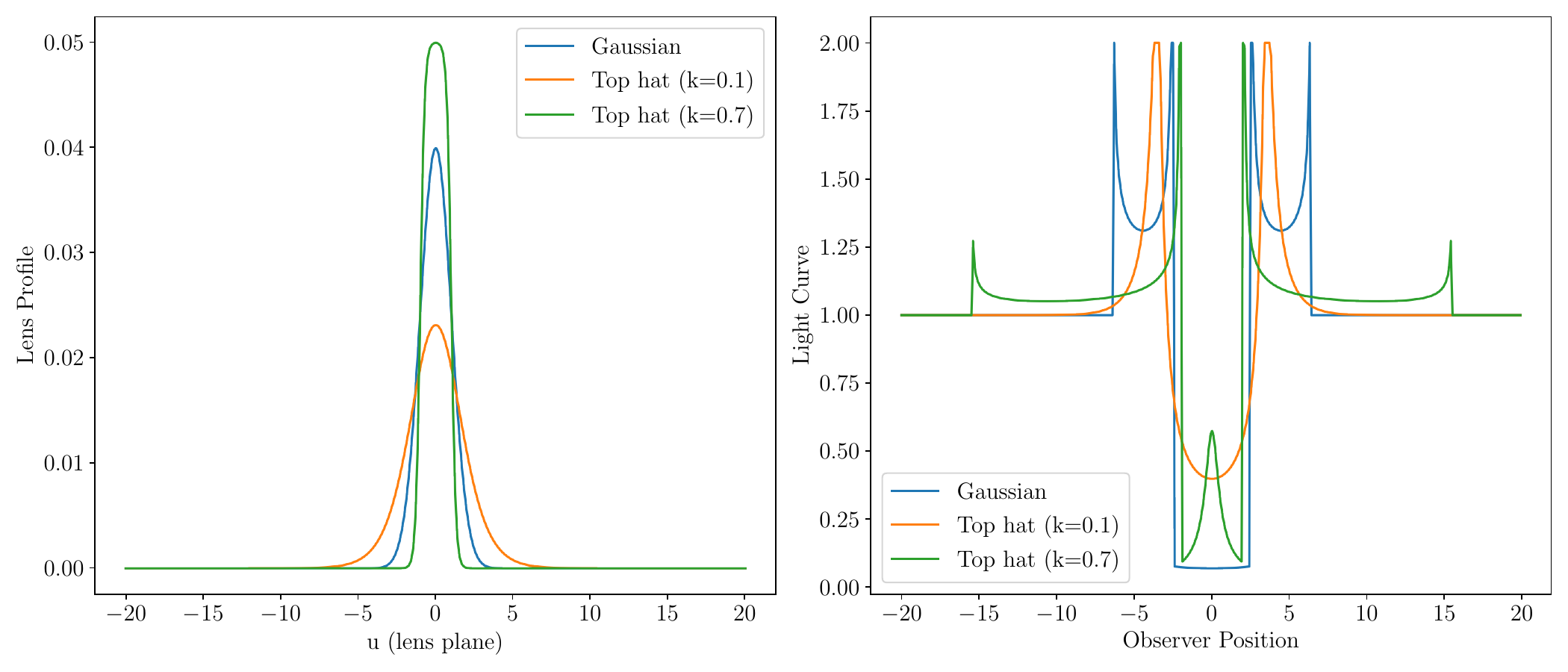}
    \caption{Figure shows the lens profiles (\textit{left}) and the observed magnifications (\textit{right}) for a Gaussian and different lens top-hat profiles. Depending on the gradient of the lens profile, changes in both the caustic regions and the central suppression can be seen.}
    \label{fig:lens_profiles}
\end{figure*}

As a demonstration of this effect, we show different lens profiles in Figure~\ref{fig:lens_profiles}. We considered various top-hat lens profiles, but with the edges smoothed to make it realistic. The profile is given by 
\begin{equation}
    f(x) \propto \frac{1}{1 + \exp\{{-k (x+a)}\}} - \frac{1}{1 + \exp\{{-k (x-a)}\}} \nonumber
\end{equation}
where the top-hat lens spans from ($-$a, a) and the parameter ``k'' dictates the tapering of the edges. Smaller values of \textit{k} result in a broadened profile (see left panel of Figure~\ref{fig:lens_profiles}) while larger values will yield a lens close to a pure top-hat function.

As the lens becomes sharper, two effects can be noticed -- the gradient in the density becomes sharper, hence the rays are refracted more strongly, resulting in a wider separation in the caustics. In addition, close to the center of the lens, since the density variations are less (in the case of a pure top-hat), rays are less defocused and hence a visible excess (compared to a Gaussian lens) is seen. As the lens becomes broader, the peak strength decreases for the same total density, and hence the lens behaves similarly to a weak lens with a similar peak density. 

\section{Light curve and Spectral fitting}
Here we describe our advanced light curve fitting, assuming correlated noise, and the spectral index time series fitting methods.  
\subsection{Light curve fitting}\label{app:lc_fit_noise}
In Appendix~\ref{app:lenses}, we described our light curve fitting techniques, assuming that the data points are strictly independent. Here, we describe it for the case of correlated noise. This correlated noise can stem from multiple processes --- intrinsic variability, scintillation, or both. We model this agnostically by using a Gaussian process-based fit. Simultaneous modeling of the lens model and the noise requires prior domain knowledge of how large the noise can be. However, due to the lack of historical light curves for most of our sources, this prior information can not be used. Hence, we examined the light curves to see what the scatter in our light curves is and found that a 10\% is reasonable in many cases. Hence, we adopt a 1-$\sigma$ noise level of 2\%. While there can be both short (few days to a month) and long-term (years) noise present in the light curves, we ignore the latter since that will be a competing effect to the ESE and will absorb the power from the lens model. Hence, we restrict ourselves to short-term noise --- within two data points in the light curve, i.e 120 days. We use a Matern-3/2 kernel to model this short-term noise, parametrized by an amplitude and a timescale. We simultaneously fit for (lens + noise) models, with upper limits of 2\% on the amplitude (1-$\sigma$) and 120\,days for the timescale. We then use the post-fit correlation matrix to compute the likelihood of the fit. We translate this to a correlated $\chi^2$ = $-2\times\log$(likelihood value), and use this as a goodness-of-fit parameter.
\subsection{Spectral evolution}\label{app:spec_fit_gp}
In general, the spectral evolution is quite sensitive to the actual plasma distribution lensing the radiation and the intrinsic spectrum, the knowledge of both of which is highly uncertain. Hence, we adopt a model-agnostic approach and perform a nonparametric fit to the spectral index time series and ask the simple question -- what is the timescale of variation? To do so, we describe the spectral index time series as a Gaussian process with two components ---  a white noise process, and a correlated process. We use a white noise kernel to describe the former and a radial basis function (RBF) kernel to describe the latter. We limit the amplitude (1-$\sigma$) of the white noise kernel to the median error on the spectral index measurements. The amplitude of the RBF kernel is limited to twice the maximal difference in the measurements, and the dimensionless timescale is allowed to vary between 0 and 100. The time-span of the actual observation is normalized from 0 to 1, so that an inferred value of 100 represents an extremely long-term (i.e, non-variable) varying component; namely, variation on a time-scale that is 100 times longer than the duration of the observation. We fit the spectral index time series to get the best-fit kernel parameters and use them to classify candidates as --- stochastically varying (timescale $<$0.05), commensurately varying with the ESE (0.05$<$timescale$<$1) and not varying (timescale$>$1).


\clearpage

\bibliography{references}{}
\bibliographystyle{aasjournalv7}



\end{document}

%% file: ese_prop.tex
\startlongtable
\setlength{\tabcolsep}{2pt}
\begin{deluxetable*}{lrrccrrrrrr}
\tabletypesize{\footnotesize}
\tablecaption{\textbf{VAST ESE gold sample}: Observed and derived properties of the ESE candidates found in ASKAP VAST. Column definitions: Name -- ASKAP name, RA -- Right Ascension, DEC -- Declination, $\alpha$ -- Strength of the lens, $\beta$ -- angular size of the source in units of the lens size, $v_{\rm obs}$ -- relative velocity between the lens and the observer, $f_{\rm unlens}$ -- fraction of source's flux that is unaffected by the lens, $F_{\nu;tot}$ -- total unlensed source flux density, and $T_0$ -- estimated time of lens alignment with the observer.}
\label{tab:ese_prop}
\tablehead{
\colhead{Name} & \colhead{RA} & \colhead{DEC} & \colhead{$\alpha$} & \colhead{$\beta$} & \colhead{$v_{\rm obs}$} & \colhead{$f_{\rm unlens}$} & \colhead{$F_{\nu;tot}$} & \colhead{$T_0$} \\ 
\colhead{ASKAP-} & \colhead{(deg)} & \colhead{(deg)} & \colhead{} & \colhead{} & \colhead{(/17.3 km/s)} & \colhead{} & \colhead{(mJy)} & \colhead{(MJD)}
}
\startdata
& & & & & & & & \\
\multicolumn{9}{c}{\textbf{\textit{\large Quantitative selection}}}\\
\hline\\
J021149.1+060015 & \phantom{0}32.9546 & \phantom{$-$}\phantom{0}6.004\phantom{0} & 10.0\phantom{0}$\pm$\phantom{0}5.9\phantom{0} & 3.38\phantom{0}$\pm$0.91\phantom{0} & 17.6\phantom{0}$\pm$4.6\phantom{0} & 0.24\phantom{0}$\pm$0.11\phantom{0} & \phantom{0}5.934$\pm$0.049 & 60146.1$\pm$\phantom{0}8.1 \\
J042632.5+022300 & \phantom{0}66.6354 & \phantom{$-$}\phantom{0}2.3831 & \phantom{0}2.6\phantom{0}$\pm$\phantom{0}1.1\phantom{0} & 0.832$\pm$0.074 & \phantom{0}3.5\phantom{0}$\pm$1.3\phantom{0} & 0.311$\pm$0.054 & \phantom{0}6.334$\pm$0.072 & 60111\phantom{.0}$\pm$36\phantom{.0} \\
J043124.7+062208 & \phantom{0}67.8529 & \phantom{$-$}\phantom{0}6.369\phantom{0} & \phantom{0}9.23$\pm$\phantom{0}0.72 & 4.03\phantom{0}$\pm$0.24\phantom{0} & 10.01$\pm$0.44 & 0.077$\pm$0.064 & \phantom{0}8.154$\pm$0.044 & 60743.3$\pm$\phantom{0}3.2 \\
J051450.7+024829 & \phantom{0}78.7112 & \phantom{$-$}\phantom{0}2.8081 & \phantom{0}9.5\phantom{0}$\pm$\phantom{0}1.3\phantom{0} & 1.73\phantom{0}$\pm$0.68\phantom{0} & \phantom{0}7.22$\pm$0.91 & 0.602$\pm$0.085 & \phantom{0}5.844$\pm$0.070 & 60211\phantom{.0}$\pm$28\phantom{.0} \\
J060158.9$-$052751 & \phantom{0}90.4954 & $-$\phantom{0}5.4641 & 10.39$\pm$\phantom{0}0.44 & 1.19\phantom{0}$\pm$0.13\phantom{0} & 10.66$\pm$0.32 & 0.629$\pm$0.015 & \phantom{0}5.686$\pm$0.029 & 60311.3$\pm$\phantom{0}5.2 \\
J071328.3+065142 & 108.3679 & \phantom{$-$}\phantom{0}6.8616 & 26.62$\pm$\phantom{0}0.84 & 1.33\phantom{0}$\pm$0.23\phantom{0} & 18.82$\pm$0.54 & 0.437$\pm$0.018 & \phantom{0}5.149$\pm$0.033 & 60759.0$\pm$\phantom{0}4.8 \\
J131126.9$-$310036 & 197.8621 & $-$31.0101 & \phantom{0}6.5\phantom{0}$\pm$\phantom{0}3.9\phantom{0} & 1.99\phantom{0}$\pm$0.51\phantom{0} & 11.7\phantom{0}$\pm$2.3\phantom{0} & 0.55\phantom{0}$\pm$0.13\phantom{0} & \phantom{0}7.652$\pm$0.032 & 60218.3$\pm$\phantom{0}3.7 \\
J151446.8+015128 & 228.695\phantom{0} & \phantom{$-$}\phantom{0}1.8577 & \phantom{0}5.07$\pm$\phantom{0}0.74 & 0.67\phantom{0}$\pm$0.13\phantom{0} & \phantom{0}7.77$\pm$0.59 & 0.579$\pm$0.023 & \phantom{0}6.452$\pm$0.059 & 60923.0$\pm$\phantom{0}8.4 \\
J154244.0+011458 & 235.6833 & \phantom{$-$}\phantom{0}1.2493 & 11.16$\pm$\phantom{0}0.98 & 0.92\phantom{0}$\pm$0.49\phantom{0} & \phantom{0}9.35$\pm$0.26 & 0.631$\pm$0.051 & \phantom{0}8.501$\pm$0.063 & 60842.4$\pm$\phantom{0}5.6 \\
J154908.8$-$011628 & 237.2867 & $-$\phantom{0}1.2744 & \phantom{0}9.1\phantom{0}$\pm$\phantom{0}2.7\phantom{0} & 0.82\phantom{0}$\pm$0.14\phantom{0} & \phantom{0}3.47$\pm$0.48 & 0.679$\pm$0.016 & \phantom{0}6.28\phantom{0}$\pm$0.10\phantom{0} & 60903\phantom{.0}$\pm$88\phantom{.0} \\
J161752.7$-$060511 & 244.4696 & $-$\phantom{0}6.0865 & \phantom{0}5.2\phantom{0}$\pm$\phantom{0}1.2\phantom{0} & 1.75\phantom{0}$\pm$0.40\phantom{0} & \phantom{0}4.1\phantom{0}$\pm$5.3\phantom{0} & 0.601$\pm$0.012 & 17.726$\pm$0.076 & 60093\phantom{.0}$\pm$27\phantom{.0} \\
J164405.3+021409 & 251.0221 & \phantom{$-$}\phantom{0}2.2358 & \phantom{0}3.6\phantom{0}$\pm$\phantom{0}1.4\phantom{0} & 1.74\phantom{0}$\pm$0.23\phantom{0} & 11.2\phantom{0}$\pm$1.5\phantom{0} & 0.378$\pm$0.032 & 23.73\phantom{0}$\pm$0.19\phantom{0} & 60703.6$\pm$\phantom{0}2.8 \\
J164958.4$-$044847 & 252.4933 & $-$\phantom{0}4.8131 & 19.6\phantom{0}$\pm$\phantom{0}4.9\phantom{0} & 2.73\phantom{0}$\pm$0.28\phantom{0} & \phantom{0}5.34$\pm$0.22 & 0.606$\pm$0.064 & 13.952$\pm$0.082 & 60098.1$\pm$\phantom{0}4.1 \\
J165844.7+081906 & 254.6862 & \phantom{$-$}\phantom{0}8.3182 & \phantom{0}9.18$\pm$\phantom{0}0.72 & 3.86\phantom{0}$\pm$0.16\phantom{0} & 17.51$\pm$0.80 & 0.047$\pm$0.037 & \phantom{0}9.354$\pm$0.048 & 60811.2$\pm$\phantom{0}2.1 \\
J170433.6$-$085552 & 256.14\phantom{00} & $-$\phantom{0}8.9312 & 18\phantom{.00}$\pm$14\phantom{.00} & 4.48\phantom{0}$\pm$0.84\phantom{0} & \phantom{0}9.1\phantom{0}$\pm$2.2\phantom{0} & 0.35\phantom{0}$\pm$0.10\phantom{0} & \phantom{0}5.63\phantom{0}$\pm$0.10\phantom{0} & 60080\phantom{.0}$\pm$14\phantom{.0} \\
J170543.5+082728 & 256.4312 & \phantom{$-$}\phantom{0}8.4576 & 17.4\phantom{0}$\pm$\phantom{0}2.4\phantom{0} & 4.0\phantom{00}$\pm$1.3\phantom{00} & 10.63$\pm$0.82 & 0.28\phantom{0}$\pm$0.13\phantom{0} & \phantom{0}6.320$\pm$0.060 & 60240.6$\pm$\phantom{0}9.4 \\
J170715.7$-$013503 & 256.8154 & $-$\phantom{0}1.5841 & \phantom{0}2.6\phantom{0}$\pm$\phantom{0}1.8\phantom{0} & 1.11\phantom{0}$\pm$0.36\phantom{0} & \phantom{0}4.8\phantom{0}$\pm$1.2\phantom{0} & 0.34\phantom{0}$\pm$0.14\phantom{0} & \phantom{0}4.382$\pm$0.063 & 60880.1$\pm$\phantom{0}7.1 \\
J173500.8$-$073321 & 263.7533 & $-$\phantom{0}7.5558 & 33.4\phantom{0}$\pm$\phantom{0}1.1\phantom{0} & 5.19\phantom{0}$\pm$0.43\phantom{0} & \phantom{0}8.05$\pm$0.36 & 0.282$\pm$0.030 & 10.330$\pm$0.075 & 60801.9$\pm$\phantom{0}6.3 \\
J174349.3+061031 & 265.9554 & \phantom{$-$}\phantom{0}6.1753 & \phantom{0}9.41$\pm$\phantom{0}0.54 & 5.21\phantom{0}$\pm$0.87\phantom{0} & \phantom{0}7.49$\pm$0.16 & 0.03\phantom{0}$\pm$0.12\phantom{0} & 14.487$\pm$0.070 & 60629.9$\pm$\phantom{0}4.2 \\
J180044.3+043159 & 270.1846 & \phantom{$-$}\phantom{0}4.5331 & \phantom{0}3.79$\pm$\phantom{0}0.93 & 1.35\phantom{0}$\pm$0.18\phantom{0} & \phantom{0}4.20$\pm$0.49 & 0.590$\pm$0.051 & \phantom{0}4.087$\pm$0.048 & 60248\phantom{.0}$\pm$30\phantom{.0} \\
J180430.7+085857 & 271.1279 & \phantom{$-$}\phantom{0}8.9826 & \phantom{0}8.40$\pm$\phantom{0}0.46 & 0.75\phantom{0}$\pm$0.32\phantom{0} & \phantom{0}6.72$\pm$0.19 & 0.686$\pm$0.034 & \phantom{0}4.905$\pm$0.057 & 60802\phantom{.0}$\pm$22\phantom{.0} \\
J193155.5$-$004537 & 292.9812 & $-$\phantom{0}0.7603 & \phantom{0}8.0\phantom{0}$\pm$\phantom{0}2.6\phantom{0} & 3.45\phantom{0}$\pm$0.71\phantom{0} & 11.0\phantom{0}$\pm$2.1\phantom{0} & 0.31\phantom{0}$\pm$0.11\phantom{0} & \phantom{0}8.416$\pm$0.084 & 60530.4$\pm$\phantom{0}4.6 \\
J195006.8$-$071643 & 297.5283 & $-$\phantom{0}7.2785 & 32\phantom{.00}$\pm$19\phantom{.00} & 2.41\phantom{0}$\pm$0.17\phantom{0} & \phantom{0}8.15$\pm$0.73 & 0.552$\pm$0.036 & \phantom{0}7.685$\pm$0.069 & 60559.5$\pm$\phantom{0}5.3 \\
J201133.7$-$335857 & 302.8904 & $-$33.9826 & \phantom{0}8.21$\pm$\phantom{0}0.66 & 2.202$\pm$0.097 & \phantom{0}9.66$\pm$0.29 & 0.387$\pm$0.019 & 13.333$\pm$0.035 & 60561.8$\pm$\phantom{0}1.5 \\
J203515.8$-$290340 & 308.8158 & $-$29.0611 & 13.7\phantom{0}$\pm$\phantom{0}1.2\phantom{0} & 5.19\phantom{0}$\pm$0.39\phantom{0} & 23.6\phantom{0}$\pm$1.1\phantom{0} & 0.123$\pm$0.043 & 18.091$\pm$0.063 & 60560.5$\pm$\phantom{0}1.2 \\
 & & & & & & & & \\
\hline
\multicolumn{9}{c}{\textbf{\textit{\large Qualitative selection}}}\\
\hline\\
J040242.0+082524 & \phantom{0}60.675\phantom{0} & \phantom{$-$}\phantom{0}8.4234 & \phantom{0}6.8\phantom{0}$\pm$\phantom{0}1.7\phantom{0} & 1.99\phantom{0}$\pm$0.38\phantom{0} & \phantom{0}4.8\phantom{00}$\pm$\phantom{0}6.7\phantom{00} & 0.581\phantom{00}$\pm$0.049\phantom{00} & \phantom{00}6.45\phantom{0}$\pm$0.24\phantom{0} & 60666\phantom{.00}$\pm$\phantom{0}13\phantom{.00} \\
J044937.4+081134 & \phantom{0}72.4058 & \phantom{$-$}\phantom{0}8.1928 & \phantom{0}9.2\phantom{0}$\pm$\phantom{0}2.3\phantom{0} & 3.69\phantom{0}$\pm$0.86\phantom{0} & \phantom{0}8.3\phantom{00}$\pm$\phantom{0}1.0\phantom{00} & 0.28\phantom{000}$\pm$0.15\phantom{000} & \phantom{00}4.147$\pm$0.049 & 60815.2\phantom{0}$\pm$\phantom{00}8.3\phantom{0} \\
J050831.9+060012 & \phantom{0}77.1329 & \phantom{$-$}\phantom{0}6.0033 & 15.5\phantom{0}$\pm$\phantom{0}1.2\phantom{0} & 1.46\phantom{0}$\pm$0.17\phantom{0} & 15.34\phantom{0}$\pm$\phantom{0}0.93\phantom{0} & 0.526\phantom{00}$\pm$0.023\phantom{00} & \phantom{00}5.059$\pm$0.038 & 60497.1\phantom{0}$\pm$\phantom{00}3.8\phantom{0} \\
J052803.1$-$330348 & \phantom{0}82.0129 & $-$33.0634 & 23.1\phantom{0}$\pm$\phantom{0}1.3\phantom{0} & 0.30\phantom{0}$\pm$0.15\phantom{0} & 32.4\phantom{00}$\pm$\phantom{0}2.5\phantom{00} & 0.1546\phantom{0}$\pm$0.0076\phantom{0} & \phantom{0}20.118$\pm$0.058 & 60174.3\phantom{0}$\pm$\phantom{00}8.1\phantom{0} \\
J090943.2+030922 & 137.43\phantom{00} & \phantom{$-$}\phantom{0}3.1562 & 27\phantom{.00}$\pm$27\phantom{.00} & 0.86\phantom{0}$\pm$0.13\phantom{0} & \phantom{0}8.6\phantom{00}$\pm$\phantom{0}1.2\phantom{00} & 0.586\phantom{00}$\pm$0.069\phantom{00} & \phantom{00}6.63\phantom{0}$\pm$0.18\phantom{0} & 60634.4\phantom{0}$\pm$\phantom{00}5.0\phantom{0} \\
J091708.1$-$030618 & 139.2837 & $-$\phantom{0}3.105\phantom{0} & \phantom{0}5.51$\pm$\phantom{0}0.46 & 1.04\phantom{0}$\pm$0.14\phantom{0} & 24.0\phantom{00}$\pm$\phantom{0}1.3\phantom{00} & 0.074\phantom{00}$\pm$0.034\phantom{00} & \phantom{00}9.469$\pm$0.013 & 60718.0\phantom{0}$\pm$\phantom{00}2.7\phantom{0} \\
J115029.4$-$335716 & 177.6225 & $-$33.9545 & \phantom{0}2.45$\pm$\phantom{0}0.32 & 1.39\phantom{0}$\pm$0.43\phantom{0} & \phantom{0}7.37\phantom{0}$\pm$\phantom{0}0.72\phantom{0} & 0.475\phantom{00}$\pm$0.048\phantom{00} & \phantom{00}7.948$\pm$0.064 & 60679.7\phantom{0}$\pm$\phantom{00}8.9\phantom{0} \\
J121758.4$-$363418 & 184.4933 & $-$36.5716 & 13.0\phantom{0}$\pm$\phantom{0}2.4\phantom{0} & 1.15\phantom{0}$\pm$0.23\phantom{0} & 11.35\phantom{0}$\pm$\phantom{0}0.55\phantom{0} & 0.585\phantom{00}$\pm$0.012\phantom{00} & 332.3\phantom{00}$\pm$1.1\phantom{00} & 60289.7\phantom{0}$\pm$\phantom{00}5.6\phantom{0} \\
J132450.9$-$063941 & 201.2121 & $-$\phantom{0}6.6613 & 11.28$\pm$\phantom{0}0.23 & 2.825$\pm$0.078 & 24.41\phantom{0}$\pm$\phantom{0}0.17\phantom{0} & 0.041\phantom{00}$\pm$0.019\phantom{00} & \phantom{0}18.001$\pm$0.026 & 60424.69$\pm$\phantom{00}0.60 \\
J133100.3$-$380928 & 202.7512 & $-$38.1578 & \phantom{0}5.60$\pm$\phantom{0}0.36 & 1.816$\pm$0.047 & 14.66\phantom{0}$\pm$\phantom{0}0.39\phantom{0} & 0.5631\phantom{0}$\pm$0.0083\phantom{0} & \phantom{0}40.09\phantom{0}$\pm$0.10\phantom{0} & 60616.96$\pm$\phantom{00}0.92 \\
J134946.1+090816 & 207.4421 & \phantom{$-$}\phantom{0}9.1379 & \phantom{0}6.16$\pm$\phantom{0}0.14 & 1.149$\pm$0.014 & 25.25\phantom{0}$\pm$\phantom{0}0.12\phantom{0} & 0.3805\phantom{0}$\pm$0.0057\phantom{0} & \phantom{0}26.075$\pm$0.040 & 60611.69$\pm$\phantom{00}0.33 \\
J143452.7$-$002828 & 218.7196 & $-$\phantom{0}0.4746 & 18.6\phantom{0}$\pm$\phantom{0}1.3\phantom{0} & 0.62\phantom{0}$\pm$0.25\phantom{0} & 24.19\phantom{0}$\pm$\phantom{0}0.46\phantom{0} & 0.3535\phantom{0}$\pm$0.0059\phantom{0} & \phantom{0}34.06\phantom{0}$\pm$0.18\phantom{0} & 60642.7\phantom{0}$\pm$\phantom{00}2.4\phantom{0} \\
J151331.7$-$091920 & 228.3821 & $-$\phantom{0}9.3221 & \phantom{0}1.53$\pm$\phantom{0}0.75 & 0.18\phantom{0}$\pm$0.19\phantom{0} & \phantom{0}7.2\phantom{00}$\pm$\phantom{0}1.0\phantom{00} & 0.18\phantom{000}$\pm$0.12\phantom{000} & \phantom{00}9.535$\pm$0.073 & 60133\phantom{.00}$\pm$\phantom{0}13\phantom{.00} \\
J160606.9+061949 & 241.5287 & \phantom{$-$}\phantom{0}6.3302 & \phantom{0}6.72$\pm$\phantom{0}0.15 & 1.947$\pm$0.016 & 13.210$\pm$\phantom{0}0.023 & 0.00034$\pm$0.00030 & \phantom{0}49.066$\pm$0.047 & 60043.52$\pm$\phantom{00}0.54 \\
J160856.1+091038 & 242.2337 & \phantom{$-$}\phantom{0}9.1774 & \phantom{0}8.70$\pm$\phantom{0}0.23 & 0.392$\pm$0.035 & \phantom{0}6.30\phantom{0}$\pm$\phantom{0}0.14\phantom{0} & 0.5902\phantom{0}$\pm$0.0073\phantom{0} & \phantom{0}18.853$\pm$0.094 & 60912.5\phantom{0}$\pm$\phantom{00}7.2\phantom{0} \\
J161011.9$-$034917 & 242.5496 & $-$\phantom{0}3.8214 & \phantom{0}7.6\phantom{0}$\pm$\phantom{0}2.1\phantom{0} & 1.067$\pm$0.057 & \phantom{0}3.82\phantom{0}$\pm$\phantom{0}0.42\phantom{0} & 0.465\phantom{00}$\pm$0.036\phantom{00} & \phantom{0}11.98\phantom{0}$\pm$0.40\phantom{0} & 59664\phantom{.00}$\pm$\phantom{0}11\phantom{.00} \\
J161131.8$-$034058 & 242.8825 & $-$\phantom{0}3.6829 & \phantom{0}6.4\phantom{0}$\pm$\phantom{0}2.0\phantom{0} & 0.38\phantom{0}$\pm$0.90\phantom{0} & \phantom{0}5.8\phantom{00}$\pm$\phantom{0}1.9\phantom{00} & 0.765\phantom{00}$\pm$0.097\phantom{00} & \phantom{0}11.65\phantom{0}$\pm$0.14\phantom{0} & 60645\phantom{.00}$\pm$\phantom{0}30\phantom{.00} \\
J162223.7+001329 & 245.5987 & \phantom{$-$}\phantom{0}0.2246 & \phantom{0}9.36$\pm$\phantom{0}0.62 & 3.06\phantom{0}$\pm$0.12\phantom{0} & \phantom{0}8.53\phantom{0}$\pm$\phantom{0}0.25\phantom{0} & 0.039\phantom{00}$\pm$0.029\phantom{00} & \phantom{00}4.898$\pm$0.030 & 60805.7\phantom{0}$\pm$\phantom{00}3.5\phantom{0} \\
J162346.0$-$083538 & 245.9417 & $-$\phantom{0}8.594\phantom{0} & \phantom{0}1.23$\pm$\phantom{0}0.18 & 0.101$\pm$0.066 & \phantom{0}2.7\phantom{00}$\pm$\phantom{0}2.5\phantom{00} & 0.201\phantom{00}$\pm$0.019\phantom{00} & \phantom{0}10.95\phantom{0}$\pm$0.14\phantom{0} & 60669.7\phantom{0}$\pm$\phantom{00}6.3\phantom{0} \\
J163051.5$-$081009 & 247.7146 & $-$\phantom{0}8.1692 & \phantom{0}2.63$\pm$\phantom{0}0.84 & 1.26\phantom{0}$\pm$0.12\phantom{0} & \phantom{0}5.19\phantom{0}$\pm$\phantom{0}0.56\phantom{0} & 0.547\phantom{00}$\pm$0.021\phantom{00} & 117.78\phantom{0}$\pm$0.39\phantom{0} & 60494.9\phantom{0}$\pm$\phantom{00}4.7\phantom{0} \\
J164351.8+085506 & 250.9658 & \phantom{$-$}\phantom{0}8.9183 & \phantom{0}4.80$\pm$\phantom{0}0.22 & 1.347$\pm$0.091 & \phantom{0}8.59\phantom{0}$\pm$\phantom{0}0.14\phantom{0} & 0.6176\phantom{0}$\pm$0.0074\phantom{0} & \phantom{0}44.716$\pm$0.081 & 60243.4\phantom{0}$\pm$\phantom{00}7.1\phantom{0} \\
J170023.9+052244 & 255.0996 & \phantom{$-$}\phantom{0}5.3789 & 15.74$\pm$\phantom{0}0.34 & 1.509$\pm$0.045 & 12.591$\pm$\phantom{0}0.013 & 0.4689\phantom{0}$\pm$0.0022\phantom{0} & 126.39\phantom{0}$\pm$0.11\phantom{0} & 60307.26$\pm$\phantom{00}0.94 \\
J170724.3$-$021751 & 256.8512 & $-$\phantom{0}2.2976 & 33.6\phantom{0}$\pm$\phantom{0}7.4\phantom{0} & 3.2\phantom{00}$\pm$1.6\phantom{00} & 13.9\phantom{00}$\pm$\phantom{0}1.4\phantom{00} & 0.525\phantom{00}$\pm$0.067\phantom{00} & \phantom{00}7.77\phantom{0}$\pm$0.15\phantom{0} & 60762\phantom{.00}$\pm$\phantom{0}25\phantom{.00} \\
J170751.3$-$051238 & 256.9638 & $-$\phantom{0}5.2105 & \phantom{0}3.8\phantom{0}$\pm$\phantom{0}2.0\phantom{0} & 2.02\phantom{0}$\pm$0.29\phantom{0} & \phantom{0}3.52\phantom{0}$\pm$\phantom{0}0.82\phantom{0} & 0.419\phantom{00}$\pm$0.072\phantom{00} & \phantom{0}12.53\phantom{0}$\pm$0.16\phantom{0} & 60656.1\phantom{0}$\pm$\phantom{00}6.3\phantom{0} \\
J171440.6+051238 & 258.6692 & \phantom{$-$}\phantom{0}5.2106 & 14.3\phantom{0}$\pm$\phantom{0}1.3\phantom{0} & 3.74\phantom{0}$\pm$0.31\phantom{0} & 27.9\phantom{00}$\pm$\phantom{0}1.5\phantom{00} & 0.119\phantom{00}$\pm$0.060\phantom{00} & \phantom{0}10.585$\pm$0.036 & 60732.7\phantom{0}$\pm$\phantom{00}1.3\phantom{0} \\
J173709.8+092804 & 264.2908 & \phantom{$-$}\phantom{0}9.4678 & 13.1\phantom{0}$\pm$\phantom{0}4.9\phantom{0} & 0.6\phantom{00}$\pm$1.0\phantom{00} & \phantom{0}9.60\phantom{0}$\pm$\phantom{0}0.56\phantom{0} & 0.70\phantom{000}$\pm$0.18\phantom{000} & \phantom{0}16.23\phantom{0}$\pm$0.19\phantom{0} & 60769.4\phantom{0}$\pm$\phantom{00}6.4\phantom{0} \\
J173933.1$-$034756 & 264.8879 & $-$\phantom{0}3.7988 & 32.2\phantom{0}$\pm$\phantom{0}1.0\phantom{0} & 4.62\phantom{0}$\pm$0.40\phantom{0} & \phantom{0}9.15\phantom{0}$\pm$\phantom{0}0.45\phantom{0} & 0.494\phantom{00}$\pm$0.024\phantom{00} & \phantom{0}27.25\phantom{0}$\pm$0.13\phantom{0} & 60682.1\phantom{0}$\pm$\phantom{00}2.6\phantom{0} \\
J173942.2+001018 & 264.9258 & \phantom{$-$}\phantom{0}0.1717 & \phantom{0}2.6\phantom{0}$\pm$\phantom{0}1.4\phantom{0} & 1.3\phantom{00}$\pm$1.2\phantom{00} & \phantom{0}3.5\phantom{00}$\pm$\phantom{0}9.8\phantom{00} & 0.620\phantom{00}$\pm$0.071\phantom{00} & \phantom{0}35.98\phantom{0}$\pm$0.70\phantom{0} & 60603\phantom{.00}$\pm$\phantom{0}67\phantom{.00} \\
J180006.4+014941 & 270.0267 & \phantom{$-$}\phantom{0}1.8281 & 33.62$\pm$\phantom{0}0.43 & 3.84\phantom{0}$\pm$0.29\phantom{0} & \phantom{0}8.83\phantom{0}$\pm$\phantom{0}0.17\phantom{0} & 0.421\phantom{00}$\pm$0.034\phantom{00} & \phantom{00}5.815$\pm$0.067 & 60629.6\phantom{0}$\pm$\phantom{00}7.5\phantom{0} \\
J180402.3$-$003551 & 271.0096 & $-$\phantom{0}0.5975 & \phantom{0}2.95$\pm$\phantom{0}0.58 & 2.65\phantom{0}$\pm$0.35\phantom{0} & \phantom{0}3.76\phantom{0}$\pm$\phantom{0}0.30\phantom{0} & 0.098\phantom{00}$\pm$0.060\phantom{00} & \phantom{00}9.926$\pm$0.092 & 60870.4\phantom{0}$\pm$\phantom{00}6.5\phantom{0} \\
J180928.5+011405 & 272.3687 & \phantom{$-$}\phantom{0}1.2346 & \phantom{0}4.0\phantom{0}$\pm$\phantom{0}1.8\phantom{0} & 2.05\phantom{0}$\pm$0.74\phantom{0} & 28\phantom{.000}$\pm$12\phantom{.000} & 0.349\phantom{00}$\pm$0.050\phantom{00} & 106.7\phantom{00}$\pm$2.6\phantom{00} & 60310\phantom{.00}$\pm$140\phantom{.00} \\
J193545.2+041420 & 293.9383 & \phantom{$-$}\phantom{0}4.2388 & \phantom{0}9.6\phantom{0}$\pm$\phantom{0}8.0\phantom{0} & 1.03\phantom{0}$\pm$0.19\phantom{0} & \phantom{0}3.02\phantom{0}$\pm$\phantom{0}0.65\phantom{0} & 0.53\phantom{000}$\pm$0.10\phantom{000} & \phantom{00}4.12\phantom{0}$\pm$0.12\phantom{0} & 60881\phantom{.00}$\pm$\phantom{0}79\phantom{.00} \\
J194444.6+074302 & 296.1858 & \phantom{$-$}\phantom{0}7.7174 & 15.4\phantom{0}$\pm$\phantom{0}1.2\phantom{0} & 2.36\phantom{0}$\pm$0.28\phantom{0} & 10.67\phantom{0}$\pm$\phantom{0}0.66\phantom{0} & 0.095\phantom{00}$\pm$0.065\phantom{00} & \phantom{00}4.086$\pm$0.043 & 59942.9\phantom{0}$\pm$\phantom{00}6.6\phantom{0} \\
J195055.8+020043 & 297.7325 & \phantom{$-$}\phantom{0}2.012\phantom{0} & 13.1\phantom{0}$\pm$\phantom{0}6.8\phantom{0} & 1.293$\pm$0.083 & \phantom{0}9.71\phantom{0}$\pm$\phantom{0}0.82\phantom{0} & 0.418\phantom{00}$\pm$0.064\phantom{00} & \phantom{0}16.83\phantom{0}$\pm$0.21\phantom{0} & 60891.3\phantom{0}$\pm$\phantom{00}1.7\phantom{0} \\
J195606.7+073633 & 299.0279 & \phantom{$-$}\phantom{0}7.6092 & \phantom{0}5.3\phantom{0}$\pm$\phantom{0}2.5\phantom{0} & 2.04\phantom{0}$\pm$0.42\phantom{0} & \phantom{0}7.4\phantom{00}$\pm$\phantom{0}1.4\phantom{00} & 0.266\phantom{00}$\pm$0.078\phantom{00} & \phantom{00}4.483$\pm$0.057 & 60043.6\phantom{0}$\pm$\phantom{00}9.8\phantom{0} \\
J195909.1$-$081623 & 299.7879 & $-$\phantom{0}8.2731 & \phantom{0}9.54$\pm$\phantom{0}0.69 & 4.34\phantom{0}$\pm$0.11\phantom{0} & \phantom{0}9.55\phantom{0}$\pm$\phantom{0}0.72\phantom{0} & 0.020\phantom{00}$\pm$0.052\phantom{00} & \phantom{00}8.879$\pm$0.036 & 60838.0\phantom{0}$\pm$\phantom{00}3.7\phantom{0} \\
J200150.2$-$061943 & 300.4592 & $-$\phantom{0}6.3287 & \phantom{0}2.60$\pm$\phantom{0}0.19 & 1.92\phantom{0}$\pm$0.29\phantom{0} & \phantom{0}5.10\phantom{0}$\pm$\phantom{0}0.19\phantom{0} & 0.383\phantom{00}$\pm$0.062\phantom{00} & \phantom{00}9.565$\pm$0.036 & 60647.9\phantom{0}$\pm$\phantom{00}4.0\phantom{0} \\
J200341.5$-$312952 & 300.9229 & $-$31.4978 & 11.76$\pm$\phantom{0}0.55 & 2.255$\pm$0.094 & \phantom{0}8.85\phantom{0}$\pm$\phantom{0}0.41\phantom{0} & 0.470\phantom{00}$\pm$0.011\phantom{00} & \phantom{0}13.08\phantom{0}$\pm$0.10\phantom{0} & 60436.8\phantom{0}$\pm$\phantom{00}2.1\phantom{0} \\
J201056.9$-$003427 & 302.7371 & $-$\phantom{0}0.5741 & \phantom{0}2.82$\pm$\phantom{0}0.44 & 1.47\phantom{0}$\pm$0.17\phantom{0} & \phantom{0}3.40\phantom{0}$\pm$\phantom{0}0.23\phantom{0} & 0.544\phantom{00}$\pm$0.027\phantom{00} & \phantom{00}6.915$\pm$0.049 & 60852\phantom{.00}$\pm$\phantom{0}16\phantom{.00} \\
J202108.8$-$355527 & 305.2867 & $-$35.9242 & \phantom{0}3.71$\pm$\phantom{0}0.33 & 0.907$\pm$0.021 & \phantom{0}6.19\phantom{0}$\pm$\phantom{0}0.30\phantom{0} & 0.6818\phantom{0}$\pm$0.0081\phantom{0} & \phantom{00}9.537$\pm$0.033 & 60565.3\phantom{0}$\pm$\phantom{00}1.6\phantom{0} \\
J202649.5+045957 & 306.7062 & \phantom{$-$}\phantom{0}4.9991 & \phantom{0}4.2\phantom{0}$\pm$\phantom{0}1.1\phantom{0} & 1.35\phantom{0}$\pm$0.34\phantom{0} & \phantom{0}5.29\phantom{0}$\pm$\phantom{0}0.55\phantom{0} & 0.548\phantom{00}$\pm$0.033\phantom{00} & \phantom{00}5.131$\pm$0.059 & 60200\phantom{.00}$\pm$\phantom{0}11\phantom{.00} \\
J203505.4$-$071131 & 308.7725 & $-$\phantom{0}7.1919 & \phantom{0}4.59$\pm$\phantom{0}0.50 & 1.280$\pm$0.079 & \phantom{0}7.05\phantom{0}$\pm$\phantom{0}0.31\phantom{0} & 0.619\phantom{00}$\pm$0.015\phantom{00} & \phantom{0}13.603$\pm$0.052 & 60615.4\phantom{0}$\pm$\phantom{00}4.9\phantom{0} \\
J203629.2$-$051304 & 309.1217 & $-$\phantom{0}5.2177 & \phantom{0}2.41$\pm$\phantom{0}0.16 & 1.00\phantom{0}$\pm$0.12\phantom{0} & \phantom{0}6.050$\pm$\phantom{0}0.068 & 0.534\phantom{00}$\pm$0.012\phantom{00} & \phantom{0}25.233$\pm$0.078 & 60623.1\phantom{0}$\pm$\phantom{00}1.4\phantom{0} \\
J205155.8$-$365522 & 312.9825 & $-$36.9229 & \phantom{0}3.66$\pm$\phantom{0}0.23 & 3.04\phantom{0}$\pm$0.28\phantom{0} & 11.22\phantom{0}$\pm$\phantom{0}0.43\phantom{0} & 0.002\phantom{00}$\pm$0.056\phantom{00} & \phantom{0}51.616$\pm$0.023 & 60849.65$\pm$\phantom{00}0.64 \\
J212506.8$-$053639 & 321.2783 & $-$\phantom{0}5.6109 & \phantom{0}6.5\phantom{0}$\pm$\phantom{0}3.0\phantom{0} & 1.24\phantom{0}$\pm$0.73\phantom{0} & \phantom{0}4.12\phantom{0}$\pm$\phantom{0}0.70\phantom{0} & 0.453\phantom{00}$\pm$0.048\phantom{00} & \phantom{0}12.66\phantom{0}$\pm$0.28\phantom{0} & 60978\phantom{.00}$\pm$\phantom{0}62\phantom{.00} \\
J220717.4$-$080210 & 331.8225 & $-$\phantom{0}8.0362 & 20\phantom{.00}$\pm$12\phantom{.00} & 0.21\phantom{0}$\pm$0.49\phantom{0} & 19.55\phantom{0}$\pm$\phantom{0}0.81\phantom{0} & 0.510\phantom{00}$\pm$0.064\phantom{00} & \phantom{0}25.546$\pm$0.074 & 60801\phantom{.00}$\pm$\phantom{0}68\phantom{.00} \\
J220755.3+093305 & 331.9804 & \phantom{$-$}\phantom{0}9.5514 & \phantom{0}9.0\phantom{0}$\pm$\phantom{0}3.2\phantom{0} & 0.796$\pm$0.094 & \phantom{0}7.71\phantom{0}$\pm$\phantom{0}0.79\phantom{0} & 0.519\phantom{00}$\pm$0.011\phantom{00} & \phantom{0}16.76\phantom{0}$\pm$0.22\phantom{0} & 60919\phantom{.00}$\pm$\phantom{0}46\phantom{.00} \\
\enddata
\end{deluxetable*}

%% file: ese_prop_extended.tex
\startlongtable
\setlength{\tabcolsep}{2pt}
\begin{deluxetable*}{lrrccrrrrrr}
\tabletypesize{\footnotesize}
\tablecaption{\textbf{VAST ESE silver sample}: Observed and derived properties of the ESE candidates found in ASKAP VAST. Column definitions: Name -- ASKAP name, RA -- Right Ascension, DEC -- Declination, $\alpha$ -- Strength of the lens, $\beta$ -- angular size of the source in units of the size of the lens, $v_{\rm obs}$ -- relative velocity between the lens and the observer, $f_{\rm unlens}$ -- fraction of the source's flux that is unaffected by the lens, $F_{\nu;tot}$ -- total unlensed source flux density, and $T_0$ -- estimated time of lens alignment with the observer.}
\label{tab:ese_prop_ext}
\tablehead{
\colhead{Name} & \colhead{RA} & \colhead{DEC} & \colhead{$\alpha$} & \colhead{$\beta$} & \colhead{$v_{\rm obs}$} & \colhead{$f_{\rm unlens}$} & \colhead{$F_{\nu;tot}$} & \colhead{$T_0$} \\ 
\colhead{ASKAP-} & \colhead{(deg)} & \colhead{(deg)} & \colhead{} & \colhead{} & \colhead{(/17.3 km/s)} & \colhead{} & \colhead{(mJy)} & \colhead{(MJD)}
}
\startdata
J035855.3+063440 & \phantom{0}59.7304 & \phantom{$-$}\phantom{0}6.5778 & \phantom{0}7.7\phantom{0}$\pm$\phantom{0}5.9\phantom{0} & 0.45\phantom{0}$\pm$0.17\phantom{0} & 14.3\phantom{00}$\pm$\phantom{0}5.4\phantom{00} & 0.445\phantom{0}$\pm$0.070\phantom{0} & \phantom{0}2.774$\pm$0.097 & 60720\phantom{.0}$\pm$120\phantom{.0} \\
J040358.8+064539 & \phantom{0}60.995\phantom{0} & \phantom{$-$}\phantom{0}6.7608 & 11.1\phantom{0}$\pm$\phantom{0}4.7\phantom{0} & 3.59\phantom{0}$\pm$0.90\phantom{0} & \phantom{0}6.1\phantom{00}$\pm$\phantom{0}1.4\phantom{00} & 0.21\phantom{00}$\pm$0.12\phantom{00} & \phantom{0}4.238$\pm$0.078 & 59972\phantom{.0}$\pm$\phantom{0}15\phantom{.0} \\
J040800.2$-$325002 & \phantom{0}62.0008 & $-$32.834\phantom{0} & \phantom{0}9.4\phantom{0}$\pm$\phantom{0}1.5\phantom{0} & 3.3\phantom{00}$\pm$1.5\phantom{00} & \phantom{0}8.2\phantom{00}$\pm$\phantom{0}1.4\phantom{00} & 0.43\phantom{00}$\pm$0.21\phantom{00} & \phantom{0}6.166$\pm$0.093 & 60706\phantom{.0}$\pm$\phantom{0}24\phantom{.0} \\
J045127.2+081254 & \phantom{0}72.8633 & \phantom{$-$}\phantom{0}8.2149 & \phantom{0}5.3\phantom{0}$\pm$\phantom{0}3.2\phantom{0} & 0.97\phantom{0}$\pm$0.23\phantom{0} & 21.4\phantom{00}$\pm$\phantom{0}5.5\phantom{00} & 0.323\phantom{0}$\pm$0.084\phantom{0} & \phantom{0}3.000$\pm$0.052 & 60611\phantom{.0}$\pm$\phantom{0}45\phantom{.0} \\
J050004.8+001421 & \phantom{0}75.02\phantom{00} & \phantom{$-$}\phantom{0}0.2391 & \phantom{0}4.77$\pm$\phantom{0}0.12 & 1.350$\pm$0.015 & 10.60\phantom{0}$\pm$\phantom{0}0.14\phantom{0} & 0.592\phantom{0}$\pm$0.015\phantom{0} & \phantom{0}7.589$\pm$0.042 & 60628.9$\pm$\phantom{00}2.4 \\
J051232.0$-$025829 & \phantom{0}78.1333 & $-$\phantom{0}2.9748 & 12.82$\pm$\phantom{0}0.92 & 0.99\phantom{0}$\pm$0.14\phantom{0} & \phantom{0}9.64\phantom{0}$\pm$\phantom{0}0.21\phantom{0} & 0.493\phantom{0}$\pm$0.026\phantom{0} & \phantom{0}4.279$\pm$0.056 & 60484.6$\pm$\phantom{00}5.1 \\
J051341.0$-$092215 & \phantom{0}78.4208 & $-$\phantom{0}9.3707 & 10.9\phantom{0}$\pm$\phantom{0}1.7\phantom{0} & 1.5\phantom{00}$\pm$1.4\phantom{00} & \phantom{0}7.2\phantom{00}$\pm$\phantom{0}4.7\phantom{00} & 0.56\phantom{00}$\pm$0.16\phantom{00} & \phantom{0}6.27\phantom{0}$\pm$0.11\phantom{0} & 59978\phantom{.0}$\pm$\phantom{0}92\phantom{.0} \\
J052325.7+015837 & \phantom{0}80.8571 & \phantom{$-$}\phantom{0}1.977\phantom{0} & \phantom{0}1.05$\pm$\phantom{0}0.54 & 0.10\phantom{0}$\pm$0.44\phantom{0} & \phantom{0}2.98\phantom{0}$\pm$\phantom{0}0.58\phantom{0} & 0.472\phantom{0}$\pm$0.071\phantom{0} & \phantom{0}9.564$\pm$0.081 & 60813.5$\pm$\phantom{00}8.5 \\
J055202.1$-$300936 & \phantom{0}88.0087 & $-$30.16\phantom{00} & \phantom{0}5.69$\pm$\phantom{0}0.19 & 0.492$\pm$0.089 & \phantom{0}5.284$\pm$\phantom{0}0.076 & 0.818\phantom{0}$\pm$0.010\phantom{0} & \phantom{0}6.210$\pm$0.044 & 60739.9$\pm$\phantom{00}4.4 \\
J074004.9+041529 & 115.0204 & \phantom{$-$}\phantom{0}4.2579 & \phantom{0}9.5\phantom{0}$\pm$\phantom{0}6.2\phantom{0} & 0.79\phantom{0}$\pm$0.47\phantom{0} & 12.6\phantom{00}$\pm$\phantom{0}2.1\phantom{00} & 0.58\phantom{00}$\pm$0.11\phantom{00} & \phantom{0}4.279$\pm$0.035 & 60173\phantom{.0}$\pm$\phantom{0}10\phantom{.0} \\
J085439.5+084326 & 133.6646 & \phantom{$-$}\phantom{0}8.7238 & \phantom{0}8.8\phantom{0}$\pm$\phantom{0}3.1\phantom{0} & 0.54\phantom{0}$\pm$0.58\phantom{0} & 12\phantom{.000}$\pm$11\phantom{.000} & 0.44\phantom{00}$\pm$0.11\phantom{00} & \phantom{0}3.28\phantom{0}$\pm$0.11\phantom{0} & 60440\phantom{.0}$\pm$190\phantom{.0} \\
J094215.5+043622 & 145.5646 & \phantom{$-$}\phantom{0}4.6061 & \phantom{0}9.48$\pm$\phantom{0}0.23 & 5.35\phantom{0}$\pm$0.29\phantom{0} & \phantom{0}8.49\phantom{0}$\pm$\phantom{0}0.44\phantom{0} & 0.185\phantom{0}$\pm$0.034\phantom{0} & 69.39\phantom{0}$\pm$0.44\phantom{0} & 60813.2$\pm$\phantom{00}1.3 \\
J102443.5$-$373347 & 156.1812 & $-$37.5631 & \phantom{0}7.2\phantom{0}$\pm$\phantom{0}2.6\phantom{0} & 1.07\phantom{0}$\pm$0.45\phantom{0} & \phantom{0}9.7\phantom{00}$\pm$\phantom{0}4.1\phantom{00} & 0.519\phantom{0}$\pm$0.073\phantom{0} & \phantom{0}3.606$\pm$0.044 & 60232\phantom{.0}$\pm$\phantom{0}32\phantom{.0} \\
J104424.0$-$304058 & 161.1\phantom{000} & $-$30.6828 & 12.33$\pm$\phantom{0}0.50 & 3.32\phantom{0}$\pm$0.12\phantom{0} & 14.44\phantom{0}$\pm$\phantom{0}0.32\phantom{0} & 0.6485$\pm$0.0086 & 49.771$\pm$0.032 & 60573.8$\pm$\phantom{00}1.1 \\
J110220.7+014206 & 165.5862 & \phantom{$-$}\phantom{0}1.7015 & 11.48$\pm$\phantom{0}0.59 & 1.355$\pm$0.070 & 18.06\phantom{0}$\pm$\phantom{0}0.29\phantom{0} & 0.455\phantom{0}$\pm$0.016\phantom{0} & \phantom{0}8.243$\pm$0.033 & 60225.3$\pm$\phantom{00}1.9 \\
J134831.0$-$303536 & 207.1292 & $-$30.5934 & 14.88$\pm$\phantom{0}0.77 & 0.114$\pm$0.070 & 15.96\phantom{0}$\pm$\phantom{0}0.62\phantom{0} & 0.649\phantom{0}$\pm$0.034\phantom{0} & \phantom{0}3.136$\pm$0.041 & 60560.4$\pm$\phantom{00}3.6 \\
J134922.2+071218 & 207.3425 & \phantom{$-$}\phantom{0}7.2049 & \phantom{0}8.5\phantom{0}$\pm$\phantom{0}1.8\phantom{0} & 1.24\phantom{0}$\pm$0.24\phantom{0} & \phantom{0}4.61\phantom{0}$\pm$\phantom{0}0.57\phantom{0} & 0.503\phantom{0}$\pm$0.045\phantom{0} & \phantom{0}3.464$\pm$0.066 & 59997\phantom{.0}$\pm$\phantom{0}40\phantom{.0} \\
J141945.9$-$082706 & 214.9412 & $-$\phantom{0}8.4518 & \phantom{0}8.2\phantom{0}$\pm$\phantom{0}1.3\phantom{0} & 3.21\phantom{0}$\pm$0.34\phantom{0} & \phantom{0}7.90\phantom{0}$\pm$\phantom{0}0.70\phantom{0} & 0.133\phantom{0}$\pm$0.083\phantom{0} & \phantom{0}2.769$\pm$0.031 & 60661.0$\pm$\phantom{00}8.0 \\
J150637.4+002540 & 226.6558 & \phantom{$-$}\phantom{0}0.4279 & 23\phantom{.00}$\pm$11\phantom{.00} & 0.30\phantom{0}$\pm$0.51\phantom{0} & 12.48\phantom{0}$\pm$\phantom{0}0.70\phantom{0} & 0.698\phantom{0}$\pm$0.078\phantom{0} & \phantom{0}5.260$\pm$0.071 & 60711.0$\pm$\phantom{00}5.2 \\
J151738.6+034528 & 229.4108 & \phantom{$-$}\phantom{0}3.7579 & 12.2\phantom{0}$\pm$\phantom{0}1.2\phantom{0} & 2.28\phantom{0}$\pm$0.56\phantom{0} & \phantom{0}9.60\phantom{0}$\pm$\phantom{0}0.36\phantom{0} & 0.51\phantom{00}$\pm$0.11\phantom{00} & 14.827$\pm$0.072 & 60830.7$\pm$\phantom{00}3.1 \\
J151951.3$-$091849 & 229.9637 & $-$\phantom{0}9.3136 & 20.51$\pm$\phantom{0}0.56 & 2.27\phantom{0}$\pm$0.36\phantom{0} & 22.01\phantom{0}$\pm$\phantom{0}0.79\phantom{0} & 0.464\phantom{0}$\pm$0.042\phantom{0} & \phantom{0}8.193$\pm$0.057 & 60642.2$\pm$\phantom{00}4.1 \\
J152914.4$-$041408 & 232.31\phantom{00} & $-$\phantom{0}4.2354 & 20.53$\pm$\phantom{0}0.49 & 4.69\phantom{0}$\pm$0.55\phantom{0} & 44.6\phantom{00}$\pm$\phantom{0}1.6\phantom{00} & 0.162\phantom{0}$\pm$0.066\phantom{0} & \phantom{0}6.840$\pm$0.031 & 60678.3$\pm$\phantom{00}1.9 \\
J154236.1$-$054755 & 235.6504 & $-$\phantom{0}5.7987 & \phantom{0}2.58$\pm$\phantom{0}0.47 & 1.32\phantom{0}$\pm$0.32\phantom{0} & 12.22\phantom{0}$\pm$\phantom{0}0.77\phantom{0} & 0.172\phantom{0}$\pm$0.054\phantom{0} & \phantom{0}3.964$\pm$0.042 & 60625.9$\pm$\phantom{00}4.4 \\
J160220.5$-$042537 & 240.5854 & $-$\phantom{0}4.4268 & \phantom{0}5.2\phantom{0}$\pm$\phantom{0}3.1\phantom{0} & 1.88\phantom{0}$\pm$0.40\phantom{0} & \phantom{0}4.38\phantom{0}$\pm$\phantom{0}0.90\phantom{0} & 0.44\phantom{00}$\pm$0.11\phantom{00} & \phantom{0}4.583$\pm$0.059 & 60100\phantom{.0}$\pm$\phantom{0}11\phantom{.0} \\
J160556.9$-$010949 & 241.4871 & $-$\phantom{0}1.1637 & \phantom{0}2.83$\pm$\phantom{0}0.53 & 0.34\phantom{0}$\pm$0.18\phantom{0} & \phantom{0}2.4\phantom{00}$\pm$\phantom{0}1.8\phantom{00} & 0.217\phantom{0}$\pm$0.035\phantom{0} & \phantom{0}9.20\phantom{0}$\pm$0.39\phantom{0} & 61030\phantom{.0}$\pm$130\phantom{.0} \\
J162712.9+041835 & 246.8038 & \phantom{$-$}\phantom{0}4.3096 & \phantom{0}5.6\phantom{0}$\pm$\phantom{0}3.0\phantom{0} & 2.00\phantom{0}$\pm$0.44\phantom{0} & \phantom{0}6.1\phantom{00}$\pm$\phantom{0}1.2\phantom{00} & 0.474\phantom{0}$\pm$0.094\phantom{0} & \phantom{0}4.993$\pm$0.042 & 60140\phantom{.0}$\pm$\phantom{0}12\phantom{.0} \\
J162747.9$-$090106 & 246.9496 & $-$\phantom{0}9.0184 & \phantom{0}8.2\phantom{0}$\pm$\phantom{0}4.0\phantom{0} & 3.3\phantom{00}$\pm$1.1\phantom{00} & \phantom{0}8.0\phantom{00}$\pm$\phantom{0}1.9\phantom{00} & 0.33\phantom{00}$\pm$0.14\phantom{00} & \phantom{0}4.031$\pm$0.049 & 60773\phantom{.0}$\pm$\phantom{0}11\phantom{.0} \\
J163007.9$-$045136 & 247.5329 & $-$\phantom{0}4.86\phantom{00} & \phantom{0}6.1\phantom{0}$\pm$\phantom{0}3.4\phantom{0} & 1.10\phantom{0}$\pm$0.29\phantom{0} & \phantom{0}7.3\phantom{00}$\pm$\phantom{0}3.5\phantom{00} & 0.594\phantom{0}$\pm$0.077\phantom{0} & \phantom{0}3.368$\pm$0.089 & 60455\phantom{.0}$\pm$\phantom{0}23\phantom{.0} \\
J163155.3$-$083008 & 247.9804 & $-$\phantom{0}8.5023 & \phantom{0}4.17$\pm$\phantom{0}0.63 & 1.278$\pm$0.093 & \phantom{0}1.73\phantom{0}$\pm$\phantom{0}0.15\phantom{0} & 0.611\phantom{0}$\pm$0.023\phantom{0} & \phantom{0}8.18\phantom{0}$\pm$0.10\phantom{0} & 61375\phantom{.0}$\pm$\phantom{0}65\phantom{.0} \\
J163939.0+090737 & 249.9125 & \phantom{$-$}\phantom{0}9.1268 & 23.5\phantom{0}$\pm$\phantom{0}5.4\phantom{0} & 2.9\phantom{00}$\pm$1.1\phantom{00} & \phantom{0}8.5\phantom{00}$\pm$\phantom{0}1.5\phantom{00} & 0.548\phantom{0}$\pm$0.096\phantom{0} & \phantom{0}3.453$\pm$0.062 & 60785\phantom{.0}$\pm$\phantom{0}27\phantom{.0} \\
J163951.1+011650 & 249.9629 & \phantom{$-$}\phantom{0}1.2807 & \phantom{0}3.1\phantom{0}$\pm$\phantom{0}2.4\phantom{0} & 1.41\phantom{0}$\pm$0.66\phantom{0} & \phantom{0}4.7\phantom{00}$\pm$\phantom{0}1.3\phantom{00} & 0.41\phantom{00}$\pm$0.12\phantom{00} & \phantom{0}2.770$\pm$0.036 & 60558.4$\pm$\phantom{00}9.2 \\
J164043.3$-$013342 & 250.1804 & $-$\phantom{0}1.5615 & \phantom{0}6.7\phantom{0}$\pm$\phantom{0}1.0\phantom{0} & 1.18\phantom{0}$\pm$0.15\phantom{0} & \phantom{0}6.44\phantom{0}$\pm$\phantom{0}0.38\phantom{0} & 0.611\phantom{0}$\pm$0.039\phantom{0} & \phantom{0}4.000$\pm$0.075 & 60087\phantom{.0}$\pm$\phantom{0}28\phantom{.0} \\
J164100.2+055811 & 250.2508 & \phantom{$-$}\phantom{0}5.9696 & \phantom{0}8.0\phantom{0}$\pm$\phantom{0}1.7\phantom{0} & 3.77\phantom{0}$\pm$0.70\phantom{0} & \phantom{0}8.8\phantom{00}$\pm$\phantom{0}1.2\phantom{00} & 0.20\phantom{00}$\pm$0.13\phantom{00} & \phantom{0}7.301$\pm$0.047 & 60016.9$\pm$\phantom{00}9.1 \\
J164613.0+051750 & 251.5542 & \phantom{$-$}\phantom{0}5.2972 & \phantom{0}4.05$\pm$\phantom{0}0.43 & 1.297$\pm$0.038 & \phantom{0}6.08\phantom{0}$\pm$\phantom{0}0.30\phantom{0} & 0.731\phantom{0}$\pm$0.013\phantom{0} & 15.012$\pm$0.095 & 60542.0$\pm$\phantom{00}6.1 \\
J164918.4+092144 & 252.3267 & \phantom{$-$}\phantom{0}9.3621 & \phantom{0}2.5\phantom{0}$\pm$\phantom{0}1.9\phantom{0} & 1.26\phantom{0}$\pm$0.22\phantom{0} & \phantom{0}5.66\phantom{0}$\pm$\phantom{0}0.72\phantom{0} & 0.470\phantom{0}$\pm$0.088\phantom{0} & \phantom{0}6.85\phantom{0}$\pm$0.12\phantom{0} & 60227\phantom{.0}$\pm$\phantom{0}21\phantom{.0} \\
J165107.1+012406 & 252.7796 & \phantom{$-$}\phantom{0}1.4017 & \phantom{0}3.5\phantom{0}$\pm$\phantom{0}1.1\phantom{0} & 1.42\phantom{0}$\pm$0.27\phantom{0} & \phantom{0}3.4\phantom{00}$\pm$\phantom{0}3.6\phantom{00} & 0.417\phantom{0}$\pm$0.046\phantom{0} & 12.22\phantom{0}$\pm$0.30\phantom{0} & 60026\phantom{.0}$\pm$\phantom{0}33\phantom{.0} \\
J165922.1$-$003959 & 254.8421 & $-$\phantom{0}0.6662 & \phantom{0}6.9\phantom{0}$\pm$\phantom{0}4.1\phantom{0} & 3.3\phantom{00}$\pm$1.1\phantom{00} & \phantom{0}6.1\phantom{00}$\pm$\phantom{0}1.4\phantom{00} & 0.44\phantom{00}$\pm$0.13\phantom{00} & \phantom{0}6.181$\pm$0.060 & 60713.1$\pm$\phantom{00}9.0 \\
J170207.7$-$074646 & 255.5321 & $-$\phantom{0}7.7794 & \phantom{0}2.42$\pm$\phantom{0}0.11 & 1.08\phantom{0}$\pm$0.11\phantom{0} & \phantom{0}3.90\phantom{0}$\pm$\phantom{0}0.11\phantom{0} & 0.442\phantom{0}$\pm$0.030\phantom{0} & \phantom{0}5.596$\pm$0.085 & 60354\phantom{.0}$\pm$\phantom{0}10\phantom{.0} \\
J170429.0+042035 & 256.1208 & \phantom{$-$}\phantom{0}4.343\phantom{0} & 20.6\phantom{0}$\pm$\phantom{0}5.8\phantom{0} & 4.76\phantom{0}$\pm$0.61\phantom{0} & 10.8\phantom{00}$\pm$\phantom{0}1.2\phantom{00} & 0.408\phantom{0}$\pm$0.064\phantom{0} & \phantom{0}6.522$\pm$0.079 & 60737.5$\pm$\phantom{00}8.6 \\
J171130.9+092128 & 257.8787 & \phantom{$-$}\phantom{0}9.3577 & \phantom{0}3.57$\pm$\phantom{0}0.98 & 0.52\phantom{0}$\pm$0.12\phantom{0} & \phantom{0}2.36\phantom{0}$\pm$\phantom{0}0.40\phantom{0} & 0.695\phantom{0}$\pm$0.044\phantom{0} & \phantom{0}4.481$\pm$0.071 & 61020\phantom{.0}$\pm$130\phantom{.0} \\
J171258.3+010200 & 258.2429 & \phantom{$-$}\phantom{0}1.0333 & \phantom{0}2.6\phantom{0}$\pm$\phantom{0}1.8\phantom{0} & 0.28\phantom{0}$\pm$0.17\phantom{0} & \phantom{0}6.9\phantom{00}$\pm$\phantom{0}1.0\phantom{00} & 0.446\phantom{0}$\pm$0.082\phantom{0} & 14.64\phantom{0}$\pm$0.26\phantom{0} & 60028\phantom{.0}$\pm$\phantom{0}13\phantom{.0} \\
J171409.6+033321 & 258.54\phantom{00} & \phantom{$-$}\phantom{0}3.5558 & \phantom{0}6.67$\pm$\phantom{0}0.29 & 1.33\phantom{0}$\pm$0.12\phantom{0} & \phantom{0}5.99\phantom{0}$\pm$\phantom{0}0.15\phantom{0} & 0.617\phantom{0}$\pm$0.011\phantom{0} & \phantom{0}7.632$\pm$0.049 & 60825.1$\pm$\phantom{00}5.3 \\
J171640.5$-$071122 & 259.1688 & $-$\phantom{0}7.1894 & \phantom{0}7.8\phantom{0}$\pm$\phantom{0}1.4\phantom{0} & 1.225$\pm$0.085 & 10.92\phantom{0}$\pm$\phantom{0}0.78\phantom{0} & 0.684\phantom{0}$\pm$0.012\phantom{0} & \phantom{0}8.646$\pm$0.034 & 60492.8$\pm$\phantom{00}4.9 \\
J172135.0+054909 & 260.3958 & \phantom{$-$}\phantom{0}5.8192 & \phantom{0}2.71$\pm$\phantom{0}0.33 & 1.12\phantom{0}$\pm$0.13\phantom{0} & \phantom{0}5.47\phantom{0}$\pm$\phantom{0}0.80\phantom{0} & 0.236\phantom{0}$\pm$0.053\phantom{0} & \phantom{0}3.544$\pm$0.068 & 60421.5$\pm$\phantom{00}7.2 \\
J173816.0+012539 & 264.5667 & \phantom{$-$}\phantom{0}1.4275 & \phantom{0}3.8\phantom{0}$\pm$\phantom{0}2.2\phantom{0} & 3.59\phantom{0}$\pm$0.71\phantom{0} & \phantom{0}6.8\phantom{00}$\pm$\phantom{0}1.4\phantom{00} & 0.27\phantom{00}$\pm$0.15\phantom{00} & \phantom{0}4.805$\pm$0.091 & 60735\phantom{.0}$\pm$\phantom{0}11\phantom{.0} \\
J173950.6+031810 & 264.9608 & \phantom{$-$}\phantom{0}3.3027 & \phantom{0}6.5\phantom{0}$\pm$\phantom{0}3.1\phantom{0} & 0.87\phantom{0}$\pm$0.69\phantom{0} & \phantom{0}3.0\phantom{00}$\pm$\phantom{0}1.1\phantom{00} & 0.67\phantom{00}$\pm$0.14\phantom{00} & \phantom{0}3.820$\pm$0.070 & 60790\phantom{.0}$\pm$160\phantom{.0} \\
J174054.7+075102 & 265.2279 & \phantom{$-$}\phantom{0}7.8505 & 15.9\phantom{0}$\pm$\phantom{0}2.7\phantom{0} & 1.79\phantom{0}$\pm$0.95\phantom{0} & 15.2\phantom{00}$\pm$\phantom{0}1.8\phantom{00} & 0.63\phantom{00}$\pm$0.11\phantom{00} & \phantom{0}5.002$\pm$0.050 & 60706.6$\pm$\phantom{00}5.4 \\
J174430.9+014321 & 266.1287 & \phantom{$-$}\phantom{0}1.7225 & \phantom{0}3.91$\pm$\phantom{0}0.23 & 1.28\phantom{0}$\pm$0.18\phantom{0} & \phantom{0}4.41\phantom{0}$\pm$\phantom{0}0.15\phantom{0} & 0.663\phantom{0}$\pm$0.017\phantom{0} & \phantom{0}6.388$\pm$0.050 & 60760.7$\pm$\phantom{00}7.5 \\
J175148.9+080525 & 267.9538 & \phantom{$-$}\phantom{0}8.0904 & \phantom{0}2.61$\pm$\phantom{0}0.53 & 2.46\phantom{0}$\pm$0.63\phantom{0} & \phantom{0}8.6\phantom{00}$\pm$\phantom{0}1.3\phantom{00} & 0.35\phantom{00}$\pm$0.13\phantom{00} & \phantom{0}5.978$\pm$0.062 & 60154\phantom{.0}$\pm$\phantom{0}12\phantom{.0} \\
J175556.8+003441 & 268.9867 & \phantom{$-$}\phantom{0}0.5781 & 29.4\phantom{0}$\pm$\phantom{0}2.6\phantom{0} & 4.97\phantom{0}$\pm$0.51\phantom{0} & 14.9\phantom{00}$\pm$\phantom{0}1.0\phantom{00} & 0.399\phantom{0}$\pm$0.042\phantom{0} & \phantom{0}5.330$\pm$0.047 & 60460.3$\pm$\phantom{00}5.8 \\
J175736.1+031939 & 269.4004 & \phantom{$-$}\phantom{0}3.3274 & 19.2\phantom{0}$\pm$\phantom{0}3.2\phantom{0} & 3.4\phantom{00}$\pm$1.2\phantom{00} & 23.0\phantom{00}$\pm$\phantom{0}2.9\phantom{00} & 0.50\phantom{00}$\pm$0.11\phantom{00} & \phantom{0}5.628$\pm$0.048 & 60207.5$\pm$\phantom{00}5.1 \\
J175742.7$-$013552 & 269.4279 & $-$\phantom{0}1.5978 & 18.3\phantom{0}$\pm$\phantom{0}4.8\phantom{0} & 4.99\phantom{0}$\pm$0.55\phantom{0} & \phantom{0}7.51\phantom{0}$\pm$\phantom{0}0.80\phantom{0} & 0.354\phantom{0}$\pm$0.057\phantom{0} & \phantom{0}6.333$\pm$0.065 & 60604.5$\pm$\phantom{00}7.7 \\
J180347.2$-$011626 & 270.9467 & $-$\phantom{0}1.274\phantom{0} & 13.7\phantom{0}$\pm$\phantom{0}2.7\phantom{0} & 1.35\phantom{0}$\pm$0.24\phantom{0} & \phantom{0}6.81\phantom{0}$\pm$\phantom{0}0.18\phantom{0} & 0.696\phantom{0}$\pm$0.063\phantom{0} & \phantom{0}7.347$\pm$0.062 & 60559.2$\pm$\phantom{00}7.5 \\
J182644.5+090406 & 276.6854 & \phantom{$-$}\phantom{0}9.0683 & \phantom{0}6.78$\pm$\phantom{0}0.94 & 1.85\phantom{0}$\pm$0.40\phantom{0} & \phantom{0}5.8\phantom{00}$\pm$\phantom{0}1.1\phantom{00} & 0.444\phantom{0}$\pm$0.093\phantom{0} & \phantom{0}5.200$\pm$0.060 & 60049\phantom{.0}$\pm$\phantom{0}52\phantom{.0} \\
J182941.0+035533 & 277.4208 & \phantom{$-$}\phantom{0}3.9258 & \phantom{0}8.2\phantom{0}$\pm$\phantom{0}3.4\phantom{0} & 2.7\phantom{00}$\pm$1.3\phantom{00} & 11.2\phantom{00}$\pm$\phantom{0}2.5\phantom{00} & 0.46\phantom{00}$\pm$0.18\phantom{00} & \phantom{0}3.066$\pm$0.040 & 60548\phantom{.0}$\pm$\phantom{0}13\phantom{.0} \\
J190855.9$-$050528 & 287.2329 & $-$\phantom{0}5.091\phantom{0} & \phantom{0}6.6\phantom{0}$\pm$\phantom{0}2.0\phantom{0} & 1.36\phantom{0}$\pm$0.40\phantom{0} & \phantom{0}9.5\phantom{00}$\pm$\phantom{0}1.5\phantom{00} & 0.610\phantom{0}$\pm$0.058\phantom{0} & \phantom{0}3.494$\pm$0.036 & 60485.1$\pm$\phantom{00}6.0 \\
J191533.4$-$051109 & 288.8892 & $-$\phantom{0}5.1859 & 22.6\phantom{0}$\pm$\phantom{0}2.7\phantom{0} & 1.26\phantom{0}$\pm$0.73\phantom{0} & 16.6\phantom{00}$\pm$\phantom{0}1.9\phantom{00} & 0.478\phantom{0}$\pm$0.091\phantom{0} & \phantom{0}3.870$\pm$0.038 & 60507.9$\pm$\phantom{00}5.6 \\
J191934.4$-$074043 & 289.8933 & $-$\phantom{0}7.6787 & \phantom{0}3.4\phantom{0}$\pm$\phantom{0}1.0\phantom{0} & 0.95\phantom{0}$\pm$0.56\phantom{0} & \phantom{0}2.7\phantom{00}$\pm$\phantom{0}9.4\phantom{00} & 0.544\phantom{0}$\pm$0.040\phantom{0} & 45.13\phantom{0}$\pm$0.63\phantom{0} & 59610\phantom{.0}$\pm$\phantom{0}18\phantom{.0} \\
J191959.0$-$075726 & 289.9958 & $-$\phantom{0}7.9571 & \phantom{0}3.65$\pm$\phantom{0}0.14 & 0.905$\pm$0.016 & \phantom{0}4.385$\pm$\phantom{0}0.073 & 0.7782$\pm$0.0072 & 13.080$\pm$0.046 & 60782.9$\pm$\phantom{00}6.2 \\
J192025.0$-$051943 & 290.1042 & $-$\phantom{0}5.3287 & 16\phantom{.00}$\pm$11\phantom{.00} & 1.85\phantom{0}$\pm$0.43\phantom{0} & \phantom{0}6.61\phantom{0}$\pm$\phantom{0}0.44\phantom{0} & 0.61\phantom{00}$\pm$0.16\phantom{00} & \phantom{0}5.05\phantom{0}$\pm$0.13\phantom{0} & 60041\phantom{.0}$\pm$\phantom{0}41\phantom{.0} \\
J192755.8+064249 & 291.9825 & \phantom{$-$}\phantom{0}6.7136 & \phantom{0}3.0\phantom{0}$\pm$\phantom{0}1.4\phantom{0} & 1.177$\pm$0.092 & \phantom{0}6.1\phantom{00}$\pm$\phantom{0}1.3\phantom{00} & 0.560\phantom{0}$\pm$0.071\phantom{0} & \phantom{0}5.95\phantom{0}$\pm$0.11\phantom{0} & 60635.3$\pm$\phantom{00}5.5 \\
J192834.3+064907 & 292.1429 & \phantom{$-$}\phantom{0}6.8186 & \phantom{0}9.1\phantom{0}$\pm$\phantom{0}1.5\phantom{0} & 3.47\phantom{0}$\pm$0.46\phantom{0} & \phantom{0}5.13\phantom{0}$\pm$\phantom{0}0.45\phantom{0} & 0.22\phantom{00}$\pm$0.11\phantom{00} & \phantom{0}6.014$\pm$0.066 & 60570.5$\pm$\phantom{00}8.0 \\
J193316.7$-$082240 & 293.3196 & $-$\phantom{0}8.3777 & \phantom{0}1.40$\pm$\phantom{0}0.57 & 0.30\phantom{0}$\pm$0.20\phantom{0} & \phantom{0}2.7\phantom{00}$\pm$\phantom{0}1.4\phantom{00} & 0.543\phantom{0}$\pm$0.089\phantom{0} & \phantom{0}3.276$\pm$0.071 & 60714\phantom{.0}$\pm$\phantom{0}46\phantom{.0} \\
J193446.2$-$052903 & 293.6925 & $-$\phantom{0}5.4841 & \phantom{0}3.2\phantom{0}$\pm$\phantom{0}1.2\phantom{0} & 1.36\phantom{0}$\pm$0.16\phantom{0} & \phantom{0}4.57\phantom{0}$\pm$\phantom{0}0.66\phantom{0} & 0.457\phantom{0}$\pm$0.078\phantom{0} & \phantom{0}3.372$\pm$0.058 & 60418.9$\pm$\phantom{00}7.5 \\
J193907.0$-$041146 & 294.7792 & $-$\phantom{0}4.1961 & \phantom{0}2.95$\pm$\phantom{0}0.27 & 1.11\phantom{0}$\pm$0.10\phantom{0} & \phantom{0}6.96\phantom{0}$\pm$\phantom{0}0.23\phantom{0} & 0.509\phantom{0}$\pm$0.026\phantom{0} & \phantom{0}6.262$\pm$0.066 & 60425.7$\pm$\phantom{00}4.7 \\
J194132.9$-$081655 & 295.3871 & $-$\phantom{0}8.2818 & \phantom{0}7.4\phantom{0}$\pm$\phantom{0}1.4\phantom{0} & 1.14\phantom{0}$\pm$0.16\phantom{0} & 12.21\phantom{0}$\pm$\phantom{0}0.70\phantom{0} & 0.691\phantom{0}$\pm$0.030\phantom{0} & \phantom{0}4.134$\pm$0.041 & 60541.9$\pm$\phantom{00}5.1 \\
J194652.3$-$050941 & 296.7179 & $-$\phantom{0}5.1614 & \phantom{0}3.98$\pm$\phantom{0}0.79 & 0.90\phantom{0}$\pm$0.15\phantom{0} & \phantom{0}5.25\phantom{0}$\pm$\phantom{0}0.36\phantom{0} & 0.767\phantom{0}$\pm$0.033\phantom{0} & \phantom{0}5.948$\pm$0.048 & 60660\phantom{.0}$\pm$\phantom{0}10\phantom{.0} \\
J195111.1+060319 & 297.7962 & \phantom{$-$}\phantom{0}6.0551 & \phantom{0}9.4\phantom{0}$\pm$\phantom{0}1.8\phantom{0} & 4.93\phantom{0}$\pm$0.66\phantom{0} & 12.1\phantom{00}$\pm$\phantom{0}1.0\phantom{00} & 0.023\phantom{0}$\pm$0.073\phantom{0} & \phantom{0}8.812$\pm$0.033 & 60608.5$\pm$\phantom{00}3.3 \\
J195457.9$-$084029 & 298.7412 & $-$\phantom{0}8.6746 & \phantom{0}8.31$\pm$\phantom{0}0.95 & 3.03\phantom{0}$\pm$0.52\phantom{0} & 10.02\phantom{0}$\pm$\phantom{0}0.65\phantom{0} & 0.32\phantom{00}$\pm$0.12\phantom{00} & \phantom{0}5.528$\pm$0.031 & 60712.9$\pm$\phantom{00}4.8 \\
J201326.5$-$045058 & 303.3604 & $-$\phantom{0}4.8496 & 23.3\phantom{0}$\pm$\phantom{0}1.0\phantom{0} & 1.33\phantom{0}$\pm$0.12\phantom{0} & \phantom{0}6.05\phantom{0}$\pm$\phantom{0}0.20\phantom{0} & 0.568\phantom{0}$\pm$0.015\phantom{0} & \phantom{0}5.547$\pm$0.052 & 60641.7$\pm$\phantom{00}7.6 \\
J201931.1+054141 & 304.8796 & \phantom{$-$}\phantom{0}5.6947 & \phantom{0}7.45$\pm$\phantom{0}0.48 & 0.222$\pm$0.040 & \phantom{0}2.71\phantom{0}$\pm$\phantom{0}0.44\phantom{0} & 0.706\phantom{0}$\pm$0.012\phantom{0} & \phantom{0}6.840$\pm$0.096 & 61000\phantom{.0}$\pm$130\phantom{.0} \\
J202840.7$-$305610 & 307.1696 & $-$30.9362 & 20.7\phantom{0}$\pm$\phantom{0}3.8\phantom{0} & 2.91\phantom{0}$\pm$0.92\phantom{0} & \phantom{0}5.54\phantom{0}$\pm$\phantom{0}0.67\phantom{0} & 0.563\phantom{0}$\pm$0.079\phantom{0} & \phantom{0}5.509$\pm$0.063 & 60071\phantom{.0}$\pm$\phantom{0}13\phantom{.0} \\
J211429.8$-$005218 & 318.6242 & $-$\phantom{0}0.8716 & 13.2\phantom{0}$\pm$\phantom{0}3.9\phantom{0} & 1.09\phantom{0}$\pm$0.30\phantom{0} & 41.48\phantom{0}$\pm$\phantom{0}0.61\phantom{0} & 0.093\phantom{0}$\pm$0.026\phantom{0} & 13.362$\pm$0.063 & 60710.9$\pm$\phantom{00}5.5 \\
J211751.1$-$091909 & 319.4629 & $-$\phantom{0}9.3191 & 24.9\phantom{0}$\pm$\phantom{0}2.1\phantom{0} & 1.83\phantom{0}$\pm$0.85\phantom{0} & 16.7\phantom{00}$\pm$\phantom{0}1.2\phantom{00} & 0.616\phantom{0}$\pm$0.061\phantom{0} & \phantom{0}6.938$\pm$0.040 & 60205.6$\pm$\phantom{00}7.3 \\
J214148.2+043820 & 325.4508 & \phantom{$-$}\phantom{0}4.6388 & \phantom{0}9.2\phantom{0}$\pm$\phantom{0}4.2\phantom{0} & 3.48\phantom{0}$\pm$0.99\phantom{0} & 13.9\phantom{00}$\pm$\phantom{0}3.4\phantom{00} & 0.29\phantom{00}$\pm$0.12\phantom{00} & \phantom{0}5.817$\pm$0.059 & 60537.9$\pm$\phantom{00}5.0 \\
J220242.1+043638 & 330.6754 & \phantom{$-$}\phantom{0}4.6105 & \phantom{0}2.8\phantom{0}$\pm$\phantom{0}1.0\phantom{0} & 0.40\phantom{0}$\pm$0.50\phantom{0} & \phantom{0}6.8\phantom{00}$\pm$\phantom{0}1.3\phantom{00} & 0.635\phantom{0}$\pm$0.066\phantom{0} & \phantom{0}8.888$\pm$0.087 & 60515\phantom{.0}$\pm$\phantom{0}26\phantom{.0} \\
\enddata
\end{deluxetable*}